\documentclass[a4paper,fleqn,usenatbib]{mnras}

\usepackage{graphicx}
\usepackage[percent]{overpic}
\usepackage{lipsum}
\usepackage{multirow}
\usepackage{color}
\usepackage{rotating}
\usepackage{mwe,tikz}
\usepackage[percent]{overpic}
\usepackage{mathtools}
\usepackage{physics}
\usepackage{amsmath}
\usepackage[utf8x]{inputenc}
\usepackage{hyperref}
\usepackage{longtable}
\usepackage{wasysym}
\usepackage[normalem]{ulem}
\hypersetup{
    colorlinks=true,
    linkcolor=blue,
    filecolor=magenta,      
    urlcolor=blue,
}
\usepackage{soul}
\usepackage{pdflscape}
\usepackage{longtable}

\newcommand{\msun}{${\rm M_{\sun}}$}

\def\ltsima{$\; \buildrel < \over \sim \;$}
\def\simlt{\lower.5ex\hbox{\ltsima}}
\def\gtsima{$\; \buildrel > \over \sim \;$}
\def\simgt{\lower.5ex\hbox{\gtsima}}
\def\kms{{\rm\,km\,s^{-1}}}
\def\pc{{\rm\,pc}}
\def\kpc{{\rm\,kpc}}

\def\msun{{\rm\,M_\odot}}
\def\lsun{{\rm\,L_\odot}}


\def\deg{^\circ}

\def\Gyr{{\rm\,Gyr}}

\def\sLz{{\sigma_{\rm L_z}}}
\def\sv{{\sigma_{\rm v_{los}}}}

\def\ltsima{$\; \buildrel < \over \sim \;$}
\def\gtsima{$\; \buildrel > \over \sim \;$}

\defcitealias{MalhanCocoonDetection2019}{M19a}
\defcitealias{Malhan2019_GD1_Kshir}{M19b}
\defcitealias{Bonaca2019Jhelum}{B19}
\defcitealias{Carlberg2018Density_Structure}{C18}
\defcitealias{Malhan2018PotentialGD1}{MI19}

\def\cocoon{{\it\,cocoon\,}}
\def\insitu{{\it\,in situ\,\,}}
\definecolor{indiagreen}{rgb}{0.07, 0.53, 0.03}
\definecolor{darkspringgreen}{rgb}{0.09, 0.45, 0.27}

\title[Probing the nature of dark matter with streams]{Probing the nature of dark matter with accreted globular cluster streams}

\author[Malhan, Valluri \& Freese]{
Khyati Malhan$^{1}$\thanks{E-mail: khyati.malhan@fysik.su.se},
Monica Valluri$^{2}$,
Katherine Freese$^{1,3,4}$
\\
$^{1}$The Oskar Klein Centre, Department of Physics, Stockholm University, AlbaNova, SE-10691 Stockholm, Sweden\\
$^{2}$Department of Astronomy, University of Michigan, Ann Arbor, MI, 48109, USA\\
$^{3}$Theory Group, Department of Physics, The University of Texas at Austin, 2515 Speedway, C1600, Austin, TX 78712-0264, USA\\
$^{4}$Leinweber Center for Theoretical Physics, Department of Physics, University of Michigan, Ann Arbor, MI 48109, USA
}

\date{Accepted 2020 November 12. Received 2020 September 16; in original form 2020 May 28}

\begin{document}
\label{firstpage}
\pagerange{\pageref{firstpage}--\pageref{lastpage}}
\maketitle

\begin{abstract}
The steepness of the central density profiles of dark matter (DM) in low-mass galaxy halos (e.g. dwarf galaxies) is a powerful probe of the nature of DM. We propose a novel scheme to probe the inner profiles of galaxy subhalos using stellar streams. We show that the present day morphological and dynamical properties of {\it accreted} globular cluster (GC) streams -- those produced from tidal stripping of GCs that initially evolved within satellite galaxies and later merged with the Milky Way (MW) -- are sensitive to the central DM density profile and mass of their parent satellites. GCs that accrete within   {\it cuspy} CDM subhalos produce streams that are physically wider and dynamically hotter than streams that accrete inside {\it cored} subhalos. A first comparison of MW streams ``GD-1'' and ``Jhelum'' (likely of accreted GC origin) with our simulations indicates a preference for cored subhalos. If these results hold up in future data, the implication is that either the DM cusps were erased by baryonic feedback, or their subhalos naturally possessed cored density profiles implying particle physics models beyond CDM. Moreover, accreted GC streams are highly structured and exhibit complex morphological features (e.g., parallel structures and ``spurs"). This implies that the accretion scenario can naturally explain the recently observed peculiarities in some of the MW streams. We also propose a novel mechanism for forming ``gaps" in stellar streams when the remnant of the parent subhalo (which hosted the GC) later passes through the GC stream. This encounter can last a longer time (and have more of an impact) than the random encounters with DM subhalos previously considered, because the GC stream and its parent subhalo are on similar orbits with small relative velocities. Current and future surveys of the MW halo will uncover numerous faint stellar streams and provide the data needed to substantiate our preliminary tests with this new probe of DM.

\end{abstract}

\section{Introduction}\label{sec:Introduction}

A very strong prediction of the $\Lambda$CDM cosmological framework is that galaxy halos (irrespective of their sizes) should possess dark matter (DM) distributions with very steeply rising inner density profiles of the form $\rho_{\rm DM} \propto r^{-\gamma}$ (with $\gamma \approx1$, \citealt{Dubinski1991, NFW1996}). If the existence of {\it cuspy} DM density profiles are confirmed by observations, it would support the hypothesis that the DM particle is non-relativistic (``cold''), collisionless and weakly interacting \citep{White_CDM_candidate1978, Blumenthal1984}. 

One of the best cosmic sites to test this prediction of CDM is in Local Group dwarf galaxies (c.f. \citealt{Evans2004, Koch2007, Walker2010, Hooper2015}). Dwarf galaxies are observed to be extremely DM dominated (with $M/L \approx 10$--$1000 \msun/\lsun$, \citealt{Mateo1998, Simon2007, Battaglia_2008, Koposov_2011}).  Therefore one can measure and subtract off the stellar component from the total dynamical mass to obtain good estimates of the DM density profiles of these galaxies. While a few dwarfs have been found to possess cuspy DM profiles (e.g., Draco, \citealt{Jardel2013DracoCusps, Read2018Draco} and Sculptor,  \citealt{Amorisco2012Sculptor}), the majority of them strongly favour DM distributions with much shallower inner density profiles. Such  {\it cored} central density distributions ($\rho_{\rm DM}(0) \approx {\rm constant}$) have now been measured in low surface brightness disks \citep{Moore1994, Burkert1995, deBlok2001}, and various dwarf satellites of the Milky Way (MW). The latter include, for example, Ursa Minor \citep{Kleyna2003UrsaMinorCore}, Fornax \citep{Goerdt2006Fornax, Walker2011, Cole_2012_fornax, Pascale2018}, NGC 6822 \citep{Weldrake2003} and Eridanus II \citep{Contenta2018EridanusII}. 

Hydrodynamic simulations have shown that this apparent discrepancy between theoretical prediction and observations can possibly be explained, within the CDM paradigm, by taking into account baryonic processes such as episodic (``bursty") star formation in the early history of a galaxy that can effectively `heat-up' DM and transform primordial cusps into cores \citep{Navarro1996_cusp2core, Read2005cusp2core, Pontzen2012, Nipoti2015, Read_core_2016, Wheeler2018}. However, classical and ultra faint dwarf galaxies ($M_*/M_{\rm halo}\simlt 10^{-3}$) have formed so few stars over their lifetimes, that it is challenging to invoke baryonic feedback (e.g. from stellar winds and supernovae) as the main mechanism responsible for drastically transforming their inner DM distributions \citep{lazar_etal_20}. 

An alternative solution to this problem is to invoke physics beyond the standard CDM paradigm, since some DM theories naturally predict cored density profiles on galactic/sub-galactic scales, while remaining consistent with CDM on larger scales. These dark sector models
 include warm DM \citep{Bond_WDM_1980, Boyarsky_WDM_cores_2009, Avila_Reese_2001_WDM} and ultra-light DM, a.k.a. fuzzy DM \citep{Hui2017_ULDM_FDM}. Other theories consider differences in the  dynamical behaviour of DM (e.g. superfluid DM, \citealt{Berezhiani2018}), or its interaction strength (e.g. self-interacting DM, \citealt{Spergel2000SIDM, Elbert2015, SIDM_review2018}). 

Although the measurement of the density profiles of DM halos can provide strong constraints on the nature of the DM particle, most observed dwarf galaxies are too far away for us to obtain accurate stellar proper motions with currently available technology. With only line-of-sight velocities for hundreds (or at most a few thousand) of stars per system, stellar dynamical determinations of the DM density profiles are limited by the well known mass-anisotropy degeneracy \citep{binney_mamon_82}. Therefore, finding novel ways to probe the central DM densities of low-mass galaxy halos is important. {\it In this paper we propose to use the structural and dynamical properties of stellar streams --- in particular those produced by the tidal stripping of globular  clusters (GCs) \footnote{GCs are dense and old star clusters (formed at redshifts $z\sim 2-4$) with $M\sim10^5\msun$ and a physical sizes of a few tens of pc \citep{Whitmore1999} that reside in the halos of galaxies.} --- as one such novel tracer. }

More than $70$ stellar streams have been discovered in the halo of the MW (many lying within galactic distances of $\sim50\kpc$, c.f. \citealt{Helmi2020_review} and references therein), and for most of these systems we now have accurate proper motion information \citep{Malhan_Ghostly_2018, Shipp2019, Ibata_Norse_streams2019} from the ESA/Gaia catalogue \citep{GaiaDR2_2018_Brown, GaiaCollab2018kinematics}. Streams are generally categorised as stellar debris produced by the tidal disruption of either dwarf galaxies  \citep[e.g. the Sagittarius stream,][]{Ibata2001Sgr, Majewski2003} or GCs   \citep[e.g., Palomar-5, NGC 5466 streams,][]{Odenkirchen2001_Pal5,Grillmair_NGC5466_2006} as they orbit the MW potential. Since GCs are compact systems, they result in streams that are narrow and nearly one dimensional (a few tens of pc wide, c.f. \citealt{GrillmairCarlin2016} and references therein), in contrast to dwarf galaxy streams which are broader (a few hundreds of pc wide). Quite surprisingly, recent observations have  revealed that some GC streams possess multiple  morphological features: parallel structures,  ``spurs'' and broad ``cocoon'' components \citep{Malhan_2018_PS1, WhelanBonacaGD12018, MalhanCocoonDetection2019, Bonaca2019Jhelum, Malhan2019_GD1_Kshir, deBoer2020, Shipp2020_Pal13spur, Li2020_Atlas_Aliqa}. These complex features were not previously observed or predicted, and are very hard to reconcile with the models of tidal disruption of \insitu GCs (GCs that were likely formed early in the MW's history and whose evolution is entirely determined by the MW potential). Simultaneously, cosmologically motivated simulations suggest that these structures could arise if the progenitor GC of these streams were accreted along with their parent dwarf galaxies in which they {evolved}, where they experienced ``pre-accretion'' tidal stripping \citep{Carlberg2018-StreamSimulation, Carlberg2018Density_Structure}. {\it As we will show, this latter scenario provides a novel tracer of the nature of DM since the ``pre-accretion'' phase of tidal stripping of the GC is extremely sensitive to the central DM density profile of its parent dwarf galaxy.} 

The idea of GCs accreting onto the MW within their parent galaxy subhalos $(M_{\rm halo}\sim10^{8-10}\msun)$ is a well motivated scenario. While GCs have long been known to exist around all intermediate to high mass galaxies, recent deep imaging surveys have detected bonafide GC systems in many dwarf galaxies as well (e.g., \citealt{Leaman2013, Phipps2019}, also \citealt{Forbes2018_GC_halo-mass} and references therein). Under the hierarchical formation scenario, several tens of dwarf galaxies (many hosting GCs) must have accreted onto the MW galaxy in the past \citep{Searle1978, Garrison-Kimmel2014, Renaud2017, Kruijssen2020_Kraken}. Evidence for this includes the ongoing merging of the Sagittarius dwarf galaxy \citep{Ibata2001Sgr, Majewski2003}, where the GCs that were formerly members of this dwarf are now distributed along its stream in the Galactic halo \citep{Bellazzini2020}. There is also evidence for the possible merger of two dwarf galaxies ``Gaia-Enceladus/Gaia-Sausage'' \citep{Belokurov2018, Helmi2018} and ``Sequoia'', both found to be associated with numerous GCs in action space \citep{Myeong2019}, and the dynamical association between the massive Cetus stream and the star cluster NGC 5824 \citep{Chang2020_C}. All of these observations provide smoking gun evidence that a significant fraction of GCs in the MW were accreted from dwarf satellites. Under this framework, it is therefore reasonable to conjecture that several of the dynamically cold stellar streams observed in the Galactic halo (of likely GC origin) were also originally accreted onto the MW as part of the hierarchical buildup of the galaxy. 

Motivated by these recent developments, we propose a new way to probe DM using the structural and dynamical properties of the accreted GC streams. This paper is organized as follows. Section~\ref{sec:f_stream} illustrates, as a baseline reference, the formation of tidal streams from \insitu GCs.  Section~\ref{sec:Accretion_model} describes the method and setup used to simulate accreted GC streams. The main results of our simulations are presented in Section~\ref{sec:Results} (and summarized in Figs.~\ref{fig:Fig_comparison} and \ref{fig:Fig_stream_profiles}). A few additional simulations are presented in Section~\ref{sec:discussion} and a comparison with some observations of the MW streams is given in Section~\ref{sec:comp_with_obs}. Finally, we conclude in Section~\ref{sec:summary_Conclusion}. 

\begin{figure}
\begin{center}
\includegraphics[width=\hsize]{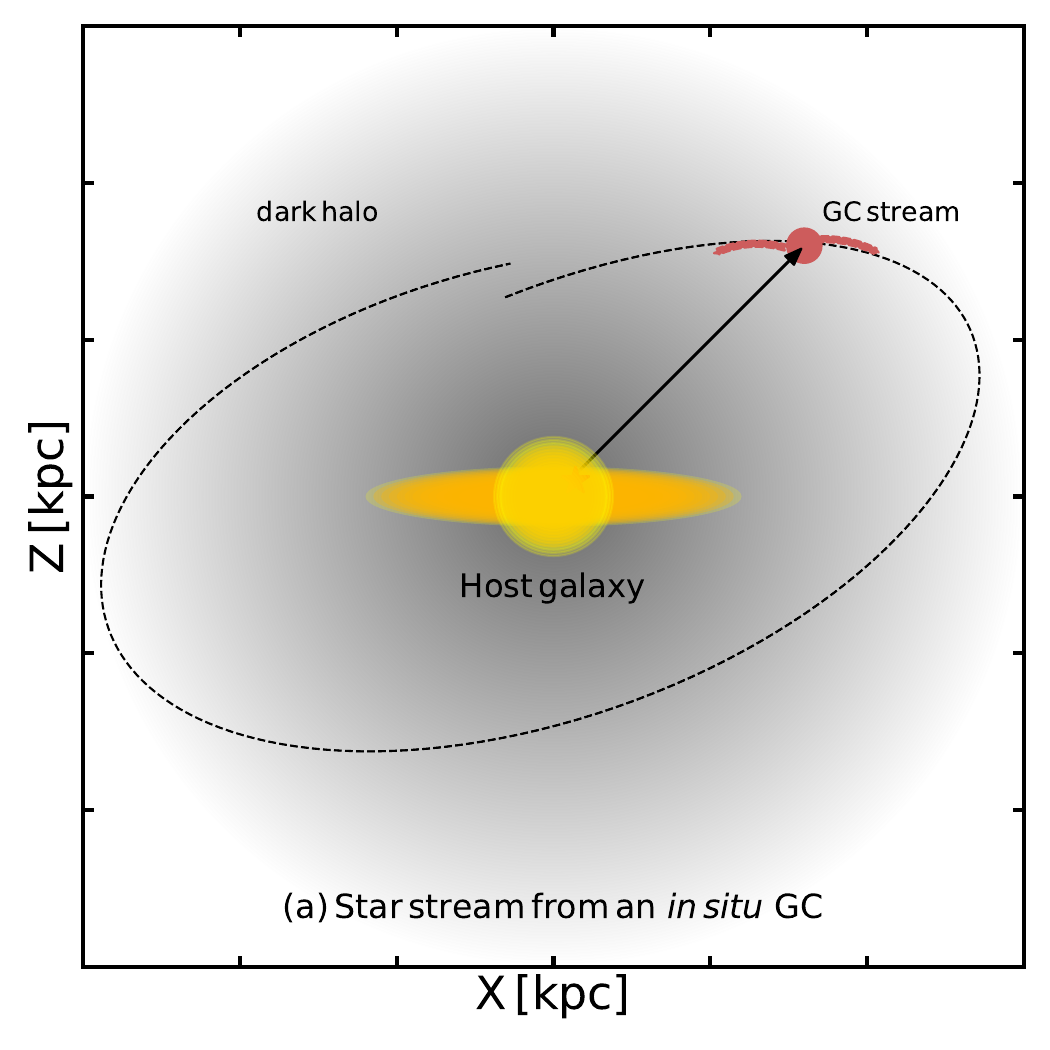}
\includegraphics[angle=0, viewport= 10 0 575 210, clip,width=1.0\hsize]{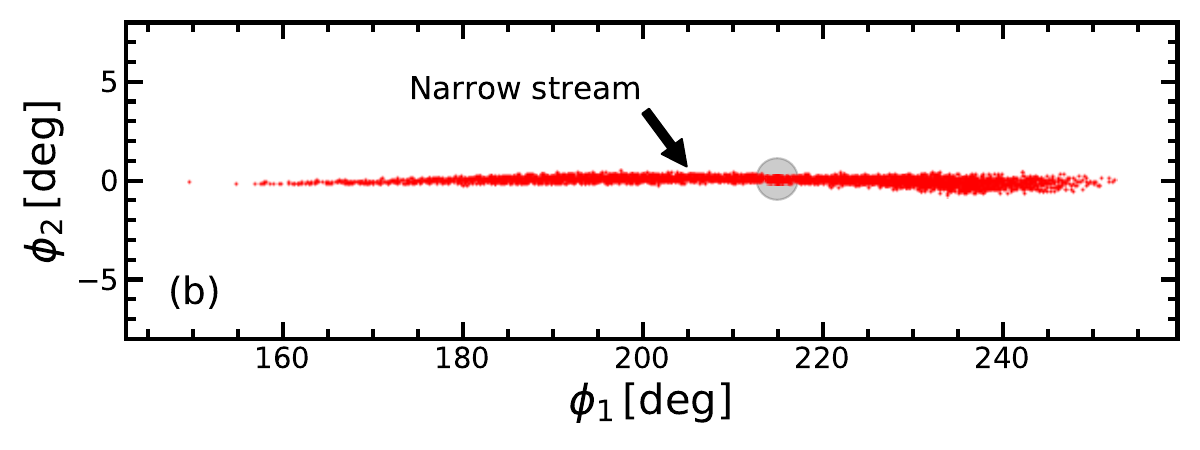}
\end{center}
\vspace{-0.5cm}
\caption{a) Schematic diagram showing the formation of a stellar stream from an \insitu GC orbiting a MW type host galaxy potential (Sec.~\ref{sec:f_stream}). b) Simulated N-body stream inside this host galaxy, with the plot shown in a rotated coordinate system that is aligned along the stream. The gray circle marks the location of the surviving GC progenitor. Tidal disruption of an \insitu GC gives rise to a narrow and dynamically-cold stream. }
\label{fig:Fig_simpleNbody}
\end{figure}
\section{Streams from {\it in situ} globular clusters}\label{sec:f_stream}

We begin with a brief discussion of the formation and characterization of tidal stellar streams formed by \insitu GCs -- those that have always resided in our Galaxy -- as they orbit the halo. This discussion will serve as a baseline for the comparison with streams from accreted GCs, presented in subsequent sections.

Most GC streams are observed to be narrow, nearly one-dimensional structures on the sky. Their morphologies and kinematics are reasonably well reproduced by N-body simulations in which the escaping stars end up lying close to the orbit of the progenitor \insitu GC, with only a small spread in their energies and angular momenta that depends on the mass of the GC and the nature of the orbit \citep{Dehnen2004, Varghese2011, Bowden2015}.  
 
For our N-body simulations of the formation of \insitu GC streams we adopted a realistic Galactic potential from \citet{Dehnen1998Massmodel} (their model 1), which contains a thin disk, a thick disk, interstellar medium, bulge and DM halo. Throughout this paper, we use the same Galactic potential to model the host Galaxy. The GC was modeled with a King profile \citep{King1966}, as these profiles are broadly consistent with the observed GC systems \citep{Sollima2017}. The mass of our fiducial GC was set at $M_{\rm GC}=5\times10^4\msun$\footnote{ The model also used a central potential depth of $W=2.5$ and tidal radius of $r_{\rm t}= 0.1 \kpc$.}, comparable to the values of  known clusters \citep{Simpson2017_Lowwmass_GC, Baumgardt_2016}, and consistent with the values previously adopted to model GC streams \citep{Bowden2015, Thomas_Pal52016}. The star particles had individual masses of $5\msun$ and softenings of $2\pc$ - suitable for such simulations \citep{Thomas_Pal52016}. The GC was launched from an initial galactocentric radius of $60\kpc$ with a random Galactic longitude and latitude. The reason for adopting this particular initial galactocentric distance is provided in Section~\ref{subsec:Nbody_simulation_setup}. The galactocentric velocity of the GC was chosen such that the resulting orbit is approximately circular, however the tangential direction of motion was chosen randomly. The simulation was evolved for $T=8\Gyr$ using the \texttt{GyrfalcON} integrator \citep{Dehnen_NEMO_2002} from the NEMO package \citep{Teuben_1995}. 

Figure~\ref{fig:Fig_simpleNbody}a is a schematic diagram showing the formation of one such stream and Figure~\ref{fig:Fig_simpleNbody}b shows the star particles in the N-body stream plotted in a coordinate system with $\phi_1$ being the angle along a great-circle that is aligned with the stream, and $\phi_2$ the angle perpendicular to the stream on the plane of the sky. The transverse physical width of this stream was estimated by fitting a polynomial to the star particles in  $\phi_1 - \phi_2$ space and then computing the dispersion in the $\phi_2$ direction. This angular dispersion was multiplied by the mean distance to the stream to obtain a physical width $w$. This calculation was done in small segments along the entire length of the stream, and averaged to yield a width $w \approx 40\pc$. During this process, the region containing the remnant of the GC was masked, and hence ignored in the computation. 

To characterise the stream in terms of its dynamical properties it is also useful to determine the dispersion in the $z-$component of angular momentum ($\sLz$) and the velocity dispersion in the line-of-sight direction ($\sv$). As expected (for this axisymmetric Galactic potential) $L_z$ is conserved for all the particles so the stream shows a tiny dispersion of $\sLz\approx 8\kms\kpc$ and dispersion in los velocity of $\sv\approx 0.5\kms$. These dispersions come from the velocity dispersion of stars in the progenitor GC. Hereafter, we will refer to $\sLz$ and $\sv$ as {\it dynamical width estimators}. As with the computation of $w$, these values were calculated locally in small segments along the entire length of the stream, and then averaged over all the segments (with the remnant cluster masked). 

For \insitu GC streams, the resulting phase-space distribution of stars (inside the host galaxy) will depend only on the gravitational potential of the host, the initial mass and core radius of the GC, and its orbit inside the Galaxy. To explore the dependence on these parameters, additional simulations were undertaken employing progenitors of different masses ($M_{\rm GC}=3-10\times10^4\msun$) and sampling over orbits of different eccentricities. The parameters $(w, \sLz, \sv)$ of all these simulated \insitu GC streams are shown as the red points in Figure~\ref{fig:Fig_comparison}, and the corresponding uncertainties in the measurements are the standard deviations in the estimates of the parameters in different segments of the stream. 

It can be readily observed that, overall, the \insitu streams form narrow and dynamical cold structures and differ only slightly in terms of their physical  properties. In order to quantify the variance in $(w, \sLz, \sv)$  for all the simulated streams, we fitted a simple Gaussian function with mean $\langle{x}\rangle$ and intrinsic dispersion $\sigma_{x}$ to each of the estimators using an MCMC exploration for a generative model. Here, $x$ refers to one of the quantities $(w,\sLz,\sv)$, $x_i$ and $\delta_i$ are the measurements and associated uncertainties for that quantity obtained for each of the $n$ simulated streams. The log-likelihood function used to obtain the parameters of the Gaussian function can be expressed as 
\begin{equation}\label{eq:likelihood}
\begin{aligned}
    \ln \mathcal{L} &= \sum_i^n \left[ -\ln (\sqrt{2\pi}\sigma_i) -0.5\dfrac{(x_i - \langle{x}\rangle)^2}{\sigma_i^2}\right]\,,\\
     {\rm with}\,\, \sigma^2_i &=\sigma^2_{x} + \delta^2_i.
\end{aligned}
\end{equation}
For the \insitu\, GC streams (red points in Figure~\ref{fig:Fig_comparison}), we found $w\sim45\pm15 \pc, \sLz\sim 9\pm1 \kpc\kms,\sv\sim0.7\pm0.2\kms$. These best fit Gaussian functions are also shown (in red) in the lower part of each panel of Figure~\ref{fig:Fig_comparison}.

\section{Streams from Globular Clusters accreted from dwarf galaxies}\label{sec:Accretion_model}

\begin{figure}
\begin{center}
\includegraphics[width=\hsize]{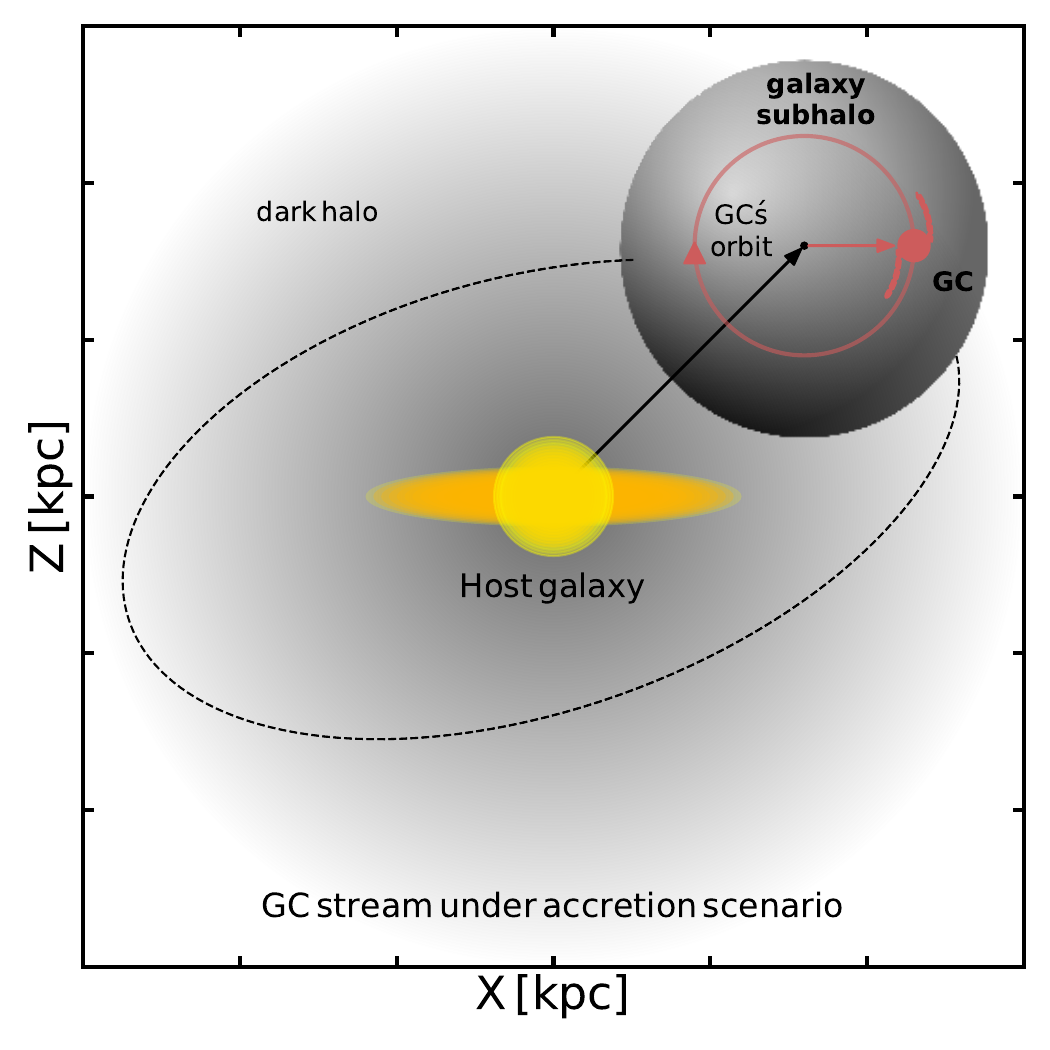}
\end{center}
\vspace{-0.5cm}
\caption{Schematic diagram showing the formation of a stellar stream produced by a GC that accretes within its parent subhalo (Sec.~\ref{sec:Accretion_model}). In contrast to the case of the \insitu GC presented in Figure~\ref{fig:Fig_simpleNbody}, streams produced under this accretion scenario possess complex structural morphologies, as shown in Figure~\ref{fig:Fig_evolution}.}
\label{fig:Fig_accretion_scheamtic}
\end{figure}

$\Lambda$CDM motivated cosmological simulations have recently shown that streams produced from GCs that evolve within their parent subhalos, and later accrete onto the host galaxy (schematically shown in Fig.~\ref{fig:Fig_accretion_scheamtic}), possess complex spatial morphologies and density structures (\citealt{Carlberg2018Density_Structure}, \citetalias{Carlberg2018Density_Structure} hereafter). In particular, a narrow stream component (like the one shown in Fig.~\ref{fig:Fig_simpleNbody}b) is expected to be surrounded by a substantial number of stars distributed in the form of a broader and more diffuse stream component. The broader component arises from the stars that were tidally stripped from the GC while it was still inside the subhalo, prior to accretion onto the host galaxy. In particular, the \citetalias{Carlberg2018Density_Structure} study implies that the gravitational potential of the parent subhalo significantly influences the final morphology of the accreted GC stream. Going a step further, our motivation in this paper is to test whether the present day physical properties of accreted GC streams can in fact encode information about the gravitational potential (or the DM distributions) inside their parent galaxy subhalos. 

In contrast with the case of the \insitu GC presented in Section~\ref{sec:f_stream}, the overall evolution and phase-space distribution of the stream stars in this case is expected to depend on the initial physical conditions of the subhalo (e.g., its mass, inner density distribution), the orbit of the GC within the subhalo and the orbit along which the subhalo accretes onto the host. A comprehensive investigation of this scenario requires iteration over different subhalo models (varying mass and inner DM density distributions), and various orbital configurations for both GCs (inside subhalos) and subhalos (inside host galaxy). A complete exploration of this enormous parameter space is a Herculean task, and therefore, for this first study we restrict ourselves to a limited number of N-body simulations. Despite not exhaustively exploring all possible parameters, the limited  set of controlled experiments we present is highly informative about whether or not the physical properties of accreted GC streams can be used to probe the density profiles (e.g., cusp or core) of their parent subhalos.

\begin{figure}
\begin{center}
\includegraphics[width=\hsize]{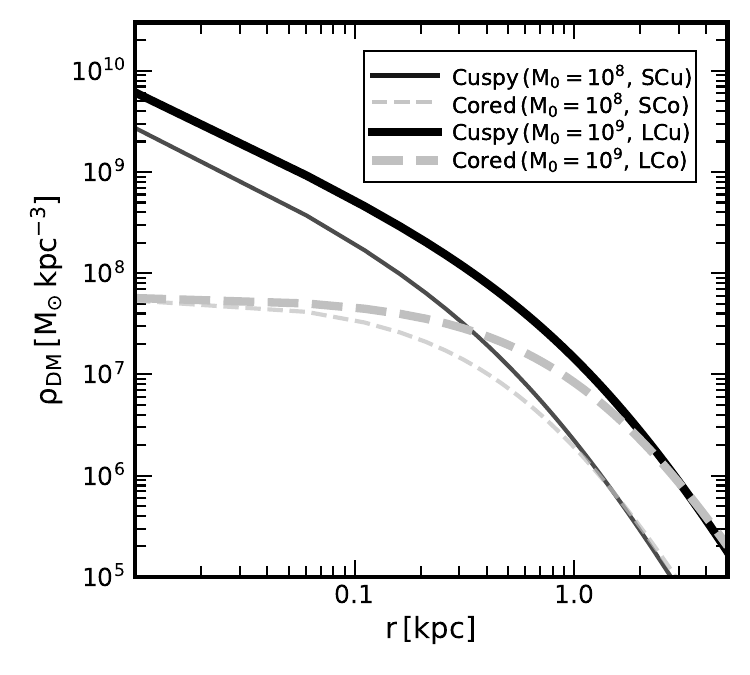}
\end{center}
\vspace{-0.6cm}
\caption{Radial density profiles for the subhalo models studied in our simulations (Sec.~\ref{subsec:subhalo_models}).}
\label{fig:Fig_subhalo_models}
\end{figure}
\subsection{Mass models for subhalos}\label{subsec:subhalo_models}

Two different DM density profiles for subhalo are studied -- cuspy profiles (such as those predicted by CDM simulations, \citealt{Dubinski1991, NFW1996}) and  cored profiles (that have been suggested by, for instance, \citealt{Gilmore_dsph_core2007} based on modelling of observations in dSph galaxies). To construct these subhalos models, we use spherically symmetric Dehnen models \citep{Dehnen1993} that are expressed as
\begin{equation}\label{eq:density_dehnen}
\rho (r) =\frac{(3-\gamma)M_0}{4\pi r^3_0} \Big( \frac{r}{r_0} \Big)^{-\gamma} \Big(1+ \frac{r}{r_0} \Big)^{\gamma-4},
\end{equation}
where $M_0, r_0,- \gamma$ are the mass, scale radius, and the logarithmic slope of the inner density profile of the subhalo, respectively. 

We use Dehnen spheres because these models can accommodate both cuspy and cored DM density profiles. The cuspy subhalos are modeled by setting $\gamma=1$, while for cored subhalos we set $\gamma=0$. From equation~\ref{eq:density_dehnen} it is clear that the former results in $\rho \propto r^{-1}$ for $r<<r_0$, much like in the NFW profile \citep{NFW1996}, while the latter has a nearly constant density region at small radii. At larger radii ($r>>r_0$), the  Dehnen density profile goes as $\rho \propto r^{-4}$, in contrast to the NFW profile for which $\rho \propto r^{-3}$, implying that, for the same halo mass, the outer regions of cuspy Dehnen models would disrupt at a slightly different rate inside the host galaxy, than the actual NFW subhalos. We expect this difference not to effect our study significantly since most GCs in our simulations orbit the subhalos with radii smaller than the half-mass radii of the subhalos.

To create a cuspy subhalo of a given mass $M_0$, we find a value for $r_0$  in equation~\ref{eq:density_dehnen} for which the resulting density profile is close to the NFW profile (for $M_0 \equiv M_{200}$). To obtain the corresponding cored subhalo model, we simply set $\gamma=0$ in equation~\ref{eq:density_dehnen} and adopt the previously chosen values of $M_0,r_0$. The half mass radius of cuspy Dehnen models is $r_{1/2} \approx 2.414r_0$, while for cored models it is $r_{1/2} \approx 3.8473r_0$.  GCs were typically launched at distances smaller than  $r_{1/2}$, but in a few cases had apocenter radii beyond this. The condition of launching GCs within $r_{1/2}$ was inspired by the present day spatial distributions of GCs around the dwarfs Fornax and Eridanus II (the only known dwarfs in the Local Group that host GC population and have masses similar to the $M_0$ values we try). In these systems,  the GCs are observed to be situated close-to, or within, the central regions of the parent dwarfs (\citealt{Mackey_Gilmore2003, Crnojevi2016}). The orbital conditions of the GCs inside the subhalos are further detailed in Section~\ref{subsec:Nbody_simulation_setup}.

In total we study four subhalo models, 1) SCu (small/cuspy) model: $\{M_0,r_0,\gamma\}=\{10^8\msun, 0.75\kpc,1\}$; 2) SCo (small/cored) model: $\{M_0,r_0,\gamma\}=\{10^8\msun, 0.75\kpc,0\}$; 3) LCu (large/cuspy) model: $\{M_0,r_0,\gamma\}=\{10^9\msun, 1.60\kpc,1\}$ and 4) LCo (large/cored) model: $\{M_0,r_0,\gamma\}=\{10^9\msun, 1.60\kpc,0\}$. The density profiles of these subhalos are shown in Figure~\ref{fig:Fig_subhalo_models} (also see Table~\ref{tab:parameters}). 

The chosen range of $M_0(\approx 10^{8-9}\msun)$ is similar to the measured masses of some of the dwarf galaxies that host GCs (c.f. \citealt{Forbes2018_GC_halo-mass}). For example,  Eridanus II \citep{Bechtol2015}, a MW dwarf (D$\approx 410\kpc$) with an estimated halo mass of $M_{\rm halo}\sim 10^{8-9}\msun$ \citep{Contenta2018EridanusII} also hosts a star cluster \citep{Crnojevi2016}. At present, the lowest mass dwarf galaxy known to host a GC is And XXV  (satellite of Andromeda, \citealt{Cusano2016}), with $M_{\rm halo}\simgt 0.13\times10^8\msun$ \citep{Forbes2018_GC_halo-mass} -- smaller than the lowest subhalo mass value we use for our tests.  Another reason for adopting this range of $M_0$ values is motivated by the recent study of the phase-space correlated ``Kshir'' and ``GD-1'' streams (\citealt{Malhan2019_GD1_Kshir}, \citetalias{Malhan2019_GD1_Kshir} hereafter), where the conjoint analysis of these structures, under the accretion framework, constrained the mass of their parent subhalo at $\approx 10^{8-9}\msun$.

\subsection{Modeling the Globular Cluster}\label{subsec:GC_system}

For simplicity, we model the accretion scenario by simulating evolution of only a single GC of mass $M_{\rm GC}=5\times 10^4\msun$ inside a given DM subhalo. Similar GC models have been previously used in the studies that analyzed the dynamical evolution of GCs in (isolated) subhalo systems \citep{Amorisco2017, Contenta2018EridanusII}. In addition, this choice of mass is small enough to not significantly alter the density profile of the subhalo.  

Admittedly, while it is not realistic to assume that a DM subhalo could contain only a GC and no extended stellar distribution (since any subhalo with enough gas to form a GC would also form an extended stellar population), for the majority of our simulations we ignore a possible extended stellar component. Since most classical dwarfs (including those with GCs) are highly DM dominated (M/L$ >100\msun/\lsun$), ignoring extended baryonic components is unlikely to significantly alter our results. Nevertheless, in Section~\ref{sec:discussion} we run a few simulations where we also add an extended stellar component inside these subhalos, similar to those observed in dwarf galaxies.

\subsection{Theoretical expectations for streams from accreted GCs \label{subsec:theory}}

Before proceeding to execute and analyse the N-body simulations, it is useful to formulate theoretical expectation on the physical properties of the accreted GC streams. The understanding of the dynamics that we present in this section will also hold utility in carefully choosing the parameters to perform simulations in Section~\ref{subsec:Nbody_simulation_setup}. Here we review some basic ideas of tidal disruption. Readers familiar with this topic may skip to Section~\ref{subsec:Nbody_simulation_setup}.

The spatial distribution and dynamical evolution of a GC in {\it any} galaxy (whether MW-like as in Section \ref{sec:f_stream}, or a dwarf subhalo as in Section 3) is sensitive to the underlying DM distribution of the galaxy hosting the GC. In either case, two factors affect the evolution of a GC: (i) the tidal forces from the galaxy hosting the GC and (ii) dynamical friction from the host's DM distribution, which causes the orbit of the GC to decay and the GC to sink to the center of the host \citep{gnedin_ostriker_97}. 

To understand the tidal effects of the host galaxy on the GC, we can estimate the GC's tidal radius $\tilde{r}_J$ (the spherical distance from the center of the GC beyond which the GC's material is tidally stripped away by its host.) More generally, consider any satellite system in a circular orbit with radius $r_p$ about the center of any host galaxy, where the average mass density of the host in the region interior to the satellite's orbit is $\rho_H(r<r_p)$.
It is straightforward to show that  the tidal radius of the satellite $\tilde{r}_J$ is given by the relation $\rho_S(\tilde{r}<\tilde{r}_J) \simeq \alpha\rho_H(r<r_p)$. Here, $\rho_S(\tilde{r}<\tilde{r}_J)$ is the average density of the satellite within the tidal radius, and $\alpha$ is a constant between $2-4$ that depends  on the mass distribution of the host and, for non-circular orbits, the orbit of the satellite, e.g. \citealt{sparke_gallagher_07}). This general rule of thumb applies for any host galaxy and satellite system -- a GC orbiting in a cuspy or cored subhalo, or a dwarf galaxy orbiting the potential of the MW.  

This principle helps to explain why GCs evolving in cuspy subhalos are expected to disrupt more significantly than those evolving in cored subhalos, as suggested previously. Since the mean DM density within a cuspy subhalo is always greater than or equal to the density in cored subhalo (see Fig.~\ref{fig:Fig_subhalo_models}), the tidal radius $\tilde{r}_J$ of a GC in a cuspy subhalo at a given radius is always smaller than in a cored subhalo. This causes severe disruption of a GC on a short timescale (a few Gyrs). Moreover, cuspy subhalos also impose high dynamical friction that can cause the GC to sink towards the center of the dwarf galaxy (e.g., \citealt{Hernandez1998, Goerdt2006Fornax, Amorisco2017}). In contrast, the tidal forces in cored DM halos are compressive in nature \citep{valluri_93}, and therefore not tidally disruptive, allowing GC to survive for timescales $\simgt$ Hubble time \citep{Read2006, Inoue2009, Petts2016, Contenta2018EridanusII}. In addition, dynamical friction in cored halos is suppressed because $\rho_{\rm DM} \approx {\rm constant} \implies \nabla^2\Phi \approx {\rm constant}.$ \footnote{For $\rho_{\rm DM} \approx {\rm constant}$ the potential resembles a harmonic oscillator potential in which all the objects (GC and DM particles) have similar frequencies resulting in resonant interactions between the GCs and background DM particles that diminishes dynamical friction.} The observations that the GCs in dwarf galaxies like Fornax \citep{Goerdt2006Fornax, Walker2011, Cole_2012_fornax} and Eridanus II \citep{Contenta2018EridanusII} are not centrally concentrated, and the fact that they have survived for tens of Gyrs (typical ages of GCs), has been used to argue that these galaxies must have cored DM density distributions. 

Along the same lines, we expect GC streams in cuspy subhalos to be more significantly disrupted and fluffier than those evolving in cored subhalos. Furthermore, the accreted streams are in turn expected to be more significantly disrupted than \insitu GC streams, since the DM density in subhalos is always higher than the DM density of a host (MW-like) galaxy. Therefore, an \insitu GC traveling freely in the {host} galaxy experiences less tidal disruption.

The expected widths of tidal streams from a virialized distribution of stars in a satellite are determined by the both the virial theorem, (which sets the initial velocity dispersion of the stars), and the principle of conservation of phase space density required by the Liouville theorem. By the virial theorem, the velocity dispersion in a spherical subhalo is
\begin{equation}\label{eq:sphere_vel_disp}
  \overline{v^2} \propto \frac{GM_0}{r_0},
\end{equation}
where $M_0$ and $r_0$ are the mass and characteristic radius of the subhalo. The RMS velocity $\overline{v^2} $ of the parent subhalo sets both the physical width and velocity dispersion of the accreted GC stream because the evolution of tidal debris is governed by collisionless dynamics and must conserve its phase space density even after it is stripped (e.g., \citealt{Johnston2001}). The dynamical width estimators $\sLz$ and $\sv$ (previously described in Sec.~\ref{sec:f_stream}) should also measure the depth of the potential and its physical scale.  This implies that, in general, massive progenitors produce broader streams than low mass progenitors. 

However, the stellar streams from GCs that pre-evolve in subhalos can still be distinguished from dwarf galaxy streams. As we will show in Section~\ref{sec:Results}, although GCs are strongly disrupted during their evolution inside subhalos,  the tidal debris does not have time to virialize within the subhalo since the subhalo itself is undergoing tidal disruption. Consequently accreted  GC streams are not as broad or as dynamically hot as the streams that would arise from virialized and extended stellar populations in dwarf galaxies. Furthermore, since tidal disruption of the  GC continues after is liberated from the subhalo, the resultant streams contain both the narrow components (seen in \insitu GC streams) and the broader cocoon components (that result from the tidal disruption by the DM distribution in the subhalo). Broadening in streams can also occur if the potential of the host galaxy is aspherical (e.g., if the potential is flattened, \citealt{Erkal2016evolution_orbits}).  However, as we show in Section~\ref{subsec:accreted_streams_properties}, our final results stand robust to this effect.
\vspace{0.5cm}

\begin{table*}
\centering
\caption{Parameters for simulations (Sec.~\ref{sec:Accretion_model}). From left to right: the subhalo model, subhalo's mass $(M_0)$, the initialising radius of the GC within the subhalo ($r_p$), the local circular velocity inside the (isolated) subhalo ($v_{\rm circ}(r_p)$), the local radial velocity dispersion inside the subhalo ($\overline{v^2_r} (r_p)$), the initial velocity provided to the GC within the subhalo ($v_t$), and the ratio of the specific angular momentum of the GC's orbit within the subhalo ($L_0/L_{\rm circ}$).}
\label{tab:parameters}
\begin{tabular}{| l |c|c|c|c|c|c|c|c|c|c|c|c|}

\hline
\hline
Model & $M_0$ & $r_p$ & $v_{\rm circ}(r_p)$ & $\overline{v^2_r} (r_p)$ & $v_t$ & $L_0/L_{\rm circ}$  \\
& $[\msun]$  & $[\kpc]$ & $[\kms]$ & $[\kms]$ & $[\kms]$  \\
\hline
\hline\\

SCu (small cuspy)&$10^8$&0.5&11.5&7.3& $\pm17.6,\pm18.6,\pm19.6,\pm20.5$ & $\pm1.5,\pm1.6,\pm1.7,\pm1.8$\\\\
                &      &1.0&11.6&6.5& $\pm9.8,\pm10.8,\pm11.7,\pm12.7,$  & $\pm0.8,\pm0.9,\pm1.0,\pm1.1,$\\
                &      &   &    &   & $\pm13.7,\pm14.7,15.6$            & $\pm1.2,\pm1.3,1.4$\\\\
SCo (small cored)&$10^8$&0.5&7.3 &5.7& $\pm7.8,\pm8.8,\pm9.8,\pm10.8,$    & $\pm1.1,\pm1.2,\pm1.3,\pm1.5,$\\
                &      &   &    &   & $\pm11.7$                         & $\pm1.6$\\\\
                &      &1.0&8.8 &5.5& $\pm1.0,\pm2.0,\pm2.9,\pm3.9,$        & $\pm0.1,\pm0.2,\pm0.3,\pm0.4,$\\
                &      &   &    &   & $\pm4.9,\pm5.9,\pm6.8$              & $\pm0.6,\pm0.7,\pm0.8$  \\

\hline\\
LCu (large cuspy)&$10^9$&0.5&21.6&16.6& $\pm48.9,\pm49.9,\pm50.9,\pm51.8$& $\pm2.2,\pm2.3,\pm2.4,\pm2.4$ \\\\
                &      &1.0&24.7&16.0& $-40.1,-41.1,\pm42.1,\pm43.0$         & $-1.6,-1.6,\pm1.7,\pm1.7$\\
                &      &   &    &    & $44.0,45.0,46.0$                     & $1.8,1.8,1.9$\\\\
LCo (small cored)&$10^9$&0.5&10.5&12.0& $\pm14.7,\pm15.7,\pm16.7,\pm17.6,$ & $\pm1.4,\pm1.5,\pm1.6,\pm1.7,$\\
                &      &   &    &    & $\pm18.6,\pm19.6,\pm20.5,\pm21.5,$ & $\pm1.8,\pm1.9,\pm1.9,\pm2.0,$\\
                &      &   &    &    & $\pm22.5,\pm23.5,\pm24.5$        & $\pm2.1,\pm2.2,\pm2.3$\\\\
                &      &1.0&15.3&12.4& $\pm13.7,\pm14.7,\pm15.6,\pm16.6,$  & $\pm0.9,\pm1.0,\pm1.0,\pm1.1,$\\
                &      &   &    &    & $\pm17.6,\pm18.6,\pm19.6,\pm20.5,$  & $\pm1.2,\pm1.2,\pm1.3,\pm1.3,$\\
                &      &   &    &    & $\pm21.5,\pm22.5$              & $\pm1.4,\pm1.5$\\
\hline
\hline
\end{tabular}
\end{table*}

%
\subsection{N-body simulation setup}\label{subsec:Nbody_simulation_setup}

The simulations of accreted GCs were undertaken using the same setup as described in Section~\ref{sec:f_stream}, except that the GCs were placed inside DM subhalos which were also represented by N-body distributions. The mass and softening parameters of DM particles were  $750\msun$ and $20\pc$, respectively, while these parameters for the star particles were the same as in Section~\ref{sec:f_stream}. This choice of resolution is based on numerical tests that we undertook in order to reproduce the expected behaviour of GCs that otherwise evolve in isolated cored/cuspy systems (see Appendix~\ref{appendix:numerics}). It is worth remarking that since we employ live N-body models for both GCs and subhalos, we accurately model dynamical friction of the GC in the subhalo. However, since the host Galactic potential is smooth and static, the effects of dynamical friction on the subhalo are absent.  

GCs were launched at an initial distance $r_p$ from the center of the subhalos (only two values, $r_p = 0.5\kpc, 1\kpc,$ were used for all simulations).  These values of $r_p$ are the pericentric distances of the GCs inside the subhalo. Our trial simulations showed that placing GCs at locations $r_p << r_0$  within cuspy subhalos caused their instantaneous disruption (for the reasons provided in Sec.~\ref{subsec:theory}, also see Appendix~\ref{appendix:numerics}), either resulting only in physically broad streams (much like those observed from disrupted dwarf galaxies) or preventing GC stars from escaping the deep potential well of the subhalo altogether. Since we are interested in exploring GC streams that contain a narrow structural component (as seen in both simulations, \citetalias{Carlberg2018Density_Structure}, and observations, \citealt{MalhanCocoonDetection2019, Bonaca2019Jhelum}, \citetalias{MalhanCocoonDetection2019} and \citetalias{Bonaca2019Jhelum} hereafter), we employed those values of $r_p$ that produced thin components for both the cuspy and cored subhalos. The problem of instantaneous disruption of the GC was not encountered in cored subhalo cases even for $r_p << r_0$; however we used the same set of $r_p$ values in all the subhalo models to facilitate fairer comparisons. Appendix~\ref{appendix:small_rp} presents results of simulations for streams for $r_p << r_0$. 

After positioning the GC inside the subhalo, a tangential velocity $v_t$ is imparted to it in the rest frame of the subhalo. For a given subhalo model, we tried both positive and negative values of $v_t$. This was done to assess the dependence of stream characteristics on the sense (prograde or retrograde) of GC's orbit inside subhalo, with respect to the subhalo's orbit inside the host galaxy. This test is important as previous work has shown that tidal disruption of a satellite is sensitive to the sense of angular momentum of tracers within the satellite (see for e.g., \citealt{Read2006}). The range of $v_t$ values was set simply by demanding that in every simulation, the GC spends $\simgt3-4\Gyr$ inside the parent subhalo, before escaping into the host. This ``gestation period'' was assumed somewhat arbitrarily, however, it lets us define the upper bound on $|v_t|$ values. This upper bound (or the ``escape velocity'') is understandably different for different subhalo models, due to the difference in their gravitational potentials. The lower limit for $|v_t|$, in the cuspy models, comes from demanding that GCs do not instantly sink to the center and do not fully disrupt within the subhalo, so that post-accretion they can produce the narrow stream component. We found that to achieve the required ``gestation period'', one requires much higher values of $v_t$  for cuspy subhalos than for cored cases. For each subhalo model, we uniformly sample the values for $v_t$ at intervals of $\approx 1\kms$, resulting in $\approx 10-20$ different orbital configurations per subhalo model. These orbits can be parameterised in terms of the ratio of the specific angular momentum ($L_0=r_p \times v_t$) to the specific angular momentum of a circular orbit $(L_{\rm circ}=r_p \times v_{\rm circ} (r_p))$ at the same {\it radius, $r_p$}. This ratio at the time of initialisation of the GC inside the subhalo is provided in Table~\ref{tab:parameters}. 

Each subhalo+GC system was evolved for a time period of $T=8\Gyr$ in the host galaxy. This time interval was adopted since we are assuming a static Galactic potential  for the host, and it is generally assumed that the Galactic potential of the MW  has not changed significantly in the past $\sim8\Gyr$. 

All the subhalos were launched on the same circular orbit inside the host to facilitate comparison between different cases (however, in Section~\ref{sec:discussion} we discuss a few cases where eccentric orbits were employed  and the results were found to be similar to the circular cases). This orbit was similar to the one employed in Section~\ref{sec:f_stream} to model \insitu GCs. We undertook a few trial simulations for orbital radii less than $60\kpc$; however, we found that the cored subhalo models were quickly disrupted by galactic tides from the host\footnote{A cored subhalo with $M_0=10^9\msun$ at a galactocentric distance of $\sim15\,(40)\kpc$ was disrupted in $\sim 1\,(2) \Gyr$.}. In contrast, the cuspy subhalo models were never completely disrupted and always retained a bound remnant even after experiencing significant mass loss. This is because the steep inner density profiles of the cuspy halos makes them much more resilient to tides \citep{Kazantzidis2004, Onghia2010, Penarrubia2010, Penarrubia2017}, while the low-binding energies of cored models render them more susceptible to tidal destruction (c.f. \citealt{Du_2018}). For these reasons, we adopted an orbital radius of $60\kpc$ to ensure the survival times of both types of subhalos, at least until such time that the GC forms a tidal stream within the subhalo prior to the subhalos disruption. Even with orbits with such large mean spherical radii, the cored subhalos disrupted in $\approx 5\Gyr$ (see Appendix~\ref{appendix:mass_loss}, Fig.~\ref{fig:Fig_sub-halo_mass_evolution_eccentricites}). Nevertheless, this timescale seems to be sufficient for the subhalo to affect the dynamics of the GC (e.g., \citealt{Goerdt2006Fornax}, and also see Fig.~\ref{fig:Fig_NumericalConvergence}). 

Although we do not explore such cases in this work,  more massive (e.g., $M_0\sim10^{10}\msun$) cored subhalos could survive long enough to deliver their GCs to smaller galactic distances ($\sim10-30\kpc$), where most of the MW streams are observed. However, due to limitations in computational resources, we do not simulate subhalos with $M_0 >10^{9}\msun$ in this paper. We return to this point later in Section~\ref{sec:comp_with_obs}.

\begin{figure*}
\begin{center}
\includegraphics[width=0.97\hsize]{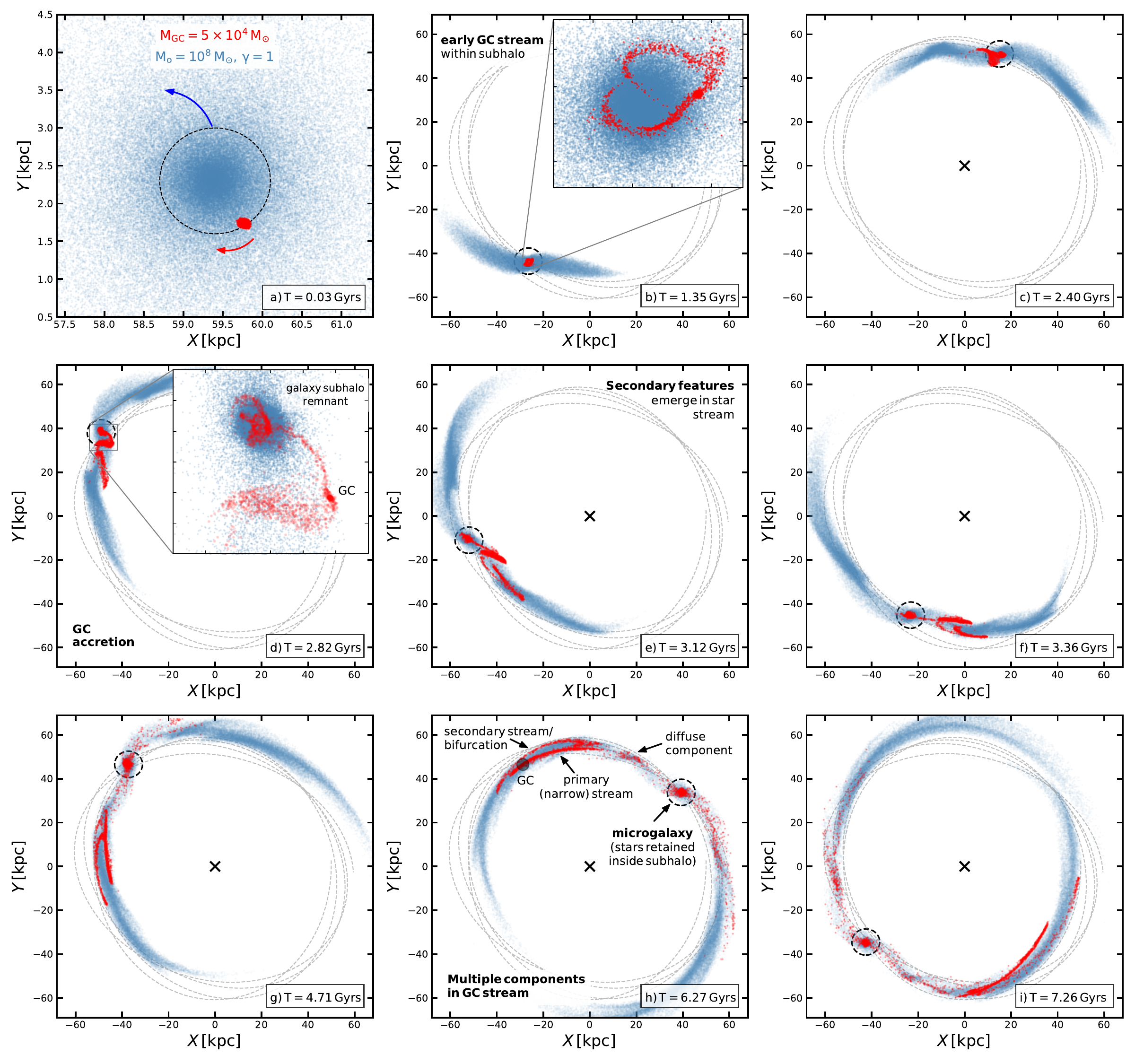}
\vspace{-0.5cm}
\end{center}
\caption{Dynamical evolution of a GC (red particles) as it accretes within its parent subhalo (blue particles) onto the MW type host. The details of this particular simulation are provided in Section~\ref{sec:Results}. The plots are shown in galactocentric Cartesian coordinates, where the panels are centered on the host galaxy; except panel `a' that is centered on the subhalo. From a)$-$i), each panel sequentially shows the formation and evolution of the GC stream. a) The GC is launched on an orbit within the subhalo (orbiting direction shown by red arrow), which is retrograde with respect to the orbit of the subhalo inside the host (orbiting direction shown by blue arrow). b)$-$c) An early star stream develops due to the disruption of the GC by the tidal forces of the subhalo. d) At some time instance, the GC escapes the subhalo along with stellar debris and accretes onto the host. e)$-$g) The accreted stellar debris and the GC are deposited on similar orbits, and now evolve under the potential of the host. h)$-$i) The surviving GC (marked by shaded gray circle) further disrupts under the potential of the host and forms the narrow stream component. On the other hand, the early stream gets dispersed as a thick component (\cocoon). Stars that move in the host galactic halo with very similar energies and momenta develop secondary narrow components. In almost every simulation employing the cuspy subhalo model, some GC stars were retained by the remnant subhalo, which at present maybe visible in the form of co-moving group of stars - a {\it micro-galaxy} (marked by dashed black ring).}
\label{fig:Fig_evolution}
\end{figure*}
\section{Results}\label{sec:Results}

In total, we executed and analysed nearly $100$ N-body simulations. Figure~\ref{fig:Fig_evolution} demonstrates the generic morphological features exhibited by accreted GC streams. The simulation in this figure is from a GC evolving in the SCu model with $(r_p,v_t)=(0.5\kpc, -19.6\kms)$. We use this representative example to briefly describe the formation and evolution of GC streams under the accretion framework.

As the GC moves inside the subhalo (Fig.~\ref{fig:Fig_evolution}a), it begins to lose mass due to the tidal forces from the parent subhalo. In a few Gyrs ($\approx 1.3\Gyr$ for this specific example), an early stellar stream develops inside the subhalo in the form of a ``ring''  (Fig.~\ref{fig:Fig_evolution}b, inset). The ring has a radius similar to the orbital radius of the GC inside the subhalo. It is worth remarking that similar ring-like stellar distributions were recently identified in the Sextans and Carina dwarf galaxies \citep{Cicuendez2018, Lora2019} that could possibly arise from disrupted GCs. We found that this early stream is more prominent in cuspy subhalos (where tidal forces are extremely strong) than in the cored subhalos. At this time the tidal force from the host galaxy is already beginning to produce tidal tails from the DM subhalo.

As this system continues to evolve inside the host galaxy, the subhalo continues to disrupt and disgorges its contents along its orbit, including whatever is left of the GC  (Fig.~\ref{fig:Fig_evolution}d, inset). Once released from the subhalo, the remnant GC primarily  experiences the tidal field of the host, which is weaker. From this time forward, the tidally liberated stars from the GC travel on orbits close to the GC's orbit, thereby producing a narrow stream (that is characteristic of \insitu GCs). Along with the GC, a substantial fraction of stars from the early (pre-accretion) stream form a {\it broad-diffuse structure surrounding the narrow-dense stream} (Fig.~\ref{fig:Fig_evolution}f-i). These findings are also in agreement with the previous studies based on cosmologically motivated simulations (\citetalias{Carlberg2018Density_Structure}). It is worth remarking that there is already observational evidence that such broad and low-surface brightness components (referred to as a \cocoon) indeed exist in some streams of the MW (\citetalias{MalhanCocoonDetection2019}, \citetalias{Bonaca2019Jhelum}). We compare our simulations with these observations in Section~\ref{sec:comp_with_obs}. 

\begin{figure*}
\begin{center}
\includegraphics[width=0.88\hsize]{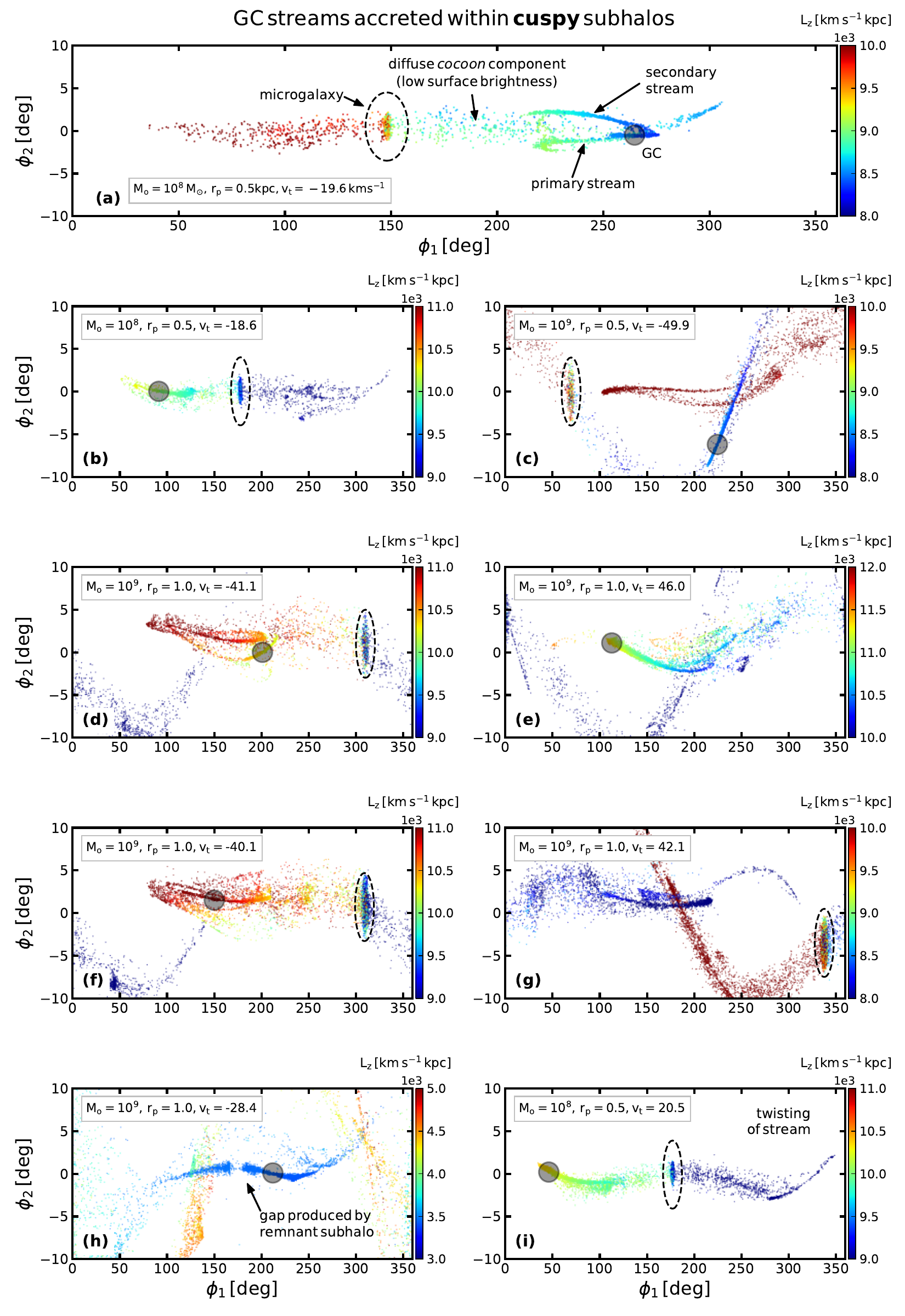}
\vspace{-0.65cm}
\end{center}
\caption{Density structure of the GC streams that accrete within cuspy subhalos (Sec.~\ref{subsec:accreted_streams_properties}). Each panel represents the final distribution of GC star particles in a given simulation, with the details of that simulation provided in the legends. Points are colored according to the value of $z-$component of angular momentum ($L_z$, calculated in Galactic coordinates). Progenitor GCs and micro-galaxies, in cases they survive, are marked by shaded circles and dashed rings, respectively. The streams can be observed as highly structured (comprising of multiple components), physically wide and dynamically hot. The physical properties of all the streams are shown in Figure~\ref{fig:Fig_comparison}.}
\label{fig:Fig_cuspy_streams}
\end{figure*}

Some of the stars from the early time stream, now moving freely in the galactic halo with very similar energies and momenta, develop into {\it secondary narrow components} (Fig.~\ref{fig:Fig_evolution}e-h). Further, in many simulations employing cuspy subhalos, bound remnants of the cusp survived with masses of $\sim 10^{6-7}\msun$ and sizes of $\simlt1\kpc$. In many cases, some GC stars were found to remain ensnared inside these remnant subhalos (e.g., see Fig.~\ref{fig:Fig_cuspy_streams}), giving rise to a low-surface brightness co-moving group of stars sometimes referred to as a {\it micro-galaxy} (also see \citealt{Errani2019}). In contrast, since the {\it cored} subhalo models disrupt completely in a few Gyr, no micro-galaxies were produced.

\begin{figure*}
\begin{center}
\includegraphics[width=0.90\hsize]{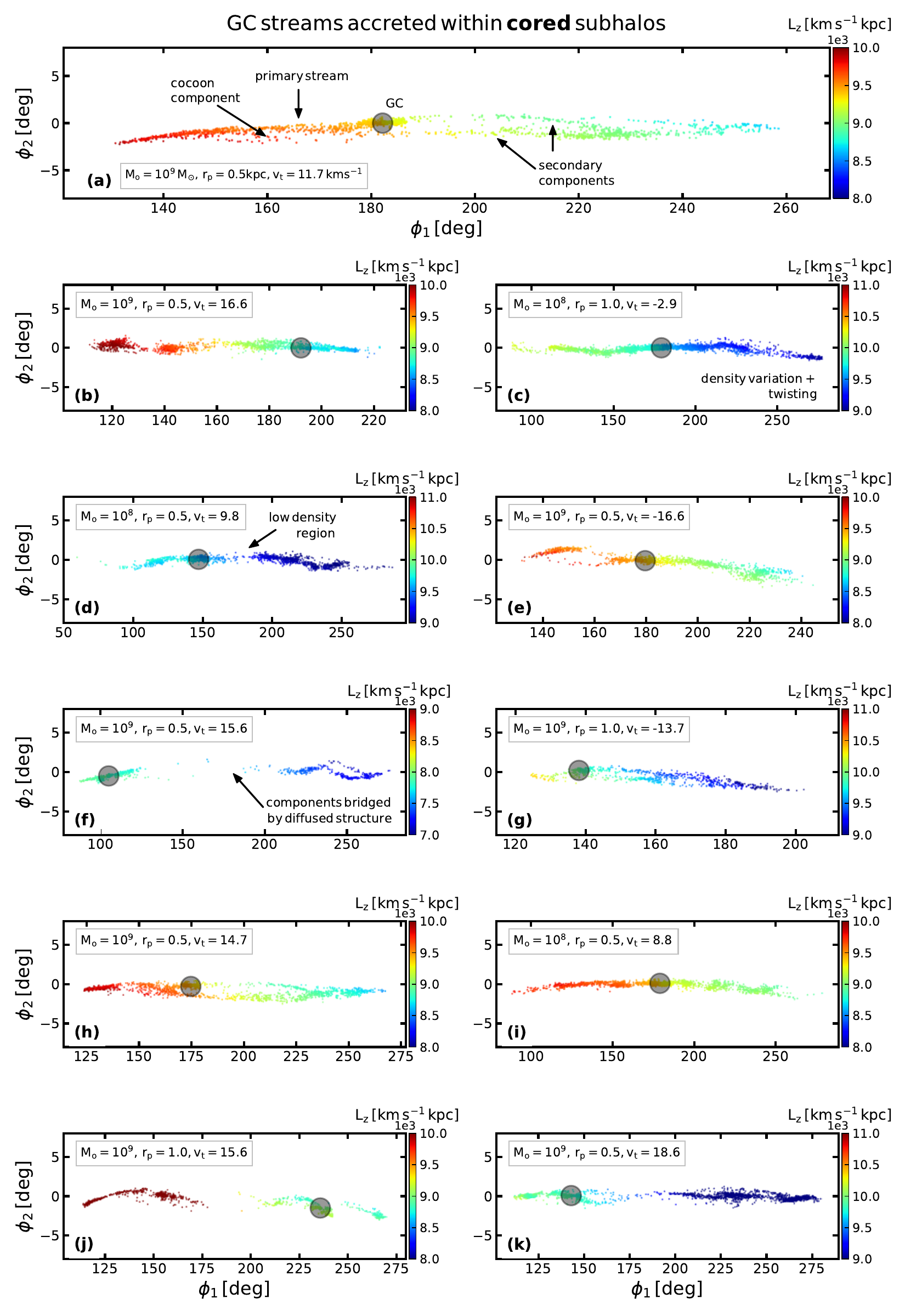}
\vspace{-0.6cm}
\end{center}
\caption{Same as Figure~\ref{fig:Fig_cuspy_streams}, but for the cored subhalo models (Sec.~\ref{subsec:accreted_streams_properties}). The range of $\phi_1$ axis is different for different panels, scaled in accordance with the length of the stream. In this case, the streams were observed to be only slightly broad in physical widths, and were found to be overall dynamically colder than in the cuspy case. Their physical properties are shown in Figure~\ref{fig:Fig_comparison}.} 
\label{fig:Fig_core_streams}
\end{figure*}
\begin{figure}
\begin{center}
\includegraphics[width=\hsize]{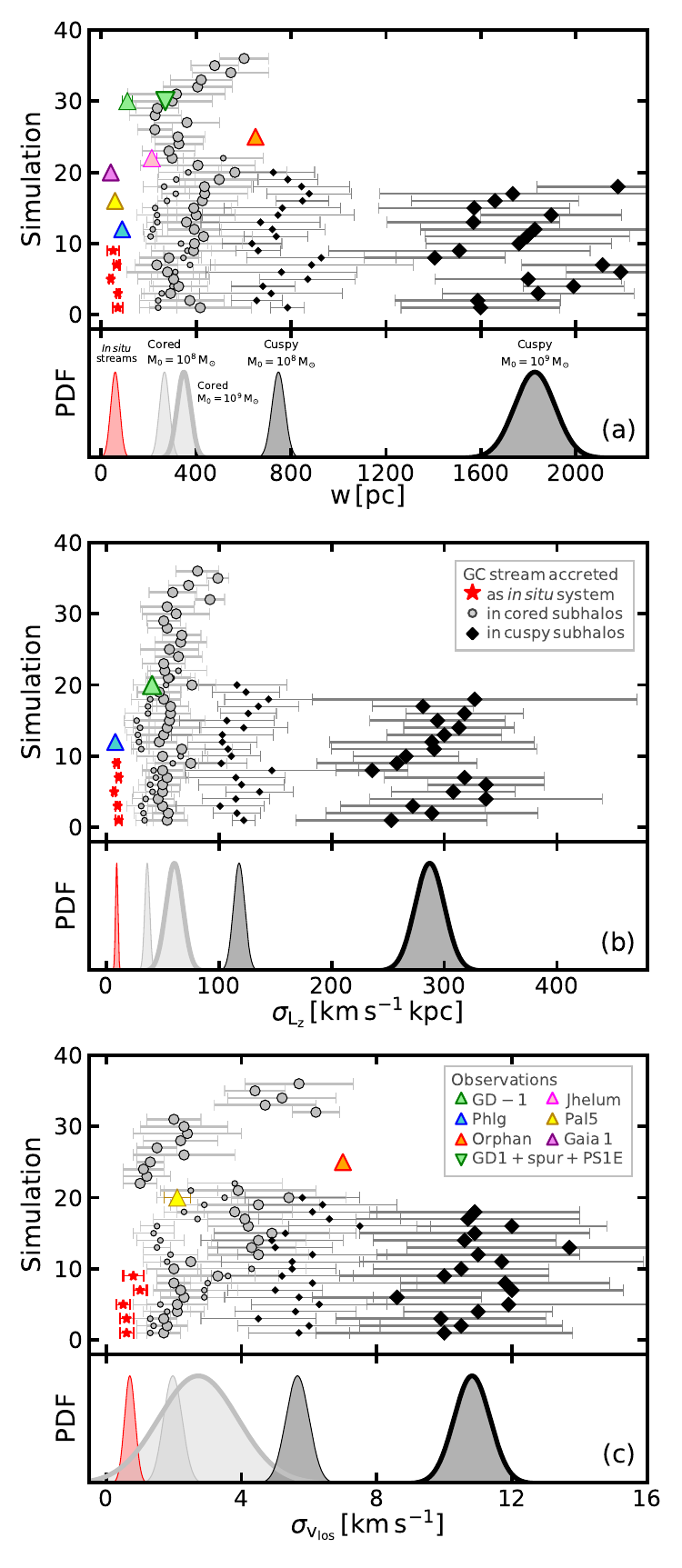}
\end{center}
\vspace{-0.75cm}
\caption{The physical widths ($w$), and dispersions in the $z-$component of angular momenta $(\sLz)$ and los velocities $(\sv)$ for GC streams analysed in our study (Sec.~\ref{subsec:accreted_streams_properties}). The Y axes denote different simulations. The red points correspond to streams produced from \insitu\, GCs (Sec.~\ref{sec:f_stream}). The black (gray) markers correspond to streams that accreted inside cuspy (cored) subhalos. The small (large) markers correspond to subhalos with $M_0=10^8\msun (10^9\msun)$. For streams formed under a given scenario, the mean and dispersion were calculated assuming Gaussian distributions. One can see that GCs that accrete inside CDM predicted cuspy subhalos produce streams that are physically broader and dynamically hotter than streams that accrete inside cored subhalos. As indicated in the legend of panel `c', measurements of some MW streams are also shown.  For the two cases of GD-1 and Jhelum, with likely accreted GC origin, the match of data to simulations favors cored subhalos (see Sec.~\ref{sec:comp_with_obs}).}
\label{fig:Fig_comparison}
\end{figure}
\begin{figure*}
\begin{center}
\includegraphics[width=\hsize]{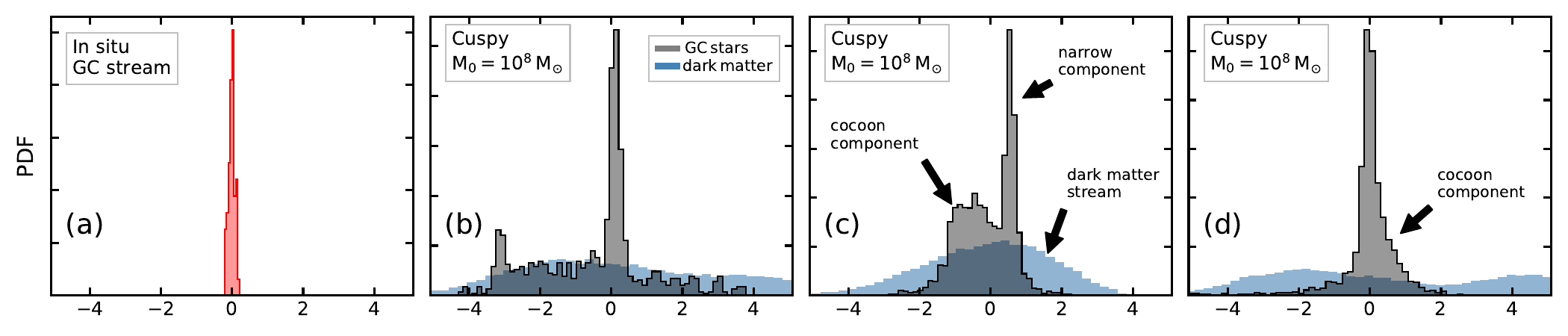}
\includegraphics[width=\hsize]{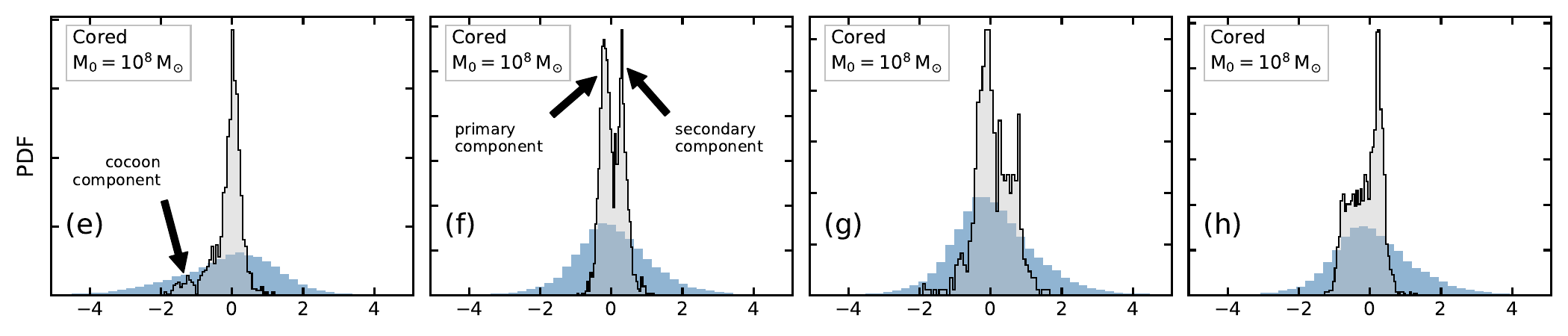}
\includegraphics[width=\hsize]{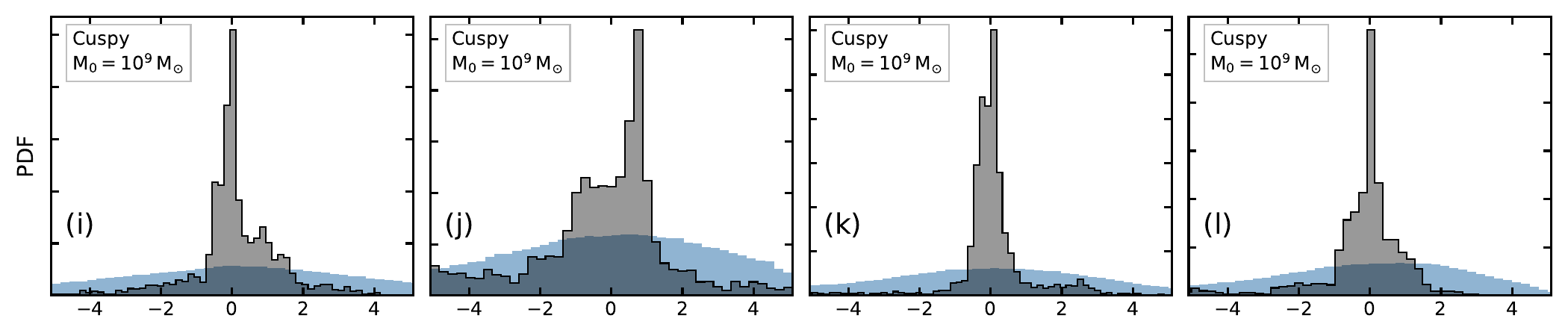}
\includegraphics[width=\hsize]{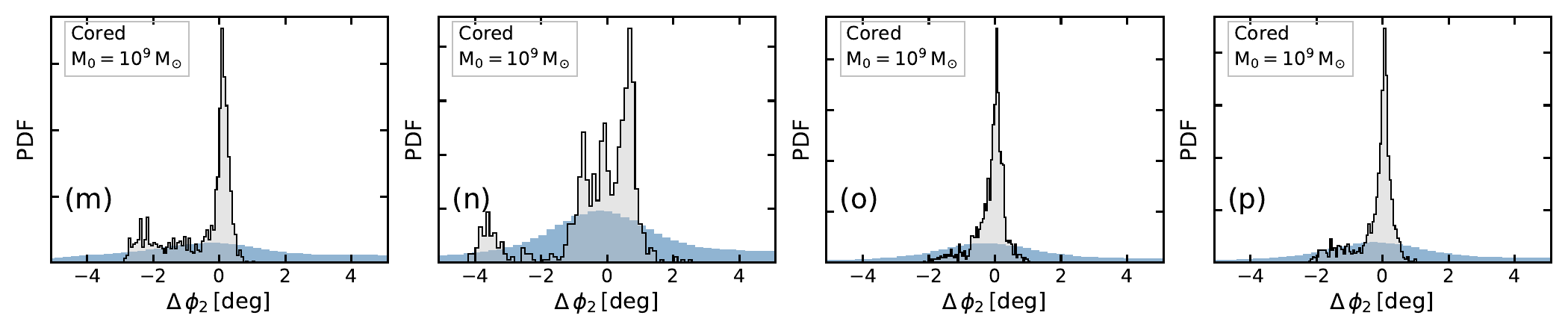}
\end{center}
\vspace{-0.48cm}
\caption{Transverse density profiles of GC streams. Panel `a' corresponds to the stream produced from \insitu\, GC (Sec.~\ref{sec:f_stream}), where the stellar distribution is uni-modal. The rest of the panels correspond to streams formed under the accretion framework (Sec.~\ref{subsec:accreted_streams_properties}). Each panel corresponds to the final distribution of GC star (gray) and DM (blue) particles in a specific simulation. The labels in the legend describe the corresonding parent subhalo. In all the panels (except `a'), the stellar distribution is multi-modal, that can be observed in the form of a narrow peak (marking the presence of the narrow-dense stream) and a broadened distribution (marking the presence of the diffuse \cocoon component). Secondary narrow peaks can be observed in some cases that mark the presence of secondary (neighbouring) narrow components (see Figs.~\ref{fig:Fig_cuspy_streams} and \ref{fig:Fig_core_streams}). Interestingly, some of the MW streams qualitatively feature similar density profiles (see Sec.~\ref{sec:comp_with_obs}).}
\label{fig:Fig_stream_profiles}
\end{figure*}
\subsection{Structural and dynamical properties of streams from accreted Globular Clusters}\label{subsec:accreted_streams_properties}

In this section we characterise the physical properties of  GC streams produced in various subhalo models. Figure~\ref{fig:Fig_cuspy_streams} and Figure~\ref{fig:Fig_core_streams} show examples of the GC streams that accreted within cuspy and cored subhalos, respectively.

{\it A) Physical widths:} 
All GC streams from cuspy subhalos exhibit complex morphologies, often comprising multiple structural components (e.g., see Fig.~\ref{fig:Fig_cuspy_streams}a). In addition to a narrow-dense primary component, most show secondary narrow component(s), as well as a broad-diffuse \cocoon component. While the primary and secondary components have physical widths similar to the narrow streams from \insitu GCs (with  $w\approx$ few tens of pc, e.g., Fig.~\ref{fig:Fig_simpleNbody}), the width of \cocoon components depended on both the mass and type (cored or cuspy) of the parent subhalo. Further, compared to cuspy subhalos, GC streams arising from cored subhalos were observed to be less structured (e.g., see Fig.~\ref{fig:Fig_core_streams}a), but more structured than the \insitu GC streams. 

We computed the physical widths of every simulated stellar stream, and also estimated the mean width and standard deviation of the widths of all streams from the same type of subhalo (using eq.~\ref{eq:likelihood}), by employing the procedure described in Section~\ref{sec:f_stream}. The regions containing the surviving GC and/or the micro-galaxy were ignored from our analysis. The estimated physical widths are shown in Figure~\ref{fig:Fig_comparison}a. For the cuspy subhalos, it can be seen that the values of the mean physical width and standard deviation are $w\approx745\pm30\,(1825\pm90)\pc$ for the SCu (LCu) model with mass $M_0=10^8\,(10^9)\msun$ (shown as Gaussians in the lower sub-panel of Figure~\ref{fig:Fig_comparison}a). Interestingly, these $w$ measurements closely approximate the adopted scale radius values ($r_0$) for the SCu and LCu subhalo models. For the cored subhalo model SCo (LCo), we found $w\approx 260\pm25 \,(348\pm30) \pc$, strikingly smaller than the corresponding cuspy cases. The mean value of the width estimators are clearly proportional to the depth of the gravitational potential of the parent subhalos (expected from the arguments in Section~\ref{subsec:theory}). The dispersions (the widths of the Gaussians) arise from the range of orbital properties of GCs inside the subhalo. 

{\it B) Dynamical widths:} 
In order to characterise the dynamics of the streams, we also computed the values of dynamical width estimators $\sLz$ and $\sv$ (previously described in Sec.~\ref{sec:f_stream}). These values are plotted in Figure~\ref{fig:Fig_comparison}b, c. The data points corresponding to a given type of subhalo can be observed to cluster around specific values, signifying that the dynamical properties of accreted GC streams are very sensitive to the nature of their parent subhalos.

For the {\it cuspy} cases, we found $\sLz\approx110\pm5\,(280\pm15)\kms\kpc$ and $\sv \approx5.5\pm0.5\,(10.5\pm0.6)\kms$ corresponding to the SCu (LCu) subhalo model. The streams arising from  {\it cored} subhalos produced dynamically colder streams with $\sLz\approx35\pm3\,(60\pm7)\kms\kpc$ and $\sv\approx2.0\pm0.3\,(2.7\pm1.2)\kms$ for the SCo (LCo) subhalo model. These values are represented as Gaussians in lower sub-panels of Figure~\ref{fig:Fig_comparison}b, c. However, streams from cuspy and cored subhalos both showed narrow components which were still measured as dynamically cold structures (with $\sv\sim 1\kms$).

{\it C) Transverse density profiles:} Figure~\ref{fig:Fig_stream_profiles} shows the transverse density profiles of the GC streams produced in different subhalo models. For every stream, a polynomial fit was made in the $\phi_1-\phi_2$ coordinate, but only to the primary narrow component. After this, the angular difference $\Delta \phi_2$, between the model fit and each star particle was calculated and histograms of $\Delta \phi_2$ were plotted. As can be seen, a feature that is common in all the plots is that each of them contain multi-modal stellar distributions: a narrow peak (marking the presence of the primary narrow-dense component) and a broader distribution (that reveals the presence of the broad-diffuse \cocoon component). In some cases, secondary narrow peaks can be observed that correspond to the presence of secondary parallel components in streams. These profiles can be explained with the reasons provided in Section~\ref{subsec:theory}. This means, that the the transverse density profiles of accreted GC streams are expected to be at least bi-modal, containing a narrow peak and a broadened distribution. We also plot the distribution of DM particles in Figure~\ref{fig:Fig_stream_profiles} that serves as a proxy for an extended stellar component, such as might arise from a virialized dwarf galaxy. It is important to note that stellar streams produced from dwarf galaxies, lacking any GC population, will be observed as broad structures with uni-modal and smooth stellar distributions, similar to the distributions of DM particles in Figure~\ref{fig:Fig_stream_profiles}. In observations, this point will be useful in correctly characterising the origin and birth sites of the observed stellar streams of the MW. This is discussed further in Section~\ref{sec:discussion}.

{\it D) Time evolution of width estimators:} It is important to assess whether the physical properties of the streams change significantly with time, since a large temporal change in physical and dynamical width estimators could affect our ability to use them as probes of the density profiles of the parent subhalos. This issue is briefly analyzed in Appendix~\ref{appendix:time_evolution} (see Fig.~\ref{fig:Fig_properties_time_evol}) where we show that once a GC stream forms under the accretion framework, the physical and dynamical width estimators $(w,\sLz, \sv)$ change by less than $10\%$ throughout the remaining  evolution time.  This occurs because dynamical widths of streams stay  nearly constant in potentials that are mostly spherical (where $q_{\phi}\sim1$). Note that although our MW mass model is axisymmetric, the potential is only slightly flattened with $q_{\phi}\sim0.94$ at the orbital radii of the simulated streams. However, most of the observed streams orbit the inner regions of the MW halo where the potential is atleast moderately flattened (due to the presence of the disk). For example, at the present day locations of GD-1 and Jhelum, we found that our potential has a flattening of $q_{\phi}\sim0.9$. This point is important because streams that orbit the non-spherical regions of the host galaxy are expected to grow in dynamical widths (e.g., in $w$ parameter) due to the precession and nutation of their angular momentum vectors (e.g., \citealt{Erkal2016evolution_orbits}). Could this effect be so strong that it may render aforementioned results invalid, namely, our ability to differentiate between core and cusp scenarios?

 To examine the possible dependence of time evolution of $w$ on $q_{\phi}$, we considered Figure 12 of \cite{Erkal2016evolution_orbits} that presents the relationship between evolution of $w$ for \insitu streams as a function of $q_{\phi}$. Their model tentatively suggests in a halo with $q_{\phi}=0.95$, a $350\pc$ wide stream (corresponding to the GC stream from $10^9\msun$ cored subhalo) will evolve into a $750\pc$ wide stream (corresponding to $10^8\msun$ cuspy subhalo case) in $\sim 23\Gyr$. Similarly, in a halo with $q_{\phi}=0.90$, the corresponding evolution time will be $\sim 13.5\Gyr$. These time scales are $\simgt$ Hubble time, and only provide lower limits since we have not added the pre-evolution times of GCs inside the parent subhalos. Therefore, it appears that our results are robust to changes in the flattening of the host potential. 

{\it E) Discriminatory power of streams:}
Over all, one concludes from  Figure~\ref{fig:Fig_comparison} that GC streams produced in different subhalo models occupy different regions on the $x$-axes representing the physical and dynamical width estimators. This signifies that GC streams accreted inside cuspy subhalos possess higher dynamical dispersions in each of the three width estimators, compared to those streams that accreted inside cored subhalos. Also, as illustrated in Figures ~\ref{fig:Fig_cuspy_streams} and \ref{fig:Fig_core_streams}, the final morphologies of accreted GC streams are sensitive to the orbital configuration of their progenitor GCs inside the parent subhalos. In different simulations, GCs move on different orbits and lose stars over different ranges of radii inside the subhalo. This dependency on orbital parameters is reflected in the widths of the Gaussians in Figure~\ref{fig:Fig_comparison}. For instance, streams from simulations where the GC was launched on a prograde orbit inside the subhalo (with respect to the orbit of the subhalo inside the Galaxy) showed relatively smaller widths, compared to the cases where GCs were launched on retrograde orbits. However, this scatter is generally much smaller than the distance between the mean values of the Gaussians for different subhalo models. For the simulations that were studied, we found that streams with $(w, \sLz, \sv)\simgt(650\pc, 95\kms\kpc, 4\kms)$ require that they must be accreted in CDM motivated cuspy subhalos. Smaller dispersion values with $(w, \sLz, \sv)\simlt(90\pc, 15\kms\kpc, 1\kms)$ would imply streams formed via disruption of \insitu GCs. Values in between this range would favour a scenario where the progenitor GCs were accreted inside cored subhalos. This makes us confident that the physical properties of accreted GC streams are extremely sensitive probes of the properties of the parent subhalos, meaning they can potentially be used to discriminate between the cusp/core scenarios. 

Some degeneracy is apparent between \insitu streams and accreted streams in $\sv$, however, can be evaded by accurately measuring the estimators $w$ and $\sLz$. Moreover, information about their multiple component morphologies (Figs.~\ref{fig:Fig_cuspy_streams} and \ref{fig:Fig_core_streams}) and transverse density profiles (Fig.~\ref{fig:Fig_stream_profiles}) further allows accreted GC streams to be distinguished from narrow \insitu GC streams. This discernment is possible irrespective of whether the GC were accreted inside cuspy or cored subhalo. From similar arguments, accreted GC streams can also be differentiated from uniformly broad dwarf galaxy streams.

While streams produced in different cuspy models (SCu and LCu) can be easily differentiated in all three width estimators, the properties of streams produced in different cored subhalos (SCo and LCo) overlap in $\sv$ (see Fig.~\ref{fig:Fig_comparison}c). This potential degeneracy implies some challenges in discriminating between cored subhalos using $\sv$. However, since cored subhalos are still well separated in  $w, \sLz$, accurate determination of these parameters can still be used to differentiate between the two cases. Similarly, some degeneracy that exists between SCu and LCo in panel `c' (implying potential challenges in discriminating between a low-mass cuspy subhalo and high-mass cored subhalo) can again be evaded using similar procedure. This second problem can further be tackled by analysing the morphology of the stream under question. For example, if a given stream is highly structured, or possesses a high contrast \cocoon component, or if a faint disrupting micro-galaxy is detected along its orbit, a cuspy parent subhalo would be favoured over a cored one. Overall, these results imply that {\it the present day physical properties of accreted GC streams can be a potentially powerful tool for infering the central DM density distribution inside their parent subhalos}.

\begin{figure}
\begin{center}
\includegraphics[width=1.02\hsize]{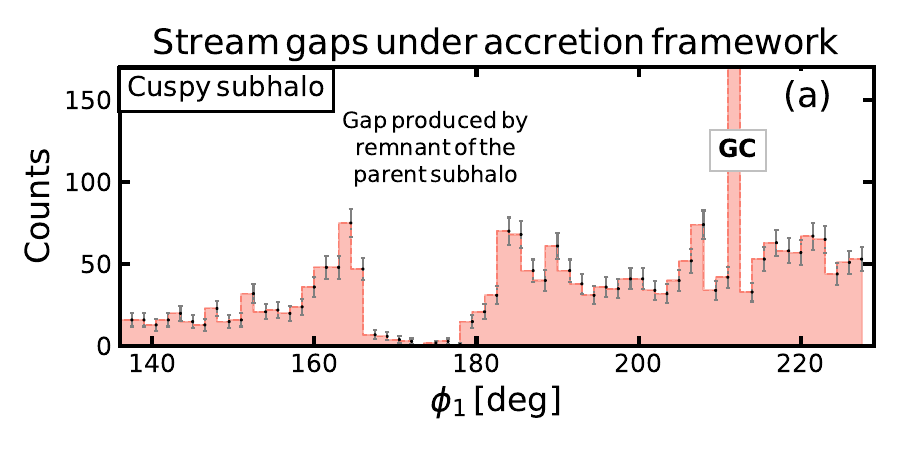}
\includegraphics[width=1.02\hsize]{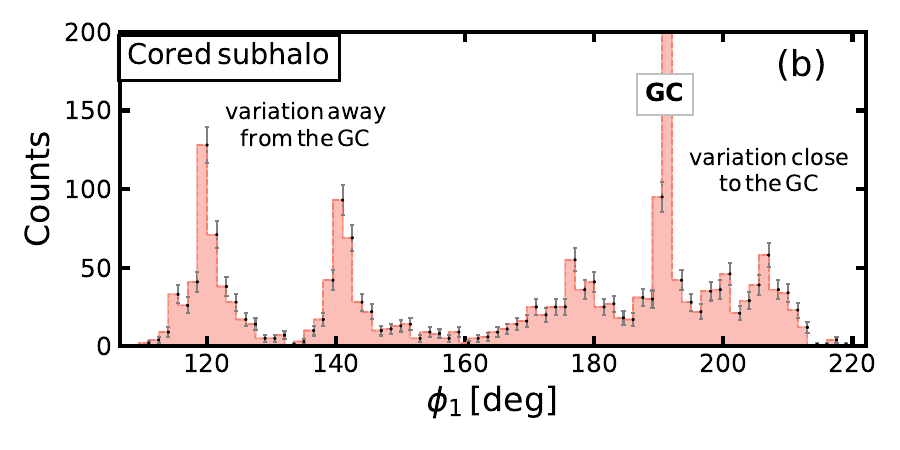}
\vspace{-0.9cm}
\end{center}
\caption{Small scale density variation and ``gaps'' in streams under the accretion framework (see Sec.~\ref{subsec:Gaps}). a) Cuspy subhalo: Counts (with $1\sigma$ uncertainties) as a function of $\phi_1$ for stream shown in Figure~\ref{fig:Fig_cuspy_streams}h. b) Cored subhalo: Same as `a' but for stream presented in Figure~\ref{fig:Fig_core_streams}b. 
}
\label{fig:Fig_density_variation}
\end{figure}
\subsection{Stream density variations and ``gap'' formation under the accretion framework}\label{subsec:Gaps}

In this section we show that the accretion of GCs in subhalos provides a novel mechanism for the formation of gaps in streams, one that  arises naturally and differs slightly from those previously considered in the literature. Density variations in narrow, dynamically-cold low-mass streams are considered potential probes of DM substructure \citep{Ibata_2002DM_TS, Johnston2002DM_TS}. CDM cosmogony predicts that hundreds or thousands of subhalos should orbit within our Galaxy \citep{Moore1999, Klypin1999, Springel2008}. The gravitational influence of subhalos with masses  $M\sim10^{6-8}\msun$, as they pass through tidal streams, can cause perturbations which  would be detectable in the form of broadening and density variations along the length of the stream. The formation of ``gaps'' in streams resulting from close passages of subhalos, are generally regarded as a promising way to detect  otherwise invisible DM substructure \citep{Carlberg2012, Ngan2014, Erkal2016_gaps}. However, so far, the application of these techniques to the observed streams of the MW have not yielded conclusive results, especially with regard to the possibility of distinguishing between various models of DM  (e.g., SIDM, WDM or CDM).

The most detailed model thus far of a gap produced by interaction with a dark subhalo is the model for the observed ($\sim 6\deg$ wide) gap and nearby ``spur'' in the GD-1 stream \citep{Bonaca_spur_2018}. This model suggests  that both features may be result of a close encounter between a DM subhalo and the GD-1 tidal stream \footnote{Also see \cite{IbataGD1gaps2020} where density variations in GD-1 are explained with simple epicyclic motion in a smooth Galactic potential.}. Their fiducial (preferred) impact model produces both the gap and the spur and requires a subhalo  of mass $\approx 5\times 10^6\msun$, and scale radius $r_s = 10\pc$ with an encounter velocity between the subhalo and stream of $\sim 250\kms$. Even their least  compact subhalo with scale radius of  $r_s = 20\pc$ subhalo is so  compact that it is only consistent with the mass-concentration relation predicted by $\Lambda$CDM at a $2-3\sigma$-level. Although they do consider a range of  relative velocities for their subhalo impact models, encounters with much lower relative velocities are ruled out since they do not produce a ``spur'', although they can produce gaps. 

Here we propose an alternative, fortuitously discovered, mechanism for the formation of both spurs and gaps in GC streams. Some of our simulated streams exhibited significant small scale density variations and gaps, despite the fact the host galaxy in our simulations is smooth, and completely devoid of any dark substructures. These density variations emerged due to the complexity involved in the formation mechanism of streams under the accretion framework, as we briefly explain below by focusing on examples from both the cuspy and cored subhalo cases. None of our simulations were designed to resemble GD-1 or any other stream -- we merely present these as a possible mechanisms that can produce gaps and spurs in streams, that should be examined further.

{\it A) Gaps in streams accreted inside cuspy subhalos}: The stream shown in Figure~\ref{fig:Fig_cuspy_streams}h possesses a prominent gap. In this case, the parent cuspy subhalo (LCu model) was launched on a slightly eccentric orbit. The corresponding star count distribution is displayed in Figure~\ref{fig:Fig_density_variation}a, produced by including only those stars that lie in the range $|\phi_2|<1.5\deg$ and $\phi_1=[100\deg,250\deg]$ (so as to focus only on the narrow-dense component of the stream). In Figure~\ref{fig:Fig_density_variation}a, the gap is visible as an under-dense region between $\phi_1\approx[170\deg,180\deg]$. This $\sim 10\deg$ wide gap was the result of a close flyby of the remnant (central cusp) of the parent subhalo that hosted the GC. Being on similar orbits, the parent subhalo and the GC tidal stream (now freely orbiting the halo) have low relative velocities, which effectively increases the interaction time and also the chances of their encounter. Thus the collision between the stream and the subhalo can last a longer time  than a random encounter, and can occasionally result in the formation of gaps in streams. Under such a scenario, a less dense subhalo (one that is consistent with CDM) can have more of an impact than a random encounter. The spur, in this scenario, would not be formed by the encounter but would result from the pre-accretion tidal stripping of the GC that produces secondary streams and spur-like features (see Sec.~\ref{subsec:accreted_streams_properties}).  It is interesting to compare this result with the recent observational study of \cite{Li2020_Atlas_Aliqa} which suggests that the gap and the ``kink'' between ATLAS and Aliqa Uma streams may have been produced by the impact of the Sagittarius dwarf galaxy (which could plausibly be the parent dwarf of these streams).

{\it B) Gaps in streams accreted inside cored subhalos}: Next we study stream gaps produced in the cored subhalo case shown in Figure~\ref{fig:Fig_core_streams}b. The corresponding star count distribution is shown in Figure~\ref{fig:Fig_density_variation}b as a function of $\phi_1$. The peak at $\phi_1 \approx 192\deg$ corresponds to the location of the surviving GC. Significant variations can be seen in counts across the length of the structure. These variations closer to the GC emerges mainly due to the epicyclic motion of the stars within the cluster (c.f. \citealt{Kupper2012, Fardal2015}) as the GC disrupts under the tidal forces from the host galaxy after escaping the parent subhalo. These epicyclic density spikes depend on the cluster's mass, its orbit, and how long the GC has been freely orbiting the host galaxy. However, density variations are also notable in the form of spikes in regions away from the GC progenitor at $\phi_1 \approx 120\deg$ and $140\deg$. These density spikes appear due to variations in mass loss rate of the subhalo as it orbits the host galaxy, that leads to the episodic deposition of GC stellar debris. The details of these observed wiggles are, however, sensitive to the details of how the subhalo merges into the main halo that can add kinks and additional breadth (see also \citetalias{Carlberg2018Density_Structure}).

Some of the streams in Figure~\ref{fig:Fig_core_streams}, that accreted inside cored subhalos, feature wide gaps (e.g. Fig.~\ref{fig:Fig_core_streams}f) which mark the positions of now completely dissolved parent subhalos.

A detailed analysis of gap formation phenomenon is beyond the scope of the present contribution, and is deferred to future study.

\begin{figure*}
\begin{center}
\includegraphics[width=0.8\hsize]{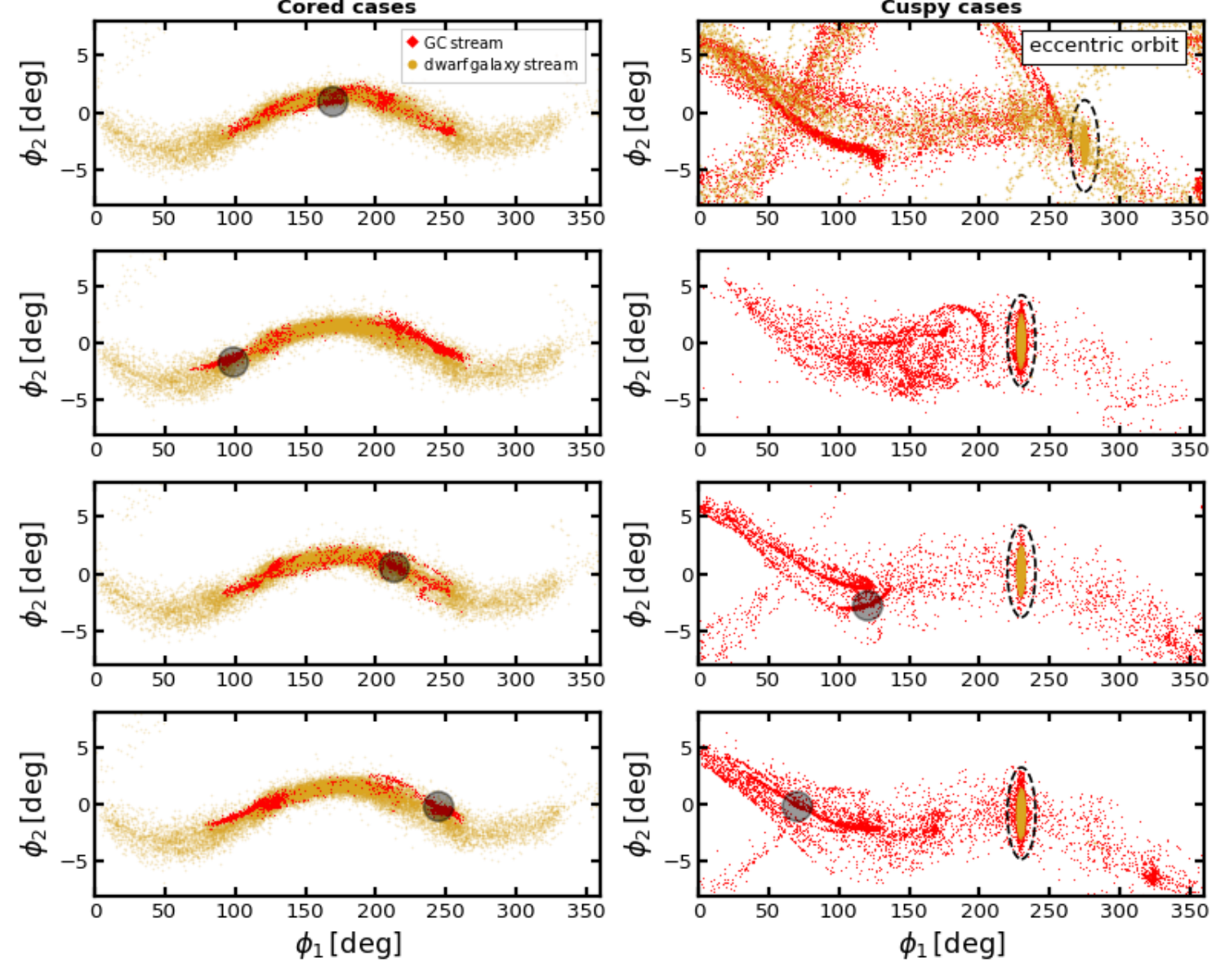}
\vspace{-0.5cm}
\end{center}
\caption{Density structure of the GC streams (red) and their parent dwarf galaxy streams (gold). The left (right) panels correspond to the cored (cuspy) cases. All the subhalos were launched on circular orbits, as described in Section~\ref{sec:discussion}, except one that is labelled with ``eccentric orbit''. Broad dwarf galaxy streams form in every cored subhalo case. However, for the cuspy cases, their formation depends on the orbit along which the subhalo accretes and the original physical extent of the stellar component inside the dwarf galaxy.}
\label{fig:GC_and_StellarHalo_streams}
\end{figure*}
\section{Additional simulations}\label{sec:discussion}

In the previous section we restricted our study to subhalos on nearly circular orbits with Galactocentric radii of $\sim 60\kpc$. However, CDM cosmological simulations show that satellites generally accrete along eccentric orbits (e.g., \citealt{vandenBosch_etal_99}) with average eccentricity $e \sim 0.6$. Recent observations show that  dwarf galaxies of the MW \citep[e.g.][]{simon_18} are on even more eccentric orbits, $ e\simgt 0.75$. In this section we carry out a limited study of subhalos on eccentric orbits. Furthermore, our models in the previous section only considered DM subhalos with a single GCs and no extended stellar component.  However, any subhalo with sufficiently vigorous star formation to have formed GC(s) would likely also contain an extended stellar distribution. This section also presents results of a few simulations which included an extended stellar distribution. The goal of these additional simulations is to confirm that the main conclusions of Section~\ref{sec:Results} hold up to these changes in parameters and assumptions.

{\it A) Subhalos on eccentric orbits:}  A sample of subhalos containing GCs were launched on eccentric orbits inside the host galaxy to test if the morphology of the accreted GC streams is sensitive to the orbit of the subhalo inside the host. The orbital eccentricity ranged from $0.1-0.9$ in steps of $0.1$ for fixed apocentric distance of $\approx 60\kpc$. These simulations were undertaken only for LCo and LCu models (where $M_0=10^9\msun$), because the SCo models disrupted completely on much shorter timescales due to their susceptibility to the substantially stronger tides they experience in the higher density inner regions of the host. For both the LCu and LCo models, the final physical properties of streams were found to be similar to those observed in Figure~\ref{fig:Fig_comparison}. However, for  a few of the  LCo models that were launched on very eccentric orbits, the stream properties deviated slightly from the general trend as they possessed smaller  physical and dynamical widths. This was because the cored subhalos, now on nearly radial orbits, lost a significant fraction of their mass after a few pericentric passages due to greater cumulative effects of gravitational shocking and tidal stripping, causing the tidal dissolution of the subhalos on a short timescale (see Appendix~\ref{appendix:mass_loss} and Fig.~\ref{fig:Fig_sub-halo_mass_evolution_eccentricites}). Thus, the duration of the ``pre-accretion'' gravitational interaction between the parent subhalo and the GC was diminished and the GC experienced the majority of its disruption only due to the tidal field of the host Galaxy. Consequently, the resultant GC stream was quite similar to an \insitu GC stream (i.e. physically narrow and dynamically colder). In contrast, the cuspy subhalos were dense enough to survive even on eccentric orbits, resulting in GC streams with physical and dynamical widths similar to the accreted GC streams on circular orbits. 

{\it B) Subhalos with extended stellar populations + GCs:} 
We ran a few simulations only for LCo and LCu models. These systems consisted of a dark subhalo halo, a GC and an extended stellar component, with the latter designed to resemble the stellar distribution of the dSph galaxy Eridanus II \citep{Contenta2018EridanusII}, which also has a single GC. The stellar halos, inside the subhalos, were modeled as per equation~\ref{eq:density_dehnen} by setting $M_0=10^5\msun,r_0=0.5\kpc,\gamma=0$. In these simulations the subhalos were once again launched on a nearly circular orbits at $\sim 60\kpc$.  The resulting GC streams possessed morphologies similar to the streams in the previous section. This is unsurprising because the diffuse population of stars inside the subhalo has such a low total mass that it has little effect on the dynamics of the GC or the formation of the GC's stream. However, the streams now  also have an additional thick and smooth stellar component that is similar to broad stream of DM particles in Figures~\ref{fig:Fig_evolution} and \ref{fig:Fig_stream_profiles}.  This is shown explicitly in Figure~\ref{fig:GC_and_StellarHalo_streams} for a few cases. Notice that dwarf galaxy streams form in every cored case. As previously stated, this occurs due to the low binding energy of the cored subhalos that results in their complete disruption. Additionally, dwarf galaxy streams lie close to the GC streams in phase-space. However, for the cuspy subhalos launched along circular orbits inside the host, the resulting dwarf galaxy streams were of very low-contrast as most of the stars remained bound to the surviving remnant subhalos (giving rise to the {\it micro-galaxy} features). This would also depend on the physical extent of the stellar component inside the subhalo (prior to its disruption). On the other hand, when cuspy subhalos were launched along eccentric orbits, this resulted in higher disruption of the subhalos, ultimately giving rise to high contrast dwarf galaxy streams. {\it This implies that the presence of an underlying dwarf galaxy stream depends on the initial conditions of the parent dwarf and the orbit along which the dwarf disrupts.}  Moreover, the detection of the dwarf galaxy streams is sensitive to their densities, which in turn depends upon the densities of the stellar components inside the parent dwarfs. In observations, this thick dwarf galaxy stream may be hard to distinguish from the broad \cocoon component of the GC stream spatially and/or dynamically,  but should be distinguishable via the chemical abundances and stellar populations (e.g. [Fe/H], [$\alpha$/Fe], Age).

It is worth remarking that we have already started discovering complex streams in the MW halo that exhibit multiple structural components that lie along similar orbits and have similar phase space properties but are characterized by fairly distinct metallicity distributions (e.g., the phase-space entangled GD-1 and Kshir streams have been measured with [Fe/H] of $-2.24\pm0.21$ dex and $-1.78\pm0.21$ dex, \citetalias{Malhan2019_GD1_Kshir}).

\section{Stellar streams of the Milky Way}\label{sec:comp_with_obs}

With the advent of ESA/Gaia \citep{GaiaDR2_2018_Brown, GaiaDR2_2018_astrometry}, the availability of large numbers of stellar proper motions has made it possible to identify and characterise many stellar streams in the MW galaxy (e.g., \citealt{Malhan_Ghostly_2018, Koposov2019Orphan, Ibata_Norse_streams2019, Palau2019M68streamFjorm, Shipp2019}). The wealth of data from Gaia and complementary ground based spectroscopic observations has allowed us to compute dynamical width estimators for many of these streams. Some of these widths are shown as colored symbols in Figure~\ref{fig:Fig_comparison} (and are identified in the legend). In this section we compare these observationally determined widths for several prominent streams with the widths predicted for various simulated streams. This will enable us to make a preliminary assessment of the implications of these observations for constraining the central density profiles of DM subhalos.

It is important to note at the outset that for the observed streams of the MW, only  $\sv$ can be directly computed from observations, in a manner similar to the one we adopted here for the simulated streams. It is difficult to compute $w$ and $\sLz$ directly from observations because both quantities require an estimate of the distances to stars in the streams. Currently, the uncertainties on the distances to typical halo stars are quite large (even after one considers the parallax information from Gaia). Nonetheless, $w$ and $\sLz$ can be computed by following an orbit-fitting procedure (in an assumed Galactic gravitational potential) that we briefly describe in Appendix~\ref{appendix:orbit_GD1_cocoon}, where we estimate $\sLz$ for the GD-1 structure\footnote{In Fig.~\ref{fig:Fig_comparison}, all values of $w$ and the value of $\sLz$  for Phlegethon are taken from previously published works and use a similar method.}.

The streams for which widths are available and marked in Figure~\ref{fig:Fig_comparison} include (in increasing $w$): 
Gaia-1 ($w\approx40\pc$, \citealt{Malhan_Ghostly_2018})
Palomar~5 (Pal~5, $w\approx58\pc$, \citealt{Ibata_2016}, $\sigma_v\approx2.1\pm0.4\kms$, \citealt{Kuzma2015_Pal5}), 
Phlegethon (Phlg, $w\approx88\pc, \sLz\approx8\kpc\kms$, \citealt{Ibata_Phlegethon_2018}), 
GD-1 ($w\approx110\pc$, \citetalias{MalhanCocoonDetection2019}, $\sLz\approx40\kpc\kms$, see Appendix~\ref{appendix:orbit_GD1_cocoon}), 
Jhelum ($w\approx 210\pc$, \citetalias{Bonaca2019Jhelum}), 
and Orphan ($w\approx 650\pc, \sigma_v\approx7\kms$, \citealt{Belokurov2007_Orphan, Fardal2019}).

The observed physical widths $w$ of Gaia-1, Pal~5 and Phlg are consistent with them having been formed by the tidal disruption of \insitu GCs (the scenario presented in Sec.~\ref{sec:f_stream}, although the value of $\sigma_v$ for Pal~5 is slightly higher than expected for an \insitu GC stream). 

At present, none of the observed GC streams exhibits the extreme widths that our simulation predict for  GCs accreted in cuspy subhalos. With its larger $w$ and higher $\sv$, the Orphan stream comes closest. However, until now, it has been observed to only have a uniform uni-modal structure, devoid of any obvious  narrow components or complex morphological features. Moreover, Orphan has a high metallicity dispersion ([Fe/H]$\approx-2.0\pm0.4$ dex, \citealt{Fardal2019}) that suggests that this stream is more likely to be the result of  tidal disruption  of a low-mass dwarf galaxy. 

GD-1 and Jhelum are the only two streams on Figure~\ref{fig:Fig_comparison} for which $w,\sLz$ are significantly larger than for \insitu streams. Interestingly,  both these streams also show the multiple narrow components and cocoon components that add the overall breadth to their structure (\citetalias{MalhanCocoonDetection2019}, \citetalias{Bonaca2019Jhelum}). Recall that Figure~\ref{fig:Fig_stream_profiles} showed the predicted transverse density profiles of accreted GC streams. These profiles are quite realistic, and in fact quite similar to the observations of GD-1 (see  Figure~6 of \citetalias{MalhanCocoonDetection2019}) and Jhelum (see  Figure~3 of \citetalias{Bonaca2019Jhelum}). This implies that the complex morphology of GD-1 and Jhelum can be explained under the accretion framework. Further, {\it a comparison of the measured values for GD-1 and Jhelum with our simulations (Figure~\ref{fig:Fig_comparison}a) favour a scenario where these two streams were accreted inside subhalos that possessed cored density profiles}. However, in reality, both streams are much closer to the Galactic centre (pericenter of $\sim14\kpc$ and $\sim10\kpc$, respectively, \citealt{Malhan2018PotentialGD1}, \citetalias{Bonaca2019Jhelum}) than any of the cored subhalo streams we have simulated. As mentioned in Section~\ref{subsec:Nbody_simulation_setup}, the reason we were unable to run simulations of GC evolution in cored halos at such small radii was that none of the parent subhalos survived long enough for a pre-accretion GC stream to form. One possible way to explain GD1's morphology, as well as its proximity to the Galactic centre, would be if it were accreted on an eccentric orbit in a  more massive cored subhalo (e.g. $M_0 \simgt 10^{10}\msun$). A more massive subhalo would not only survive longer, allowing the formation of a pre-accretion GC stream, but it would also experience much stronger dynamical friction against the Galactic DM potential that would drag it down to small pericenter radius and circularize its orbit. This point is in concordance with the highly retrograde GD-1 stream that lies along nearly tangential orbit. This proposed scenario is not unreasonable since recent observations of clustering in action space of both stars and GCs in the inner MW halo has led to the discovery of two recent fairly massive accretion events which have deposited most of their debris within $30\kpc$ (``Gaia Enceladus/Gaia Sausage'' and ``the Sequoia'', \citealt{Belokurov2018, Helmi2018, Myeong2019}).  Interestingly, \cite{deBoer2020} showed that GD-1 has similar actions/energies to the GCs associated with the Sequoia --- suggesting a possible link between GD-1 and the Sequoia itself. Similarly, \cite{Li2020_Atlas_Aliqa} have recently shown that the ATLAS stream is plausibly associated with the Sagittarius dwarf galaxy.

\begin{figure*}
\begin{center}
\includegraphics[width=0.8\hsize]{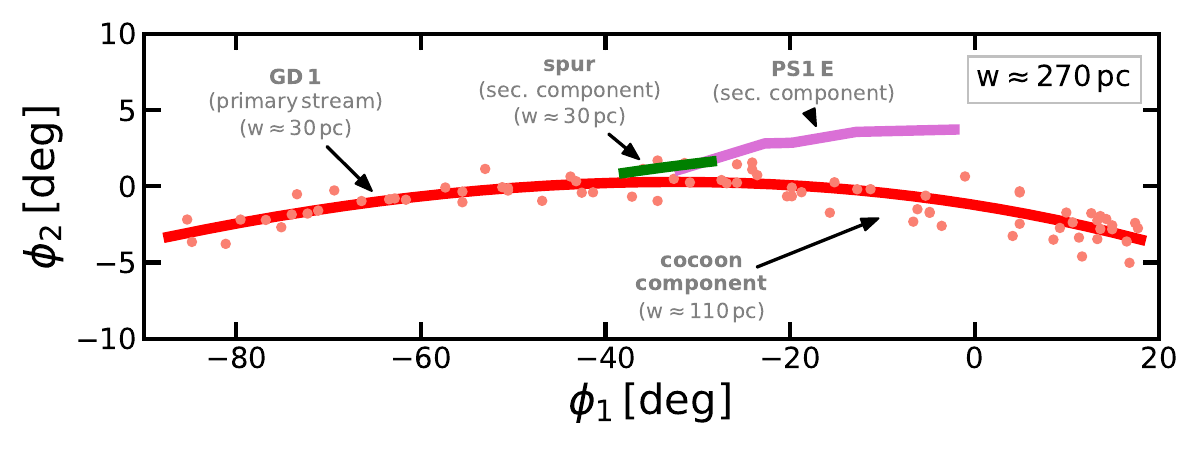}
\end{center}
\vspace{-0.6cm}
\caption{Schematic diagram of the trajectories of the MW streams GD-1 and PS1-E, and the spur and the \cocoon component associated with GD-1. These trajectories were produced by approximately curve fitting the on-sky extents of the respective structures as observed in previous studies \citep{Bernard2016, Bonaca_spur_2018, Malhan_2018_PS1}. GD-1, spur and \cocoon lie along essentially the same orbit. As for PS1-E, though it strongly overlaps in distance and proper motion space with GD-1, but the lack of los velocity measurements has left us uninformed about its orbit. Interestingly, all of these components possess similar stellar population (in terms of Age, [Fe/H]), as that of GD-1 itself, suggesting that they likely emerged from the same progenitor GC that accreted inside a dwarf galaxy.}
\label{fig:Fig_GD1_structure}
\end{figure*}

Moreover, there is already some observational evidence that GD-1 may have been accreted in a more massive subhalo. This can be understood from Figure~\ref{fig:Fig_GD1_structure} that shows the observed multiple components in the neighbourhood of this stream. These include the PS1-E stream \citep{Bernard2016, Malhan_2018_PS1}, the narrow ``spur'' \citep{WhelanBonacaGD12018} and the \cocoon component. The GD-1 stream, spur and \cocoon are remarkably correlated in the velocity space, lie along similar orbits, and possess similar stellar population. As for PS1-E, while we still do not have its line-of-sight velocity measurements, the proximity in proper-motion, distance and position space, and similarity in stellar population with GD-1, strongly suggests that PS1-E is the other secondary narrow component associated with GD-1. This implies that GD-1, \cocoon, PS1-E and spur perhaps emerged from the same progenitor GC. If true, then the true width of GD-1 is actually the overall width of the entire structure presented in Figure~\ref{fig:Fig_GD1_structure}. When we include  PS1-E in the measurement of GD1's width, it increases to $w \sim 270\pc$, which would imply that this structure was formed in a more massive, but still cored subhalo.  Since the computational expense of simulating subhalos more massive than $10^9\msun$ are beyond the scope of this paper, we defer more detailed simulations of GD-1 and its possible accretion in a more massive halo to a future study. 

It is tempting at this juncture to conclude that the observations of GD-1 and Jhelum point to them being accreted in cored DM subhalos and that this may imply alternative DM candidates. However, it has been argued that if a dwarf galaxy has a sufficiently vigorous phase of star formation to have formed GC(s), it would also have experienced episodic star bursts that could have resulted in the erasure of the DM cusp  by baryonic feedback (e.g. \citealt{Pontzen2012}). Note that a cored subhalo produced under such a scenario is still consistent with the CDM paradigm.  On the other hand, recent cosmological hydrodynamic simulations \citep[e.g.,][]{lazar_etal_20} show that ultra faint/classical dwarfs, with DM masses up to {$\sim10^{10}\msun$} and $M_*/M_{\rm halo} \simlt10^{-3}$, have formed too few stars over their lifetimes to have experienced adequate episodic baryonic feedback to unbind their DM cusps. If future spectroscopy  confirms that the parent progenitors of GD-1 and Jhelum had very low stellar mass, we may be forced to move to models beyond CDM.

\section{Summary and Conclusion}\label{sec:summary_Conclusion}

Recent observational studies have revealed interesting peculiarities and the complex morphologies of some of the low-mass streams (likely of GC origin) of the MW galaxy (\citetalias{MalhanCocoonDetection2019}, \citetalias{Bonaca2019Jhelum}, \citetalias{Malhan2019_GD1_Kshir}). These results lend credence to the {\it accretion scenario} wherein the GC progenitors of these streams initially evolved within their parent satellite galaxies and later merged with the MW (\citetalias{Carlberg2018Density_Structure}, \citetalias{MalhanCocoonDetection2019}). In this paper, motivated by recent observations and theoretical developments,  we proposed a new way to probe the nature of DM and showed that {\it the present day structural and dynamical properties of such accreted GC streams are very sensitive to the inner DM density profiles (e.g., cusp or core) of their parent satellite galaxies.} 

We generated and analysed a suite of over 100 N-body simulations, in which the host galaxy was modeled as a realistic Galactic potential (comprising of bulge, disk and DM halo), and the subhalos were modeled with cuspy and cored DM profiles with a few different values of mass and scale radius (Sec.~\ref{subsec:subhalo_models}). The main results of our simulations are presented in Section~\ref{sec:Results} and Section~\ref{sec:discussion} and summarised in Figures~\ref{fig:Fig_cuspy_streams}, \ref{fig:Fig_core_streams},  \ref{fig:Fig_comparison} and \ref{fig:Fig_stream_profiles}. 

Figure~\ref{fig:Fig_comparison} displays the physical properties of the accreted GC streams in terms of their transverse physical widths ($w$) and dynamical widths ($\sLz$, $\sv$), the latter corresponding to dispersions in $z$-component of angular momenta and line-of-sight velocities respectively. For the simulations we studied, we found that streams with $(w, \sLz, \sv)\simgt(650\pc,95\kms\kpc,4\kms)$ were only produced when GCs were accreted in CDM motivated cuspy subhalos. Somewhat smaller widths $(w, \sLz,\sv)\sim(90-500\pc,15-95\kms\kpc,<4\kms)$ resulted when GCs were accreted inside cored subhalos. This difference in the two cases emerges because the present day properties of these streams are sensitive to the depth and radial profile of the gravitational potential of the parent subhalo (i.e., $w, \sLz, \sv \propto M_0/r_0$). These physical and dynamical width estimators were substantially smaller for streams that were formed via tidal disruption of \insitu GCs (GCs that were likely formed early within the MW galaxy). 

The second main result of our study is that in contrast to the narrow, nearly one-dimensional \insitu GC streams (Fig.~\ref{fig:Fig_simpleNbody}), accreted GC streams were found to be highly structured (e.g., see Figs~\ref{fig:Fig_cuspy_streams},  \ref{fig:Fig_core_streams}). Furthermore, their transverse density profiles are at least bi-modal, containing a narrow peak (marking the presence of the primary narrow, dense component of the stream) and a broad underlying component (that reveals the presence of braoder, diffuse cocoon component of the stream). These profiles are shown in Figure~\ref{fig:Fig_stream_profiles}. Most streams also possessed secondary narrow components (e.g., ``spurs'' and parallel streams). In contrast, the \insitu GC streams are extremely narrow and uni-modal. 

These properties (dynamical widths and complex morphologies) allow us to easily differentiate accreted GC streams from \insitu streams. The complex morphologies also enable us to distinguish between accreted GC streams and tidal streams from dwarf galaxies which only have broad, smooth and uni-modal transverse density profiles (see Sec.~\ref{sec:discussion}). Intriguingly, the structure of our simulated streams seem to be quite realistic when compared with the observed streams of our Galaxy (see Sec.~\ref{sec:comp_with_obs} for details). 

While we have not run simulation to reproduce the complex morphologies of any specific observed stream, the confrontation of the measured properties of GD-1 and Jhelum streams (of likely accreted GC origin) with our simulations favour a scenario where these two streams were accreted inside {\it cored subhalos} (see Sec.~\ref{sec:comp_with_obs}, and Fig.~\ref{fig:Fig_comparison}).  This tentative result implies that either CDM cusps of the `hypothesized' parent satellites of these streams were `heated up' by episodic star formation, or we are seeing evidence for physics beyond CDM. In particular, if the analysis of stellar populations and widths of these stellar streams indicate that their parent galaxies were ultra faint/classical dwarfs with DM masses $\simlt 10^{10} M_\odot$ (and $M_*/M_{\rm halo} \simlt10^{-3}$), then we are driven to consider physics beyond CDM; in these smaller galaxies, baryonic feedback has been found to be inadequate to convert cusps to cores in their inner regions. Since our simulations show that the physical and dynamical widths of streams change only by $\simlt 10\%$ over $8\Gyr$ or more, our novel approach allows us to even probe the  density profiles of subhalos that might be partially or completely tidally dissolved by the MW galaxy at the present epoch.

We have also discovered a novel mechanism  for creating  ``gaps'' in stellar streams (see Sec.~\ref{subsec:Gaps}). Once inside the host, the accreted stellar stream and the remnant of the parent subhalo move on very similar orbits and therefore have small relative velocities. This increases the chances that the remnant of the subhalo (inside which the GC was accreted), will impact the GC stream. Such impacts, while infrequent, can cause  wide gaps to form in the stream.  There is already some evidence suggesting that such mechanisms can create gaps and ``kinks'' in the streams of the MW halo \citep{Li2020_Atlas_Aliqa}. While low-velocity impacts don't form parallel ``spur'' features \citep{Bonaca_spur_2018}, however, such structures are easily produced in the pre-accretion phase of the evolution of accreted GC streams. The accretion framework  therefore provides a new mechanism for the gap formation in streams of the MW, with DM subhalos of masses and concentrations consistent with CDM. In a future study we will explore whether  one can quantitatively distinguish between the gaps arising due to  random encounters with dark subhalos and encounters with the remnants of the cored vs. cuspy subhalos from which the streams originated.

Although our results imply a potentially powerful new probe of DM, we must note the following caveats. Modeling the GC streams under the accretion framework involves many variables, and here we have investigated only a small part of that gigantic parameter space. A comprehensive analysis that iterates over different subhalo models (varying mass and size), employs various DM density profiles and shapes for the subhalo (e.g., aspherical or triaxial models) and samples over a wide range of orbital configurations for both the GC (inside the subhalo) and the subhalo (inside the host) and considers different Galaxy halo models is necessary to assess exactly how well one can discriminate between different types of subhalo DM density profiles. It is particularly important to model GC stream in the range of galactic radii that they are currently observed, since most of our simulations have considered larger radii.  Furthermore, our results depend on how much the GC disrupts inside the parent dwarf before being stripped by the MW, and here we have tried just one GC model. A higher density GC will undergo lower tidal disruption inside the dwarf, and as a result will produce a stream with smaller dynamical widths. To examine this briefly, we ran two test simulations -- one for a cuspy subhalo and the other for a cored subhalo (with $M_0=10^9\msun$ for both the models). These subhalos contained a denser GC with tidal radius of $50\pc$ (in comparison to the value of $100\pc$ used throughout the study). We found that for the cuspy case, the resulting GC stream possessed dynamical widths of $(w,\sLz,\sv)\sim(755\pc, 159\kms\kpc,9.7\kms)$, and the corresponding values for the cored case were $(w,\sLz,\sv)\sim(86\pc, 43\kms\kpc,1.1\kms)$. Note that the values are lower than the predictions we made for $10^9\msun$ subhalos in Section~\ref{subsec:accreted_streams_properties} --- for the cuspy case it was $(w,\sLz,\sv)\sim(1825\pc, 280\kms \kpc,10.5\kms)$, and for the cored case it was $(w,\sLz,\sv)~(348\pc, 60\kms \kpc, 2.7\kms)$. However, the values corresponding to the denser GC model still lie in the regime of our general prediction for cored and cuspy subhalos. This tentatively implies that even if we lack the knowledge about the progenitor GC, the given stream should should still be useful in discerning between the core/cusp scenarios. Nonetheless, it would be interesting to extend our analysis in future by iterating over different GC models. Exploring a broader range of parameters would help us understand the sensitivity of the present day physical properties of streams (shown in Fig.~\ref{fig:Fig_comparison}) on these parameters. Also, executing higher resolution simulations may reduce some of the artifical heating that may be happening in our study. Moreover, our assumption that the Galactic potential is axisymmetric, static and smooth (ignoring the luminous and DM substructures of the host) will likely underestimate the dispersion in the physical properties of the streams \citep{Ibata_2002DM_TS}. This is because in smooth potentials, orbits are largely regular and therefore the tidal debris are expected to phase-mix close to the orbit of the progenitor system. However, if the MW's dark halo is triaxial and/or lumpy then an appreciable fraction of orbits will be chaotic \citep{Valluri2012, Fardal2015, Ngan2015,PriceWhelan2016StreamChaos}. To date, these aspects have been considered individually, but numerical methods including several of these effects simultaneously have recently opened new possibilities \citep{Renaud2015_streams_timeevolving, Carlberg2018Density_Structure, Carlberg2018-StreamSimulation}.

While $>70$ tidal streams have so far been detected to orbit in the halo of the MW (c.f. \citealt{Helmi2020_review} and references therein), very few have been studied deeply enough to detect secondary low-surface brightness features (such as the  spurs and cocoons observed in GD-1 and Jhelum streams, \citetalias{MalhanCocoonDetection2019}, \citetalias{Bonaca2019Jhelum}).  However, this is about to change due to the ongoing all-sky astrometric ESA/Gaia survey \citep{GaiaDR2_2018_Brown, GaiaDR2_2018_astrometry} and numerous recent and upcoming spectroscopic surveys (e.g., DESI, WEAVE, \citealt{DESI2016, WEAVE2014}) that will obtain both stellar radial velocities and chemical abundances, along with precise photometry from deeper  imaging surveys (e.g., CFIS, DES, LSST, \citealt{CFIS2017, DES2016, LSST2019}). Carrying out a detailed census of streams would also be extremely useful for extending the comparison between numerical efforts and observations, and for potentially unveiling the DM environment in the parent satellites of accreted GC streams. 
 
 The number of such streams, originally brought in as part of the hierarchical build of the galaxy, will also be useful to put a lower limit on past accretion events into the Galactic halo. Synergy between the advances made in the observational and simulation realm should render accreted streams as a potential direct probe of DM, and will also help us understand the birth sites of ancient stellar streams of our Galaxy. 

\section*{ACKNOWLEDGEMENTS}

We thank the referee for helpful comments and suggestions. It is a pleasure to thank P.~J.  Teuben for help with \texttt{NEMO} stellar toolbox software, and R.~G. Carlberg for extremely useful discussion. KM \& MV acknowledge useful discussions with the ``Stellar Halos group'' of Department of Astronomy (University of Michigan), and the hospitality received at the LCTP (University of Michigan) where part of this work was performed. MV is supported by NASA-ATP awards NNX15AK79G and 80NSSC20K0509 and a Catalyst grant from the Michigan Institute for Computational Discovery and Engineering (MICDE). KF and KM acknowledge support  from the $\rm{Vetenskapsr\mathring{a}de}$t (Swedish Research Council) through contract No. 638-2013-8993 and the Oskar Klein Centre for Cosmoparticle Physics. KF acknowledges support from the Jeff and Gail Kodosky Endowed Chair in Physics at the University of Texas, Austin.  She further acknowledges support from DoE grant DE- SC007859 and the LCTP at the University of Michigan.

\section*{DATA AVAILABILITY}

The data underlying this article will be shared on reasonable request to the corresponding author.

\bibliographystyle{mnras}
\bibliography{ref1} 

\begin{thebibliography}{}
\makeatletter
\relax
\def\mn@urlcharsother{\let\do\@makeother \do\$\do\&\do\#\do\^\do\_\do\%\do\~}
\def\mn@doi{\begingroup\mn@urlcharsother \@ifnextchar [ {\mn@doi@}
  {\mn@doi@[]}}
\def\mn@doi@[#1]#2{\def\@tempa{#1}\ifx\@tempa\@empty \href
  {http://dx.doi.org/#2} {doi:#2}\else \href {http://dx.doi.org/#2} {#1}\fi
  \endgroup}
\def\mn@eprint#1#2{\mn@eprint@#1:#2::\@nil}
\def\mn@eprint@arXiv#1{\href {http://arxiv.org/abs/#1} {{\tt arXiv:#1}}}
\def\mn@eprint@dblp#1{\href {http://dblp.uni-trier.de/rec/bibtex/#1.xml}
  {dblp:#1}}
\def\mn@eprint@#1:#2:#3:#4\@nil{\def\@tempa {#1}\def\@tempb {#2}\def\@tempc
  {#3}\ifx \@tempc \@empty \let \@tempc \@tempb \let \@tempb \@tempa \fi \ifx
  \@tempb \@empty \def\@tempb {arXiv}\fi \@ifundefined
  {mn@eprint@\@tempb}{\@tempb:\@tempc}{\expandafter \expandafter \csname
  mn@eprint@\@tempb\endcsname \expandafter{\@tempc}}}

\bibitem[\protect\citeauthoryear{{Amorisco}}{{Amorisco}}{2017}]{Amorisco2017}
{Amorisco} N.~C.,  2017, \mn@doi [\apj] {10.3847/1538-4357/aa745f}, \href
  {https://ui.adsabs.harvard.edu/abs/2017ApJ...844...64A} {844, 64}

\bibitem[\protect\citeauthoryear{{Amorisco} \& {Evans}}{{Amorisco} \&
  {Evans}}{2012}]{Amorisco2012Sculptor}
{Amorisco} N.~C.,  {Evans} N.~W.,  2012, \mn@doi [\mnras]
  {10.1111/j.1365-2966.2011.19684.x}, \href
  {https://ui.adsabs.harvard.edu/abs/2012MNRAS.419..184A} {419, 184}

\bibitem[\protect\citeauthoryear{Avila-Reese, Colin, Valenzuela, D'Onghia  \&
  Firmani}{Avila-Reese et~al.}{2001}]{Avila_Reese_2001_WDM}
Avila-Reese V.,  Colin P.,  Valenzuela O.,  D'Onghia E.,   Firmani C.,  2001,
  \mn@doi [The Astrophysical Journal] {10.1086/322411}, 559, 516

\bibitem[\protect\citeauthoryear{Battaglia, Helmi, Tolstoy, Irwin, Hill  \&
  Jablonka}{Battaglia et~al.}{2008}]{Battaglia_2008}
Battaglia G.,  Helmi A.,  Tolstoy E.,  Irwin M.,  Hill V.,   Jablonka P.,
  2008, \mn@doi [The Astrophysical Journal] {10.1086/590179}, 681, L13

\bibitem[\protect\citeauthoryear{Baumgardt}{Baumgardt}{2016}]{Baumgardt_2016}
Baumgardt H.,  2016, \mn@doi [Monthly Notices of the Royal Astronomical
  Society] {10.1093/mnras/stw2488}, 464, 2174

\bibitem[\protect\citeauthoryear{{Bechtol} et~al.,}{{Bechtol}
  et~al.}{2015}]{Bechtol2015}
{Bechtol} K.,  et~al., 2015, \mn@doi [\apj] {10.1088/0004-637X/807/1/50}, \href
  {https://ui.adsabs.harvard.edu/abs/2015ApJ...807...50B} {807, 50}

\bibitem[\protect\citeauthoryear{{Bellazzini}, {Ibata}, {Malhan}, {Martin},
  {Famaey}  \& {Thomas}}{{Bellazzini} et~al.}{2020}]{Bellazzini2020}
{Bellazzini} M.,  {Ibata} R.,  {Malhan} K.,  {Martin} N.,  {Famaey} B.,
  {Thomas} G.,  2020, arXiv e-prints, \href
  {https://ui.adsabs.harvard.edu/abs/2020arXiv200307871B} {p. arXiv:2003.07871}

\bibitem[\protect\citeauthoryear{{Belokurov} et~al.,}{{Belokurov}
  et~al.}{2007}]{Belokurov2007_Orphan}
{Belokurov} V.,  et~al., 2007, \mn@doi [\apj] {10.1086/511302}, \href
  {https://ui.adsabs.harvard.edu/abs/2007ApJ...658..337B} {658, 337}

\bibitem[\protect\citeauthoryear{{Belokurov}, {Erkal}, {Evans}, {Koposov}  \&
  {Deason}}{{Belokurov} et~al.}{2018}]{Belokurov2018}
{Belokurov} V.,  {Erkal} D.,  {Evans} N.~W.,  {Koposov} S.~E.,   {Deason}
  A.~J.,  2018, \mn@doi [\mnras] {10.1093/mnras/sty982}, \href
  {https://ui.adsabs.harvard.edu/abs/2018MNRAS.478..611B} {478, 611}

\bibitem[\protect\citeauthoryear{{Berezhiani}, {Famaey}  \&
  {Khoury}}{{Berezhiani} et~al.}{2018}]{Berezhiani2018}
{Berezhiani} L.,  {Famaey} B.,   {Khoury} J.,  2018, \mn@doi [\jcap]
  {10.1088/1475-7516/2018/09/021}, \href
  {https://ui.adsabs.harvard.edu/abs/2018JCAP...09..021B} {2018, 021}

\bibitem[\protect\citeauthoryear{{Bernard} et~al.,}{{Bernard}
  et~al.}{2016}]{Bernard2016}
{Bernard} E.~J.,  et~al., 2016, \mn@doi [\mnras] {10.1093/mnras/stw2134}, \href
  {http://adsabs.harvard.edu/abs/2016MNRAS.463.1759B} {463, 1759}

\bibitem[\protect\citeauthoryear{{Binney} \& {Mamon}}{{Binney} \&
  {Mamon}}{1982}]{binney_mamon_82}
{Binney} J.,  {Mamon} G.~A.,  1982, \mn@doi [\mnras] {10.1093/mnras/200.2.361},
  \href {https://ui.adsabs.harvard.edu/abs/1982MNRAS.200..361B} {200, 361}

\bibitem[\protect\citeauthoryear{{Blumenthal}, {Faber}, {Primack}  \&
  {Rees}}{{Blumenthal} et~al.}{1984}]{Blumenthal1984}
{Blumenthal} G.~R.,  {Faber} S.~M.,  {Primack} J.~R.,   {Rees} M.~J.,  1984,
  \mn@doi [\nat] {10.1038/311517a0}, \href
  {https://ui.adsabs.harvard.edu/abs/1984Natur.311..517B} {311, 517}

\bibitem[\protect\citeauthoryear{{Bonaca}, {Hogg}, {Price-Whelan}  \&
  {Conroy}}{{Bonaca} et~al.}{2019a}]{Bonaca_spur_2018}
{Bonaca} A.,  {Hogg} D.~W.,  {Price-Whelan} A.~M.,   {Conroy} C.,  2019a,
  \mn@doi [\apj] {10.3847/1538-4357/ab2873}, \href
  {https://ui.adsabs.harvard.edu/abs/2019ApJ...880...38B} {880, 38}

\bibitem[\protect\citeauthoryear{{Bonaca}, {Conroy}, {Price-Whelan}  \&
  {Hogg}}{{Bonaca} et~al.}{2019b}]{Bonaca2019Jhelum}
{Bonaca} A.,  {Conroy} C.,  {Price-Whelan} A.~M.,   {Hogg} D.~W.,  2019b,
  \mn@doi [\apjl] {10.3847/2041-8213/ab36ba}, \href
  {https://ui.adsabs.harvard.edu/abs/2019ApJ...881L..37B} {881, L37}

\bibitem[\protect\citeauthoryear{Bond, Efstathiou  \& Silk}{Bond
  et~al.}{1980}]{Bond_WDM_1980}
Bond J.~R.,  Efstathiou G.,   Silk J.,  1980, \mn@doi [Phys. Rev. Lett.]
  {10.1103/PhysRevLett.45.1980}, 45, 1980

\bibitem[\protect\citeauthoryear{{Bowden}, {Belokurov}  \& {Evans}}{{Bowden}
  et~al.}{2015}]{Bowden2015}
{Bowden} A.,  {Belokurov} V.,   {Evans} N.~W.,  2015, \mn@doi [\mnras]
  {10.1093/mnras/stv285}, \href
  {https://ui.adsabs.harvard.edu/abs/2015MNRAS.449.1391B} {449, 1391}

\bibitem[\protect\citeauthoryear{Boyarsky, Ruchayskiy  \& Iakubovskyi}{Boyarsky
  et~al.}{2009}]{Boyarsky_WDM_cores_2009}
Boyarsky A.,  Ruchayskiy O.,   Iakubovskyi D.,  2009, \mn@doi [Journal of
  Cosmology and Astroparticle Physics] {10.1088/1475-7516/2009/03/005}, 2009,
  005

\bibitem[\protect\citeauthoryear{{Burkert}}{{Burkert}}{1995}]{Burkert1995}
{Burkert} A.,  1995, \mn@doi [\apjl] {10.1086/309560}, \href
  {https://ui.adsabs.harvard.edu/abs/1995ApJ...447L..25B} {447, L25}

\bibitem[\protect\citeauthoryear{{Carlberg}}{{Carlberg}}{2012}]{Carlberg2012}
{Carlberg} R.~G.,  2012, \mn@doi [\apj] {10.1088/0004-637X/748/1/20}, \href
  {http://adsabs.harvard.edu/abs/2012ApJ...748...20C} {748, 20}

\bibitem[\protect\citeauthoryear{{Carlberg}}{{Carlberg}}{2018a}]{Carlberg2018Density_Structure}
{Carlberg} R.~G.,  2018a, arXiv e-prints, \href
  {https://ui.adsabs.harvard.edu/\#abs/2018arXiv181110084C} {p.
  arXiv:1811.10084}

\bibitem[\protect\citeauthoryear{{Carlberg}}{{Carlberg}}{2018b}]{Carlberg2018-StreamSimulation}
{Carlberg} R.~G.,  2018b, \mn@doi [\apj] {10.3847/1538-4357/aac88a}, \href
  {http://adsabs.harvard.edu/abs/2018ApJ...861...69C} {861, 69}

\bibitem[\protect\citeauthoryear{{Chang}, {Yuan}, {Xue}, {Simion}, {Kang},
  {Li}, {Zhao}  \& {Zhao}}{{Chang} et~al.}{2020}]{Chang2020_C}
{Chang} J.,  {Yuan} Z.,  {Xue} X.-X.,  {Simion} I.~T.,  {Kang} X.,  {Li} T.~S.,
   {Zhao} J.-K.,   {Zhao} G.,  2020, arXiv e-prints, \href
  {https://ui.adsabs.harvard.edu/abs/2020arXiv200302378C} {p. arXiv:2003.02378}

\bibitem[\protect\citeauthoryear{{Cicu{\'e}ndez} et~al.,}{{Cicu{\'e}ndez}
  et~al.}{2018}]{Cicuendez2018}
{Cicu{\'e}ndez} L.,  et~al., 2018, \mn@doi [\aap]
  {10.1051/0004-6361/201731450}, \href
  {https://ui.adsabs.harvard.edu/abs/2018A&A...609A..53C} {609, A53}

\bibitem[\protect\citeauthoryear{{Cole}, {Dehnen}, {Read}  \&
  {Wilkinson}}{{Cole} et~al.}{2012}]{Cole_2012_fornax}
{Cole} D.~R.,  {Dehnen} W.,  {Read} J.~I.,   {Wilkinson} M.~I.,  2012, \mn@doi
  [\mnras] {10.1111/j.1365-2966.2012.21885.x}, \href
  {https://ui.adsabs.harvard.edu/abs/2012MNRAS.426..601C} {426, 601}

\bibitem[\protect\citeauthoryear{Collaboration: et~al.,}{Collaboration:
  et~al.}{2016}]{DES2016}
Collaboration: D. E.~S.,  et~al., 2016, \mn@doi [Monthly Notices of the Royal
  Astronomical Society] {10.1093/mnras/stw641}, 460, 1270

\bibitem[\protect\citeauthoryear{{Contenta} et~al.,}{{Contenta}
  et~al.}{2018}]{Contenta2018EridanusII}
{Contenta} F.,  et~al., 2018, \mn@doi [\mnras] {10.1093/mnras/sty424}, \href
  {https://ui.adsabs.harvard.edu/abs/2018MNRAS.476.3124C} {476, 3124}

\bibitem[\protect\citeauthoryear{Crnojevi{\'{c}}, Sand, Zaritsky, Spekkens,
  Willman  \& Hargis}{Crnojevi{\'{c}} et~al.}{2016}]{Crnojevi2016}
Crnojevi{\'{c}} D.,  Sand D.~J.,  Zaritsky D.,  Spekkens K.,  Willman B.,
  Hargis J.~R.,  2016, \mn@doi [The Astrophysical Journal]
  {10.3847/2041-8205/824/1/l14}, 824, L14

\bibitem[\protect\citeauthoryear{Cusano et~al.,}{Cusano
  et~al.}{2016}]{Cusano2016}
Cusano F.,  et~al., 2016, \mn@doi [The Astrophysical Journal]
  {10.3847/0004-637x/829/1/26}, 829, 26

\bibitem[\protect\citeauthoryear{{DESI Collaboration} et~al.,}{{DESI
  Collaboration} et~al.}{2016}]{DESI2016}
{DESI Collaboration} et~al., 2016, arXiv e-prints, \href
  {https://ui.adsabs.harvard.edu/abs/2016arXiv161100036D} {p. arXiv:1611.00036}

\bibitem[\protect\citeauthoryear{{D'Onghia}, {Springel}, {Hernquist}  \&
  {Keres}}{{D'Onghia} et~al.}{2010}]{Onghia2010}
{D'Onghia} E.,  {Springel} V.,  {Hernquist} L.,   {Keres} D.,  2010, \mn@doi
  [\apj] {10.1088/0004-637X/709/2/1138}, \href
  {https://ui.adsabs.harvard.edu/abs/2010ApJ...709.1138D} {709, 1138}

\bibitem[\protect\citeauthoryear{{Dalton} et~al.,}{{Dalton}
  et~al.}{2014}]{WEAVE2014}
{Dalton} G.,  et~al., 2014, {Project overview and update on WEAVE: the next
  generation wide-field spectroscopy facility for the William Herschel
  Telescope}.
p. 91470L, \mn@doi{10.1117/12.2055132}

\bibitem[\protect\citeauthoryear{{Dehnen}}{{Dehnen}}{1993}]{Dehnen1993}
{Dehnen} W.,  1993, \mn@doi [\mnras] {10.1093/mnras/265.1.250}, \href
  {http://adsabs.harvard.edu/abs/1993MNRAS.265..250D} {265, 250}

\bibitem[\protect\citeauthoryear{{Dehnen}}{{Dehnen}}{2002}]{Dehnen_NEMO_2002}
{Dehnen} W.,  2002, \mn@doi [Journal of Computational Physics]
  {10.1006/jcph.2002.7026}, \href
  {http://adsabs.harvard.edu/abs/2002JCoPh.179...27D} {179, 27}

\bibitem[\protect\citeauthoryear{{Dehnen} \& {Binney}}{{Dehnen} \&
  {Binney}}{1998}]{Dehnen1998Massmodel}
{Dehnen} W.,  {Binney} J.,  1998, \mn@doi [\mnras]
  {10.1046/j.1365-8711.1998.01282.x}, \href
  {http://adsabs.harvard.edu/abs/1998MNRAS.294..429D} {294, 429}

\bibitem[\protect\citeauthoryear{Dehnen, Odenkirchen, Grebel  \& Rix}{Dehnen
  et~al.}{2004}]{Dehnen2004}
Dehnen W.,  Odenkirchen M.,  Grebel E.~K.,   Rix H.-W.,  2004, \mn@doi [The
  Astronomical Journal] {10.1086/383214}, 127, 2753

\bibitem[\protect\citeauthoryear{{Du}, {Schwabe}, {Niemeyer}  \&
  {B{\"u}rger}}{{Du} et~al.}{2018}]{Du_2018}
{Du} X.,  {Schwabe} B.,  {Niemeyer} J.~C.,   {B{\"u}rger} D.,  2018, \mn@doi
  [\prd] {10.1103/PhysRevD.97.063507}, \href
  {https://ui.adsabs.harvard.edu/abs/2018PhRvD..97f3507D} {97, 063507}

\bibitem[\protect\citeauthoryear{{Dubinski} \& {Carlberg}}{{Dubinski} \&
  {Carlberg}}{1991}]{Dubinski1991}
{Dubinski} J.,  {Carlberg} R.~G.,  1991, \mn@doi [\apj] {10.1086/170451}, \href
  {http://adsabs.harvard.edu/abs/1991ApJ...378..496D} {378, 496}

\bibitem[\protect\citeauthoryear{{Elbert}, {Bullock}, {Garrison-Kimmel},
  {Rocha}, {O{\~n}orbe}  \& {Peter}}{{Elbert} et~al.}{2015}]{Elbert2015}
{Elbert} O.~D.,  {Bullock} J.~S.,  {Garrison-Kimmel} S.,  {Rocha} M.,
  {O{\~n}orbe} J.,   {Peter} A. H.~G.,  2015, \mn@doi [\mnras]
  {10.1093/mnras/stv1470}, \href
  {https://ui.adsabs.harvard.edu/abs/2015MNRAS.453...29E} {453, 29}

\bibitem[\protect\citeauthoryear{{Erkal}, {Sanders}  \& {Belokurov}}{{Erkal}
  et~al.}{2016a}]{Erkal2016evolution_orbits}
{Erkal} D.,  {Sanders} J.~L.,   {Belokurov} V.,  2016a, \mn@doi [\mnras]
  {10.1093/mnras/stw1400}, \href
  {https://ui.adsabs.harvard.edu/abs/2016MNRAS.461.1590E} {461, 1590}

\bibitem[\protect\citeauthoryear{Erkal, Belokurov, Bovy  \& Sanders}{Erkal
  et~al.}{2016b}]{Erkal2016_gaps}
Erkal D.,  Belokurov V.,  Bovy J.,   Sanders J.~L.,  2016b, \mn@doi [Monthly
  Notices of the Royal Astronomical Society] {10.1093/mnras/stw1957}, 463, 102

\bibitem[\protect\citeauthoryear{{Errani} \& {Pe{\~n}arrubia}}{{Errani} \&
  {Pe{\~n}arrubia}}{2019}]{Errani2019}
{Errani} R.,  {Pe{\~n}arrubia} J.,  2019, arXiv e-prints, \href
  {https://ui.adsabs.harvard.edu/abs/2019arXiv190601642E} {p. arXiv:1906.01642}

\bibitem[\protect\citeauthoryear{{Evans}, {Ferrer}  \& {Sarkar}}{{Evans}
  et~al.}{2004}]{Evans2004}
{Evans} N.~W.,  {Ferrer} F.,   {Sarkar} S.,  2004, \mn@doi [\prd]
  {10.1103/PhysRevD.69.123501}, \href
  {https://ui.adsabs.harvard.edu/abs/2004PhRvD..69l3501E} {69, 123501}

\bibitem[\protect\citeauthoryear{{Fardal}, {Huang}  \& {Weinberg}}{{Fardal}
  et~al.}{2015}]{Fardal2015}
{Fardal} M.~A.,  {Huang} S.,   {Weinberg} M.~D.,  2015, \mn@doi [\mnras]
  {10.1093/mnras/stv1198}, \href
  {https://ui.adsabs.harvard.edu/abs/2015MNRAS.452..301F} {452, 301}

\bibitem[\protect\citeauthoryear{{Fardal}, {van der Marel}, {Sohn}  \& {del
  Pino Molina}}{{Fardal} et~al.}{2019}]{Fardal2019}
{Fardal} M.~A.,  {van der Marel} R.~P.,  {Sohn} S.~T.,   {del Pino Molina} A.,
  2019, \mn@doi [\mnras] {10.1093/mnras/stz749}, \href
  {https://ui.adsabs.harvard.edu/abs/2019MNRAS.486..936F} {486, 936}

\bibitem[\protect\citeauthoryear{{Forbes}, {Read}, {Gieles}  \&
  {Collins}}{{Forbes} et~al.}{2018}]{Forbes2018_GC_halo-mass}
{Forbes} D.~A.,  {Read} J.~I.,  {Gieles} M.,   {Collins} M. L.~M.,  2018,
  \mn@doi [\mnras] {10.1093/mnras/sty2584}, \href
  {https://ui.adsabs.harvard.edu/abs/2018MNRAS.481.5592F} {481, 5592}

\bibitem[\protect\citeauthoryear{{Gaia Collaboration}, {Brown, A. G. A.},
  {Vallenari, A.}, {Prusti, T.}, {de Bruijne, J. H. J.}  \& {et al.}}{{Gaia
  Collaboration} et~al.}{2018a}]{GaiaDR2_2018_Brown}
{Gaia Collaboration} {Brown, A. G. A.} {Vallenari, A.} {Prusti, T.} {de
  Bruijne, J. H. J.}  {et al.} 2018a, \mn@doi [A\&A]
  {10.1051/0004-6361/201833051}

\bibitem[\protect\citeauthoryear{{Gaia Collaboration} et~al.,}{{Gaia
  Collaboration} et~al.}{2018b}]{GaiaCollab2018kinematics}
{Gaia Collaboration} et~al., 2018b, \mn@doi [\aap]
  {10.1051/0004-6361/201832698}, \href
  {http://adsabs.harvard.edu/abs/2018A%26A...616A..12G} {616, A12}

\bibitem[\protect\citeauthoryear{Garrison-Kimmel, Boylan-Kolchin, Bullock  \&
  Lee}{Garrison-Kimmel et~al.}{2014}]{Garrison-Kimmel2014}
Garrison-Kimmel S.,  Boylan-Kolchin M.,  Bullock J.~S.,   Lee K.,  2014,
  \mn@doi [Monthly Notices of the Royal Astronomical Society]
  {10.1093/mnras/stt2377}, 438, 2578

\bibitem[\protect\citeauthoryear{{Gilmore}, {Wilkinson}, {Wyse}, {Kleyna},
  {Koch}, {Evans}  \& {Grebel}}{{Gilmore} et~al.}{2007}]{Gilmore_dsph_core2007}
{Gilmore} G.,  {Wilkinson} M.~I.,  {Wyse} R.~F.~G.,  {Kleyna} J.~T.,  {Koch}
  A.,  {Evans} N.~W.,   {Grebel} E.~K.,  2007, \mn@doi [\apj] {10.1086/518025},
  \href {http://adsabs.harvard.edu/abs/2007ApJ...663..948G} {663, 948}

\bibitem[\protect\citeauthoryear{{Gnedin} \& {Ostriker}}{{Gnedin} \&
  {Ostriker}}{1997}]{gnedin_ostriker_97}
{Gnedin} O.~Y.,  {Ostriker} J.~P.,  1997, \mn@doi [\apj] {10.1086/303441},
  \href {https://ui.adsabs.harvard.edu/abs/1997ApJ...474..223G} {474, 223}

\bibitem[\protect\citeauthoryear{{Goerdt}, {Moore}, {Read}, {Stadel}  \&
  {Zemp}}{{Goerdt} et~al.}{2006}]{Goerdt2006Fornax}
{Goerdt} T.,  {Moore} B.,  {Read} J.~I.,  {Stadel} J.,   {Zemp} M.,  2006,
  \mn@doi [\mnras] {10.1111/j.1365-2966.2006.10182.x}, \href
  {https://ui.adsabs.harvard.edu/abs/2006MNRAS.368.1073G} {368, 1073}

\bibitem[\protect\citeauthoryear{{Grillmair} \& {Carlin}}{{Grillmair} \&
  {Carlin}}{2016}]{GrillmairCarlin2016}
{Grillmair} C.~J.,  {Carlin} J.~L.,  2016, in {Newberg} H.~J.,  {Carlin} J.~L.,
   eds,  Astrophysics and Space Science Library Vol. 420, Tidal Streams in the
  Local Group and Beyond. p.~87 (\mn@eprint {arXiv} {1603.08936}),
  \mn@doi{10.1007/978-3-319-19336-6_4}

\bibitem[\protect\citeauthoryear{{Grillmair} \& {Johnson}}{{Grillmair} \&
  {Johnson}}{2006}]{Grillmair_NGC5466_2006}
{Grillmair} C.~J.,  {Johnson} R.,  2006, \mn@doi [\apjl] {10.1086/501439},
  \href {https://ui.adsabs.harvard.edu/abs/2006ApJ...639L..17G} {639, L17}

\bibitem[\protect\citeauthoryear{{Helmi}}{{Helmi}}{2020}]{Helmi2020_review}
{Helmi} A.,  2020, arXiv e-prints, \href
  {https://ui.adsabs.harvard.edu/abs/2020arXiv200204340H} {p. arXiv:2002.04340}

\bibitem[\protect\citeauthoryear{{Helmi}, {Babusiaux}, {Koppelman}, {Massari},
  {Veljanoski}  \& {Brown}}{{Helmi} et~al.}{2018}]{Helmi2018}
{Helmi} A.,  {Babusiaux} C.,  {Koppelman} H.~H.,  {Massari} D.,  {Veljanoski}
  J.,   {Brown} A. G.~A.,  2018, \mn@doi [\nat] {10.1038/s41586-018-0625-x},
  \href {https://ui.adsabs.harvard.edu/abs/2018Natur.563...85H} {563, 85}

\bibitem[\protect\citeauthoryear{{Hernandez} \& {Gilmore}}{{Hernandez} \&
  {Gilmore}}{1998}]{Hernandez1998}
{Hernandez} X.,  {Gilmore} G.,  1998, \mn@doi [\mnras]
  {10.1046/j.1365-8711.1998.01511.x}, \href
  {https://ui.adsabs.harvard.edu/abs/1998MNRAS.297..517H} {297, 517}

\bibitem[\protect\citeauthoryear{Hooper \& Linden}{Hooper \&
  Linden}{2015}]{Hooper2015}
Hooper D.,  Linden T.,  2015, \mn@doi [Journal of Cosmology and Astroparticle
  Physics] {10.1088/1475-7516/2015/09/016}, 2015, 016

\bibitem[\protect\citeauthoryear{{Hui}, {Ostriker}, {Tremaine}  \&
  {Witten}}{{Hui} et~al.}{2017}]{Hui2017_ULDM_FDM}
{Hui} L.,  {Ostriker} J.~P.,  {Tremaine} S.,   {Witten} E.,  2017, \mn@doi
  [\prd] {10.1103/PhysRevD.95.043541}, \href
  {https://ui.adsabs.harvard.edu/abs/2017PhRvD..95d3541H} {95, 043541}

\bibitem[\protect\citeauthoryear{{Ibata}, {Irwin}, {Lewis}  \&
  {Stolte}}{{Ibata} et~al.}{2001}]{Ibata2001Sgr}
{Ibata} R.,  {Irwin} M.,  {Lewis} G.~F.,   {Stolte} A.,  2001, \mn@doi [\apjl]
  {10.1086/318894}, \href
  {https://ui.adsabs.harvard.edu/abs/2001ApJ...547L.133I} {547, L133}

\bibitem[\protect\citeauthoryear{{Ibata}, {Lewis}, {Irwin}  \& {Quinn}}{{Ibata}
  et~al.}{2002}]{Ibata_2002DM_TS}
{Ibata} R.~A.,  {Lewis} G.~F.,  {Irwin} M.~J.,   {Quinn} T.,  2002, \mn@doi
  [\mnras] {10.1046/j.1365-8711.2002.05358.x}, \href
  {http://adsabs.harvard.edu/abs/2002MNRAS.332..915I} {332, 915}

\bibitem[\protect\citeauthoryear{Ibata, Lewis  \& Martin}{Ibata
  et~al.}{2016}]{Ibata_2016}
Ibata R.~A.,  Lewis G.~F.,   Martin N.~F.,  2016, \mn@doi [The Astrophysical
  Journal] {10.3847/0004-637x/819/1/1}, 819, 1

\bibitem[\protect\citeauthoryear{{Ibata} et~al.,}{{Ibata}
  et~al.}{2017}]{CFIS2017}
{Ibata} R.~A.,  et~al., 2017, \mn@doi [\apj] {10.3847/1538-4357/aa855c}, \href
  {https://ui.adsabs.harvard.edu/abs/2017ApJ...848..128I} {848, 128}

\bibitem[\protect\citeauthoryear{{Ibata}, {Malhan}, {Martin}  \&
  {Starkenburg}}{{Ibata} et~al.}{2018}]{Ibata_Phlegethon_2018}
{Ibata} R.~A.,  {Malhan} K.,  {Martin} N.~F.,   {Starkenburg} E.,  2018,
  \mn@doi [\apj] {10.3847/1538-4357/aadba3}, \href
  {https://ui.adsabs.harvard.edu/\#abs/2018ApJ...865...85I} {865, 85}

\bibitem[\protect\citeauthoryear{{Ibata}, {Malhan}  \& {Martin}}{{Ibata}
  et~al.}{2019}]{Ibata_Norse_streams2019}
{Ibata} R.~A.,  {Malhan} K.,   {Martin} N.~F.,  2019, \mn@doi [\apj]
  {10.3847/1538-4357/ab0080}, \href
  {https://ui.adsabs.harvard.edu/\#abs/2019ApJ...872..152I} {872, 152}

\bibitem[\protect\citeauthoryear{{Ibata}, {Thomas}, {Famaey}, {Malhan},
  {Martin}  \& {Monari}}{{Ibata} et~al.}{2020}]{IbataGD1gaps2020}
{Ibata} R.,  {Thomas} G.,  {Famaey} B.,  {Malhan} K.,  {Martin} N.,   {Monari}
  G.,  2020, arXiv e-prints, \href
  {https://ui.adsabs.harvard.edu/abs/2020arXiv200201488I} {p. arXiv:2002.01488}

\bibitem[\protect\citeauthoryear{Inoue}{Inoue}{2009}]{Inoue2009}
Inoue S.,  2009, \mn@doi [Monthly Notices of the Royal Astronomical Society]
  {10.1111/j.1365-2966.2009.15066.x}, 397, 709

\bibitem[\protect\citeauthoryear{{Ivezi{\'c}} et~al.,}{{Ivezi{\'c}}
  et~al.}{2019}]{LSST2019}
{Ivezi{\'c}} {\v{Z}}.,  et~al., 2019, \mn@doi [\apj]
  {10.3847/1538-4357/ab042c}, \href
  {https://ui.adsabs.harvard.edu/abs/2019ApJ...873..111I} {873, 111}

\bibitem[\protect\citeauthoryear{{Jardel}, {Gebhardt}, {Fabricius}, {Drory}  \&
  {Williams}}{{Jardel} et~al.}{2013}]{Jardel2013DracoCusps}
{Jardel} J.~R.,  {Gebhardt} K.,  {Fabricius} M.~H.,  {Drory} N.,   {Williams}
  M.~J.,  2013, \mn@doi [\apj] {10.1088/0004-637X/763/2/91}, \href
  {https://ui.adsabs.harvard.edu/abs/2013ApJ...763...91J} {763, 91}

\bibitem[\protect\citeauthoryear{{Johnston}, {Sackett}  \&
  {Bullock}}{{Johnston} et~al.}{2001}]{Johnston2001}
{Johnston} K.~V.,  {Sackett} P.~D.,   {Bullock} J.~S.,  2001, \mn@doi [\apj]
  {10.1086/321644}, \href
  {https://ui.adsabs.harvard.edu/abs/2001ApJ...557..137J} {557, 137}

\bibitem[\protect\citeauthoryear{{Johnston}, {Spergel}  \& {Haydn}}{{Johnston}
  et~al.}{2002}]{Johnston2002DM_TS}
{Johnston} K.~V.,  {Spergel} D.~N.,   {Haydn} C.,  2002, \mn@doi [\apj]
  {10.1086/339791}, \href {http://adsabs.harvard.edu/abs/2002ApJ...570..656J}
  {570, 656}

\bibitem[\protect\citeauthoryear{{Kazantzidis}, {Mayer}, {Mastropietro},
  {Diemand}, {Stadel}  \& {Moore}}{{Kazantzidis}
  et~al.}{2004}]{Kazantzidis2004}
{Kazantzidis} S.,  {Mayer} L.,  {Mastropietro} C.,  {Diemand} J.,  {Stadel} J.,
    {Moore} B.,  2004, \mn@doi [\apj] {10.1086/420840}, \href
  {https://ui.adsabs.harvard.edu/abs/2004ApJ...608..663K} {608, 663}

\bibitem[\protect\citeauthoryear{{King}}{{King}}{1966}]{King1966}
{King} I.~R.,  1966, \mn@doi [\aj] {10.1086/109857}, \href
  {http://adsabs.harvard.edu/abs/1966AJ.....71...64K} {71, 64}

\bibitem[\protect\citeauthoryear{{Kleyna}, {Wilkinson}, {Gilmore}  \&
  {Evans}}{{Kleyna} et~al.}{2003}]{Kleyna2003UrsaMinorCore}
{Kleyna} J.~T.,  {Wilkinson} M.~I.,  {Gilmore} G.,   {Evans} N.~W.,  2003,
  \mn@doi [\apjl] {10.1086/375522}, \href
  {https://ui.adsabs.harvard.edu/abs/2003ApJ...588L..21K} {588, L21}

\bibitem[\protect\citeauthoryear{{Klypin}, {Kravtsov}, {Valenzuela}  \&
  {Prada}}{{Klypin} et~al.}{1999}]{Klypin1999}
{Klypin} A.,  {Kravtsov} A.~V.,  {Valenzuela} O.,   {Prada} F.,  1999, \mn@doi
  [\apj] {10.1086/307643}, \href
  {https://ui.adsabs.harvard.edu/abs/1999ApJ...522...82K} {522, 82}

\bibitem[\protect\citeauthoryear{{Koch}, {Kleyna}, {Wilkinson}, {Grebel},
  {Gilmore}, {Evans}, {Wyse}  \& {Harbeck}}{{Koch} et~al.}{2007}]{Koch2007}
{Koch} A.,  {Kleyna} J.~T.,  {Wilkinson} M.~I.,  {Grebel} E.~K.,  {Gilmore}
  G.~F.,  {Evans} N.~W.,  {Wyse} R. F.~G.,   {Harbeck} D.~R.,  2007, \mn@doi
  [\aj] {10.1086/519380}, \href
  {https://ui.adsabs.harvard.edu/abs/2007AJ....134..566K} {134, 566}

\bibitem[\protect\citeauthoryear{Koposov et~al.,}{Koposov
  et~al.}{2011}]{Koposov_2011}
Koposov S.~E.,  et~al., 2011, \mn@doi [The Astrophysical Journal]
  {10.1088/0004-637x/736/2/146}, 736, 146

\bibitem[\protect\citeauthoryear{{Koposov} et~al.,}{{Koposov}
  et~al.}{2019}]{Koposov2019Orphan}
{Koposov} S.~E.,  et~al., 2019, \mn@doi [\mnras] {10.1093/mnras/stz457}, \href
  {https://ui.adsabs.harvard.edu/abs/2019MNRAS.485.4726K} {485, 4726}

\bibitem[\protect\citeauthoryear{{Kruijssen} et~al.,}{{Kruijssen}
  et~al.}{2020}]{Kruijssen2020_Kraken}
{Kruijssen} J.~M.~D.,  et~al., 2020, arXiv e-prints, \href
  {https://ui.adsabs.harvard.edu/abs/2020arXiv200301119K} {p. arXiv:2003.01119}

\bibitem[\protect\citeauthoryear{{K{\"u}pper}, {Lane}  \&
  {Heggie}}{{K{\"u}pper} et~al.}{2012}]{Kupper2012}
{K{\"u}pper} A. H.~W.,  {Lane} R.~R.,   {Heggie} D.~C.,  2012, \mn@doi [\mnras]
  {10.1111/j.1365-2966.2011.20242.x}, \href
  {https://ui.adsabs.harvard.edu/abs/2012MNRAS.420.2700K} {420, 2700}

\bibitem[\protect\citeauthoryear{{Kuzma}, {Da Costa}, {Keller}  \&
  {Maunder}}{{Kuzma} et~al.}{2015}]{Kuzma2015_Pal5}
{Kuzma} P.~B.,  {Da Costa} G.~S.,  {Keller} S.~C.,   {Maunder} E.,  2015,
  \mn@doi [\mnras] {10.1093/mnras/stu2343}, \href
  {https://ui.adsabs.harvard.edu/abs/2015MNRAS.446.3297K} {446, 3297}

\bibitem[\protect\citeauthoryear{{Lazar} et~al.,}{{Lazar}
  et~al.}{2020}]{lazar_etal_20}
{Lazar} A.,  et~al., 2020, arXiv e-prints, \href
  {https://ui.adsabs.harvard.edu/abs/2020arXiv200410817L} {p. arXiv:2004.10817}

\bibitem[\protect\citeauthoryear{{Leaman}, {VandenBerg}  \& {Mendel}}{{Leaman}
  et~al.}{2013}]{Leaman2013}
{Leaman} R.,  {VandenBerg} D.~A.,   {Mendel} J.~T.,  2013, \mn@doi [\mnras]
  {10.1093/mnras/stt1540}, \href
  {https://ui.adsabs.harvard.edu/abs/2013MNRAS.436..122L} {436, 122}

\bibitem[\protect\citeauthoryear{{Li} et~al.,}{{Li}
  et~al.}{2020}]{Li2020_Atlas_Aliqa}
{Li} T.~S.,  et~al., 2020, arXiv e-prints, \href
  {https://ui.adsabs.harvard.edu/abs/2020arXiv200610763L} {p. arXiv:2006.10763}

\bibitem[\protect\citeauthoryear{{Lindegren}, {Hernandez}, {Bombrun, A.},
  {Klioner, S.}, {Bastian, U.}  \& {Ramos-Lerate, M.}}{{Lindegren}
  et~al.}{2018}]{GaiaDR2_2018_astrometry}
{Lindegren} L.,  {Hernandez} J.,  {Bombrun, A.} {Klioner, S.} {Bastian, U.}
  {Ramos-Lerate, M.} 2018, \mn@doi [A\&A] {10.1051/0004-6361/201832727}

\bibitem[\protect\citeauthoryear{{Lora}, {Grebel}, {Schmeja}  \& {Koch}}{{Lora}
  et~al.}{2019}]{Lora2019}
{Lora} V.,  {Grebel} E.~K.,  {Schmeja} S.,   {Koch} A.,  2019, arXiv e-prints,
  \href {https://ui.adsabs.harvard.edu/abs/2019arXiv190410560L} {p.
  arXiv:1904.10560}

\bibitem[\protect\citeauthoryear{{Mackey} \& {Gilmore}}{{Mackey} \&
  {Gilmore}}{2003}]{Mackey_Gilmore2003}
{Mackey} A.~D.,  {Gilmore} G.~F.,  2003, \mn@doi [\mnras]
  {10.1046/j.1365-8711.2003.06275.x}, \href
  {https://ui.adsabs.harvard.edu/abs/2003MNRAS.340..175M} {340, 175}

\bibitem[\protect\citeauthoryear{{Majewski}, {Skrutskie}, {Weinberg}  \&
  {Ostheimer}}{{Majewski} et~al.}{2003}]{Majewski2003}
{Majewski} S.~R.,  {Skrutskie} M.~F.,  {Weinberg} M.~D.,   {Ostheimer} J.~C.,
  2003, \mn@doi [\apj] {10.1086/379504}, \href
  {https://ui.adsabs.harvard.edu/abs/2003ApJ...599.1082M} {599, 1082}

\bibitem[\protect\citeauthoryear{{Malhan} \& {Ibata}}{{Malhan} \&
  {Ibata}}{2019}]{Malhan2018PotentialGD1}
{Malhan} K.,  {Ibata} R.~A.,  2019, \mn@doi [\mnras] {10.1093/mnras/stz1035},
  \href {https://ui.adsabs.harvard.edu/abs/2019MNRAS.486.2995M} {486, 2995}

\bibitem[\protect\citeauthoryear{{Malhan}, {Ibata}, {Goldman}, {Martin},
  {Magnier}  \& {Chambers}}{{Malhan} et~al.}{2018a}]{Malhan_2018_PS1}
{Malhan} K.,  {Ibata} R.~A.,  {Goldman} B.,  {Martin} N.~F.,  {Magnier} E.,
  {Chambers} K.,  2018a, \mn@doi [\mnras] {10.1093/mnras/sty1338}, \href
  {https://ui.adsabs.harvard.edu/abs/2018MNRAS.478.3862M} {478, 3862}

\bibitem[\protect\citeauthoryear{{Malhan}, {Ibata}  \& {Martin}}{{Malhan}
  et~al.}{2018b}]{Malhan_Ghostly_2018}
{Malhan} K.,  {Ibata} R.~A.,   {Martin} N.~F.,  2018b, \mn@doi [\mnras]
  {10.1093/mnras/sty2474}, \href
  {https://ui.adsabs.harvard.edu/abs/2018MNRAS.481.3442M} {481, 3442}

\bibitem[\protect\citeauthoryear{Malhan, Ibata, Carlberg, Valluri  \&
  Freese}{Malhan et~al.}{2019a}]{MalhanCocoonDetection2019}
Malhan K.,  Ibata R.~A.,  Carlberg R.~G.,  Valluri M.,   Freese K.,  2019a,
  \mn@doi [The Astrophysical Journal] {10.3847/1538-4357/ab2e07}, 881, 106

\bibitem[\protect\citeauthoryear{Malhan, Ibata, Carlberg, Bellazzini, Famaey
  \& Martin}{Malhan et~al.}{2019b}]{Malhan2019_GD1_Kshir}
Malhan K.,  Ibata R.~A.,  Carlberg R.~G.,  Bellazzini M.,  Famaey B.,   Martin
  N.~F.,  2019b, \mn@doi [The Astrophysical Journal]
  {10.3847/2041-8213/ab530e}, 886, L7

\bibitem[\protect\citeauthoryear{Mateo}{Mateo}{1998}]{Mateo1998}
Mateo M.,  1998, \mn@doi [Annual Review of Astronomy and Astrophysics]
  {10.1146/annurev.astro.36.1.435}, 36, 435

\bibitem[\protect\citeauthoryear{{Moore}}{{Moore}}{1994}]{Moore1994}
{Moore} B.,  1994, \mn@doi [\nat] {10.1038/370629a0}, \href
  {https://ui.adsabs.harvard.edu/abs/1994Natur.370..629M} {370, 629}

\bibitem[\protect\citeauthoryear{{Moore}, {Ghigna}, {Governato}, {Lake},
  {Quinn}, {Stadel}  \& {Tozzi}}{{Moore} et~al.}{1999}]{Moore1999}
{Moore} B.,  {Ghigna} S.,  {Governato} F.,  {Lake} G.,  {Quinn} T.,  {Stadel}
  J.,   {Tozzi} P.,  1999, \mn@doi [\apjl] {10.1086/312287}, \href
  {https://ui.adsabs.harvard.edu/abs/1999ApJ...524L..19M} {524, L19}

\bibitem[\protect\citeauthoryear{{Myeong}, {Vasiliev}, {Iorio}, {Evans}  \&
  {Belokurov}}{{Myeong} et~al.}{2019}]{Myeong2019}
{Myeong} G.~C.,  {Vasiliev} E.,  {Iorio} G.,  {Evans} N.~W.,   {Belokurov} V.,
  2019, \mn@doi [\mnras] {10.1093/mnras/stz1770}, \href
  {https://ui.adsabs.harvard.edu/abs/2019MNRAS.488.1235M} {488, 1235}

\bibitem[\protect\citeauthoryear{{Navarro}, {Eke}  \& {Frenk}}{{Navarro}
  et~al.}{1996a}]{Navarro1996_cusp2core}
{Navarro} J.~F.,  {Eke} V.~R.,   {Frenk} C.~S.,  1996a, \mn@doi [\mnras]
  {10.1093/mnras/283.3.L72}, \href
  {https://ui.adsabs.harvard.edu/abs/1996MNRAS.283L..72N} {283, L72}

\bibitem[\protect\citeauthoryear{{Navarro}, {Frenk}  \& {White}}{{Navarro}
  et~al.}{1996b}]{NFW1996}
{Navarro} J.~F.,  {Frenk} C.~S.,   {White} S.~D.~M.,  1996b, \mn@doi [\apj]
  {10.1086/177173}, \href {http://adsabs.harvard.edu/abs/1996ApJ...462..563N}
  {462, 563}

\bibitem[\protect\citeauthoryear{{Ngan} \& {Carlberg}}{{Ngan} \&
  {Carlberg}}{2014}]{Ngan2014}
{Ngan} W.~H.~W.,  {Carlberg} R.~G.,  2014, \mn@doi [\apj]
  {10.1088/0004-637X/788/2/181}, \href
  {https://ui.adsabs.harvard.edu/abs/2014ApJ...788..181N} {788, 181}

\bibitem[\protect\citeauthoryear{{Ngan}, {Bozek}, {Carlberg}, {Wyse}, {Szalay}
  \& {Madau}}{{Ngan} et~al.}{2015}]{Ngan2015}
{Ngan} W.,  {Bozek} B.,  {Carlberg} R.~G.,  {Wyse} R. F.~G.,  {Szalay} A.~S.,
  {Madau} P.,  2015, \mn@doi [\apj] {10.1088/0004-637X/803/2/75}, \href
  {https://ui.adsabs.harvard.edu/abs/2015ApJ...803...75N} {803, 75}

\bibitem[\protect\citeauthoryear{{Nipoti} \& {Binney}}{{Nipoti} \&
  {Binney}}{2015}]{Nipoti2015}
{Nipoti} C.,  {Binney} J.,  2015, \mn@doi [\mnras] {10.1093/mnras/stu2217},
  \href {https://ui.adsabs.harvard.edu/abs/2015MNRAS.446.1820N} {446, 1820}

\bibitem[\protect\citeauthoryear{{Odenkirchen} et~al.,}{{Odenkirchen}
  et~al.}{2001}]{Odenkirchen2001_Pal5}
{Odenkirchen} M.,  et~al., 2001, \mn@doi [\apjl] {10.1086/319095}, \href
  {https://ui.adsabs.harvard.edu/abs/2001ApJ...548L.165O} {548, L165}

\bibitem[\protect\citeauthoryear{{Palau} \& {Miralda-Escud{\'e}}}{{Palau} \&
  {Miralda-Escud{\'e}}}{2019}]{Palau2019M68streamFjorm}
{Palau} C.~G.,  {Miralda-Escud{\'e}} J.,  2019, arXiv e-prints, \href
  {https://ui.adsabs.harvard.edu/abs/2019arXiv190501193P} {p. arXiv:1905.01193}

\bibitem[\protect\citeauthoryear{{Pascale}, {Posti}, {Nipoti}  \&
  {Binney}}{{Pascale} et~al.}{2018}]{Pascale2018}
{Pascale} R.,  {Posti} L.,  {Nipoti} C.,   {Binney} J.,  2018, \mn@doi [\mnras]
  {10.1093/mnras/sty1860}, \href
  {https://ui.adsabs.harvard.edu/abs/2018MNRAS.480..927P} {480, 927}

\bibitem[\protect\citeauthoryear{{Pe{\~n}arrubia}, {Benson}, {Walker},
  {Gilmore}, {McConnachie}  \& {Mayer}}{{Pe{\~n}arrubia}
  et~al.}{2010}]{Penarrubia2010}
{Pe{\~n}arrubia} J.,  {Benson} A.~J.,  {Walker} M.~G.,  {Gilmore} G.,
  {McConnachie} A.~W.,   {Mayer} L.,  2010, \mn@doi [\mnras]
  {10.1111/j.1365-2966.2010.16762.x}, \href
  {https://ui.adsabs.harvard.edu/abs/2010MNRAS.406.1290P} {406, 1290}

\bibitem[\protect\citeauthoryear{{Pe{\~n}arrubia}, {Varri}, {Breen}, {Ferguson}
   \& {S{\'a}nchez-Janssen}}{{Pe{\~n}arrubia} et~al.}{2017}]{Penarrubia2017}
{Pe{\~n}arrubia} J.,  {Varri} A.~L.,  {Breen} P.~G.,  {Ferguson} A. M.~N.,
  {S{\'a}nchez-Janssen} R.,  2017, \mn@doi [\mnras] {10.1093/mnrasl/slx094},
  \href {https://ui.adsabs.harvard.edu/abs/2017MNRAS.471L..31P} {471, L31}

\bibitem[\protect\citeauthoryear{{Petts}, {Read}  \& {Gualandris}}{{Petts}
  et~al.}{2016}]{Petts2016}
{Petts} J.~A.,  {Read} J.~I.,   {Gualandris} A.,  2016, \mn@doi [\mnras]
  {10.1093/mnras/stw2011}, \href
  {https://ui.adsabs.harvard.edu/abs/2016MNRAS.463..858P} {463, 858}

\bibitem[\protect\citeauthoryear{{Phipps}, {Khochfar}, {Varri}  \& {Dalla
  Vecchia}}{{Phipps} et~al.}{2019}]{Phipps2019}
{Phipps} F.,  {Khochfar} S.,  {Varri} A.~L.,   {Dalla Vecchia} C.,  2019, arXiv
  e-prints, \href {https://ui.adsabs.harvard.edu/abs/2019arXiv191009924P} {p.
  arXiv:1910.09924}

\bibitem[\protect\citeauthoryear{{Pontzen} \& {Governato}}{{Pontzen} \&
  {Governato}}{2012}]{Pontzen2012}
{Pontzen} A.,  {Governato} F.,  2012, \mn@doi [\mnras]
  {10.1111/j.1365-2966.2012.20571.x}, \href
  {https://ui.adsabs.harvard.edu/abs/2012MNRAS.421.3464P} {421, 3464}

\bibitem[\protect\citeauthoryear{{Price-Whelan} \& {Bonaca}}{{Price-Whelan} \&
  {Bonaca}}{2018}]{WhelanBonacaGD12018}
{Price-Whelan} A.~M.,  {Bonaca} A.,  2018, \mn@doi [\apjl]
  {10.3847/2041-8213/aad7b5}, \href
  {http://adsabs.harvard.edu/abs/2018ApJ...863L..20P} {863, L20}

\bibitem[\protect\citeauthoryear{{Price-Whelan}, {Johnston}, {Valluri},
  {Pearson}, {K{\"u}pper}  \& {Hogg}}{{Price-Whelan}
  et~al.}{2016}]{PriceWhelan2016StreamChaos}
{Price-Whelan} A.~M.,  {Johnston} K.~V.,  {Valluri} M.,  {Pearson} S.,
  {K{\"u}pper} A. H.~W.,   {Hogg} D.~W.,  2016, \mn@doi [\mnras]
  {10.1093/mnras/stv2383}, \href
  {https://ui.adsabs.harvard.edu/abs/2016MNRAS.455.1079P} {455, 1079}

\bibitem[\protect\citeauthoryear{{Read} \& {Gilmore}}{{Read} \&
  {Gilmore}}{2005}]{Read2005cusp2core}
{Read} J.~I.,  {Gilmore} G.,  2005, \mn@doi [\mnras]
  {10.1111/j.1365-2966.2004.08424.x}, \href
  {https://ui.adsabs.harvard.edu/abs/2005MNRAS.356..107R} {356, 107}

\bibitem[\protect\citeauthoryear{Read, Wilkinson, Evans, Gilmore  \&
  Kleyna}{Read et~al.}{2006}]{Read2006}
Read J.~I.,  Wilkinson M.~I.,  Evans N.~W.,  Gilmore G.,   Kleyna J.~T.,  2006,
  \mn@doi [Monthly Notices of the Royal Astronomical Society]
  {10.1111/j.1365-2966.2005.09861.x}, 366, 429

\bibitem[\protect\citeauthoryear{{Read}, {Agertz}  \& {Collins}}{{Read}
  et~al.}{2016}]{Read_core_2016}
{Read} J.~I.,  {Agertz} O.,   {Collins} M.~L.~M.,  2016, \mn@doi [\mnras]
  {10.1093/mnras/stw713}, \href
  {https://ui.adsabs.harvard.edu/abs/2016MNRAS.459.2573R} {459, 2573}

\bibitem[\protect\citeauthoryear{{Read}, {Walker}  \& {Steger}}{{Read}
  et~al.}{2018}]{Read2018Draco}
{Read} J.~I.,  {Walker} M.~G.,   {Steger} P.,  2018, \mn@doi [\mnras]
  {10.1093/mnras/sty2286}, \href
  {https://ui.adsabs.harvard.edu/abs/2018MNRAS.481..860R} {481, 860}

\bibitem[\protect\citeauthoryear{{Renaud} \& {Gieles}}{{Renaud} \&
  {Gieles}}{2015}]{Renaud2015_streams_timeevolving}
{Renaud} F.,  {Gieles} M.,  2015, \mn@doi [\mnras] {10.1093/mnras/stv245},
  \href {https://ui.adsabs.harvard.edu/abs/2015MNRAS.448.3416R} {448, 3416}

\bibitem[\protect\citeauthoryear{{Renaud}, {Agertz}  \& {Gieles}}{{Renaud}
  et~al.}{2017}]{Renaud2017}
{Renaud} F.,  {Agertz} O.,   {Gieles} M.,  2017, \mn@doi [\mnras]
  {10.1093/mnras/stw2969}, \href
  {https://ui.adsabs.harvard.edu/abs/2017MNRAS.465.3622R} {465, 3622}

\bibitem[\protect\citeauthoryear{{Searle} \& {Zinn}}{{Searle} \&
  {Zinn}}{1978}]{Searle1978}
{Searle} L.,  {Zinn} R.,  1978, \mn@doi [\apj] {10.1086/156499}, \href
  {http://adsabs.harvard.edu/abs/1978ApJ...225..357S} {225, 357}

\bibitem[\protect\citeauthoryear{{Shipp} et~al.,}{{Shipp}
  et~al.}{2019}]{Shipp2019}
{Shipp} N.,  et~al., 2019, arXiv e-prints, \href
  {https://ui.adsabs.harvard.edu/abs/2019arXiv190709488S} {p. arXiv:1907.09488}

\bibitem[\protect\citeauthoryear{{Shipp}, {Price-Whelan}, {Tavangar}, {Mateu}
  \& {Drlica-Wagner}}{{Shipp} et~al.}{2020}]{Shipp2020_Pal13spur}
{Shipp} N.,  {Price-Whelan} A.,  {Tavangar} K.,  {Mateu} C.,   {Drlica-Wagner}
  A.,  2020, arXiv e-prints, \href
  {https://ui.adsabs.harvard.edu/abs/2020arXiv200612501S} {p. arXiv:2006.12501}

\bibitem[\protect\citeauthoryear{{Simon}}{{Simon}}{2018}]{simon_18}
{Simon} J.~D.,  2018, \mn@doi [\apj] {10.3847/1538-4357/aacdfb}, \href
  {https://ui.adsabs.harvard.edu/abs/2018ApJ...863...89S} {863, 89}

\bibitem[\protect\citeauthoryear{{Simon} \& {Geha}}{{Simon} \&
  {Geha}}{2007}]{Simon2007}
{Simon} J.~D.,  {Geha} M.,  2007, \mn@doi [\apj] {10.1086/521816}, \href
  {https://ui.adsabs.harvard.edu/abs/2007ApJ...670..313S} {670, 313}

\bibitem[\protect\citeauthoryear{{Simpson}, {De Silva}, {Martell}, {Navin}  \&
  {Zucker}}{{Simpson} et~al.}{2017}]{Simpson2017_Lowwmass_GC}
{Simpson} J.~D.,  {De Silva} G.,  {Martell} S.~L.,  {Navin} C.~A.,   {Zucker}
  D.~B.,  2017, \mn@doi [\mnras] {10.1093/mnras/stx2174}, \href
  {https://ui.adsabs.harvard.edu/abs/2017MNRAS.472.2856S} {472, 2856}

\bibitem[\protect\citeauthoryear{{Sollima}, {Dalessandro}, {Beccari}  \&
  {Pallanca}}{{Sollima} et~al.}{2017}]{Sollima2017}
{Sollima} A.,  {Dalessandro} E.,  {Beccari} G.,   {Pallanca} C.,  2017, \mn@doi
  [\mnras] {10.1093/mnras/stw2628}, \href
  {https://ui.adsabs.harvard.edu/abs/2017MNRAS.464.3871S} {464, 3871}

\bibitem[\protect\citeauthoryear{{Sparke} \& {Gallagher}}{{Sparke} \&
  {Gallagher}}{2007}]{sparke_gallagher_07}
{Sparke} L.~S.,  {Gallagher} John~S. I.,  2007, {Galaxies in the Universe}.
{Cambridge University Press: Cambridge, UK}

\bibitem[\protect\citeauthoryear{{Spergel} \& {Steinhardt}}{{Spergel} \&
  {Steinhardt}}{2000}]{Spergel2000SIDM}
{Spergel} D.~N.,  {Steinhardt} P.~J.,  2000, \mn@doi [\prl]
  {10.1103/PhysRevLett.84.3760}, \href
  {https://ui.adsabs.harvard.edu/abs/2000PhRvL..84.3760S} {84, 3760}

\bibitem[\protect\citeauthoryear{Springel et~al.,}{Springel
  et~al.}{2008}]{Springel2008}
Springel V.,  et~al., 2008, \mn@doi [Monthly Notices of the Royal Astronomical
  Society] {10.1111/j.1365-2966.2008.14066.x}, 391, 1685

\bibitem[\protect\citeauthoryear{{Teuben}}{{Teuben}}{1995}]{Teuben_1995}
{Teuben} P.,  1995, in {Shaw} R.~A.,  {Payne} H.~E.,   {Hayes} J.~J.~E.,  eds,
  Astronomical Society of the Pacific Conference Series Vol. 77, Astronomical
  Data Analysis Software and Systems IV. p.~398

\bibitem[\protect\citeauthoryear{{Thomas}, {Ibata}, {Famaey}, {Martin}  \&
  {Lewis}}{{Thomas} et~al.}{2016}]{Thomas_Pal52016}
{Thomas} G.~F.,  {Ibata} R.,  {Famaey} B.,  {Martin} N.~F.,   {Lewis} G.~F.,
  2016, \mn@doi [\mnras] {10.1093/mnras/stw1189}, \href
  {https://ui.adsabs.harvard.edu/abs/2016MNRAS.460.2711T} {460, 2711}

\bibitem[\protect\citeauthoryear{{Tulin} \& {Yu}}{{Tulin} \&
  {Yu}}{2018}]{SIDM_review2018}
{Tulin} S.,  {Yu} H.-B.,  2018, \mn@doi [\physrep]
  {10.1016/j.physrep.2017.11.004}, \href
  {https://ui.adsabs.harvard.edu/abs/2018PhR...730....1T} {730, 1}

\bibitem[\protect\citeauthoryear{{Valluri}}{{Valluri}}{1993}]{valluri_93}
{Valluri} M.,  1993, \mn@doi [\apj] {10.1086/172569}, \href
  {https://ui.adsabs.harvard.edu/abs/1993ApJ...408...57V} {408, 57}

\bibitem[\protect\citeauthoryear{{Valluri}, {Debattista}, {Quinn},
  {Ro{\v{s}}kar}  \& {Wadsley}}{{Valluri} et~al.}{2012}]{Valluri2012}
{Valluri} M.,  {Debattista} V.~P.,  {Quinn} T.~R.,  {Ro{\v{s}}kar} R.,
  {Wadsley} J.,  2012, \mn@doi [\mnras] {10.1111/j.1365-2966.2011.19853.x},
  \href {https://ui.adsabs.harvard.edu/abs/2012MNRAS.419.1951V} {419, 1951}

\bibitem[\protect\citeauthoryear{{Varghese}, {Ibata}  \& {Lewis}}{{Varghese}
  et~al.}{2011}]{Varghese2011}
{Varghese} A.,  {Ibata} R.,   {Lewis} G.~F.,  2011, \mn@doi [\mnras]
  {10.1111/j.1365-2966.2011.19097.x}, \href
  {https://ui.adsabs.harvard.edu/abs/2011MNRAS.417..198V} {417, 198}

\bibitem[\protect\citeauthoryear{{Walker} \& {Pe{\~n}arrubia}}{{Walker} \&
  {Pe{\~n}arrubia}}{2011}]{Walker2011}
{Walker} M.~G.,  {Pe{\~n}arrubia} J.,  2011, \mn@doi [\apj]
  {10.1088/0004-637X/742/1/20}, \href
  {https://ui.adsabs.harvard.edu/abs/2011ApJ...742...20W} {742, 20}

\bibitem[\protect\citeauthoryear{{Walker}, {McGaugh}, {Mateo}, {Olszewski}  \&
  {Kuzio de Naray}}{{Walker} et~al.}{2010}]{Walker2010}
{Walker} M.~G.,  {McGaugh} S.~S.,  {Mateo} M.,  {Olszewski} E.~W.,   {Kuzio de
  Naray} R.,  2010, \mn@doi [\apj] {10.1088/2041-8205/717/2/L87}, \href
  {https://ui.adsabs.harvard.edu/abs/2010ApJ...717L..87W} {717, L87}

\bibitem[\protect\citeauthoryear{{Weldrake}, {de Blok}  \& {Walter}}{{Weldrake}
  et~al.}{2003}]{Weldrake2003}
{Weldrake} D.~T.~F.,  {de Blok} W.~J.~G.,   {Walter} F.,  2003, \mn@doi
  [\mnras] {10.1046/j.1365-8711.2003.06170.x}, \href
  {https://ui.adsabs.harvard.edu/abs/2003MNRAS.340...12W} {340, 12}

\bibitem[\protect\citeauthoryear{{Wheeler} et~al.,}{{Wheeler}
  et~al.}{2018}]{Wheeler2018}
{Wheeler} C.,  et~al., 2018, arXiv e-prints, \href
  {https://ui.adsabs.harvard.edu/abs/2018arXiv181202749W} {p. arXiv:1812.02749}

\bibitem[\protect\citeauthoryear{White \& Rees}{White \&
  Rees}{1978}]{White_CDM_candidate1978}
White S. D.~M.,  Rees M.~J.,  1978, \mn@doi [Monthly Notices of the Royal
  Astronomical Society] {10.1093/mnras/183.3.341}, 183, 341

\bibitem[\protect\citeauthoryear{{Whitmore}, {Zhang}, {Leitherer}, {Fall},
  {Schweizer}  \& {Miller}}{{Whitmore} et~al.}{1999}]{Whitmore1999}
{Whitmore} B.~C.,  {Zhang} Q.,  {Leitherer} C.,  {Fall} S.~M.,  {Schweizer} F.,
    {Miller} B.~W.,  1999, \mn@doi [\aj] {10.1086/301041}, \href
  {https://ui.adsabs.harvard.edu/abs/1999AJ....118.1551W} {118, 1551}

\bibitem[\protect\citeauthoryear{{de Blok}, {McGaugh}  \& {Rubin}}{{de Blok}
  et~al.}{2001}]{deBlok2001}
{de Blok} W.~J.~G.,  {McGaugh} S.~S.,   {Rubin} V.~C.,  2001, \mn@doi [\aj]
  {10.1086/323450}, \href
  {https://ui.adsabs.harvard.edu/abs/2001AJ....122.2396D} {122, 2396}

\bibitem[\protect\citeauthoryear{{de Boer}, {Erkal}  \& {Gieles}}{{de Boer}
  et~al.}{2020}]{deBoer2020}
{de Boer} T.~J.~L.,  {Erkal} D.,   {Gieles} M.,  2020, \mn@doi [\mnras]
  {10.1093/mnras/staa917}, \href
  {https://ui.adsabs.harvard.edu/abs/2020MNRAS.494.5315D} {494, 5315}

\bibitem[\protect\citeauthoryear{{van den Bosch}, {Lewis}, {Lake}  \&
  {Stadel}}{{van den Bosch} et~al.}{1999}]{vandenBosch_etal_99}
{van den Bosch} F.~C.,  {Lewis} G.~F.,  {Lake} G.,   {Stadel} J.,  1999,
  \mn@doi [\apj] {10.1086/307023}, \href
  {https://ui.adsabs.harvard.edu/abs/1999ApJ...515...50V} {515, 50}

\makeatother
\end{thebibliography}


\appendix
\section*{Appendix}\label{sec:Appendix1}

\section{Numerical Considerations}
\label{appendix:numerics}

\begin{figure*}
\begin{center}
\hbox{
\includegraphics[width=0.51\hsize]{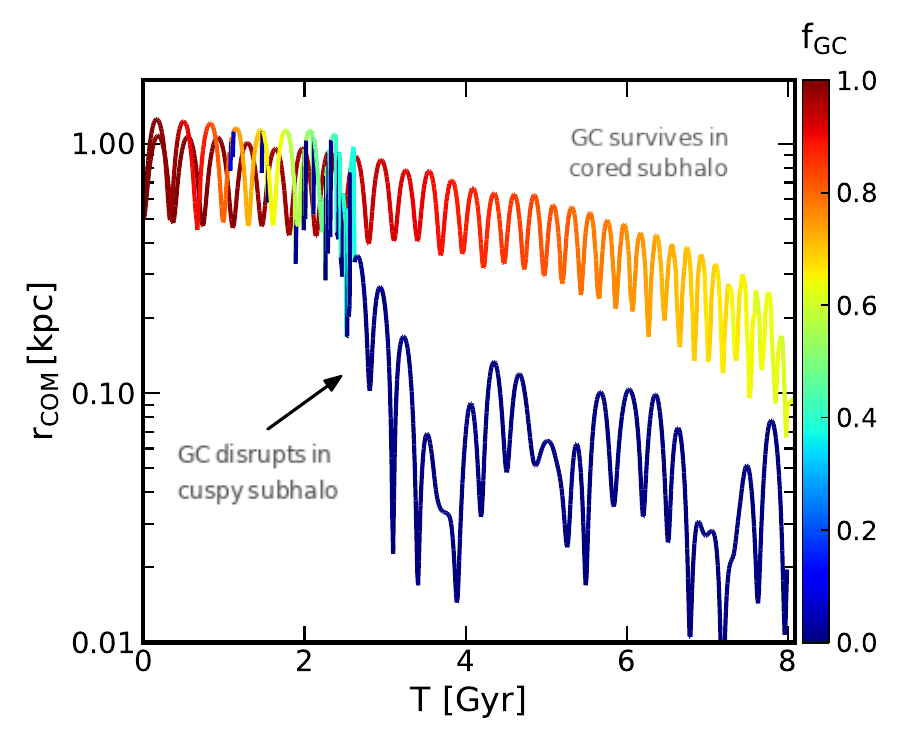}
\includegraphics[width=0.44\hsize]{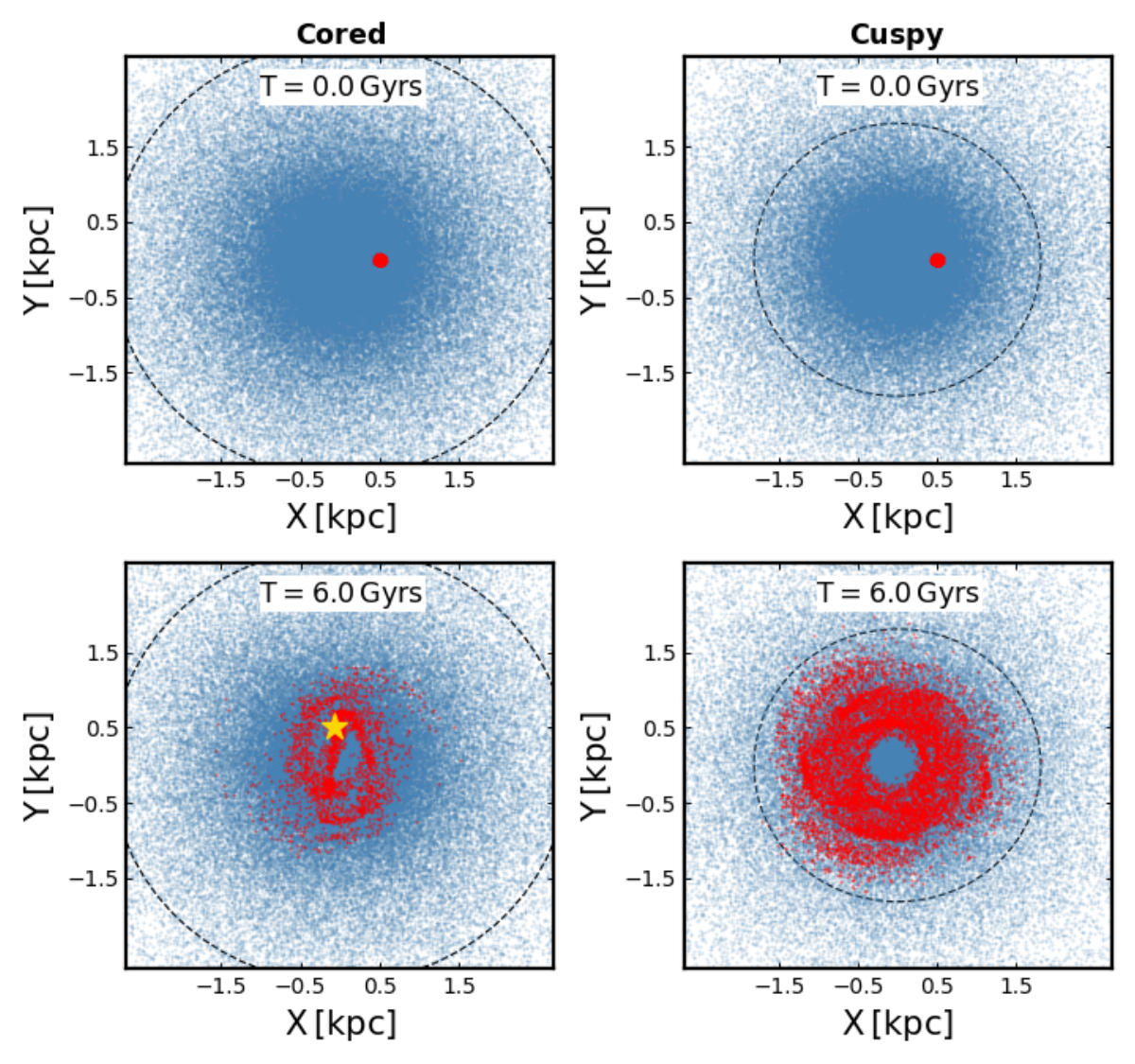}
}
\end{center}
\vspace{-0.8cm}
\caption{{\it Left panel:} The evolution of the orbital radius of the GC moving in core and cuspy subhalo models. Sinusoidal pattern is indicative of slightly eccentric orbit. Some of the fluctuations in the case of the cuspy subhalo at later times ($T>2\Gyr$) are due to the uncertainty in determining the center of the N-body GC as it has almost completely disrupted. The curves are colored according to $f_{\rm GC}$, which defines the fraction of stars bound to the GC at a given time (more precisely, the algorithm looks for an overdensity of stars). {\it Right panels}: Evolution of the GC (red particles) inside core (left) and cuspy (right) DM subhalo (blue particles). The upper and bottom panels show the distribution of GC stars at $T=0\Gyr$ and $6\Gyr$. The `star' marks the location of the surviving cluster in the case of cored subhalo, which disrupts completely in the case employing cuspy subhalo. The dashed circles mark the locus of the half mass radii $(r_{1/2})$.}
\label{fig:Fig_NumericalConvergence}
\end{figure*}

We ran some test simulations in order to ensure that our simulations are executed at appropriate resolution. This is important so as to verify that the GC in the cuspy and cored cases behave as predicted by theory and previous simulations -- especially that the rate of disruption of a GC should be higher in a cuspy subhalo compared to a cored subhalo (see the main text). To this end, we tried two different mass models, one cusped and one cored. In both cases we took $M_0=10^8\msun, r_0=0.75\kpc$, $ r_p=0.5\kpc$, but for cuspy model we adopted $v_t=18\kms$ and for cored model we chose $11\kms$. We specifically launch GCs on eccentric orbits inside the subhalos as it has been previously shown that convergence is most difficult for an eccentric orbit in a cored halo, because the two-body noise in a simulation can cause the cluster orbit to precess and cause artificial decay of the orbit once in the core \citep{Read2005cusp2core}. These simulations were executed with the same resolution as described in Section~\ref{subsec:Nbody_simulation_setup} and as isolated systems, i.e., in the absence of the tidal field of the host galaxy.

The dynamical evolution of the GC in the two subhalos is shown in Figure~\ref{fig:Fig_NumericalConvergence}. The left panel shows the evolution of the orbital radius of the center of mass of the cluster $(r_{\rm COM})$ in the two cases. Due to large tidal forces present in the cuspy model, the cluster instantly gets disrupted (in $T\approx2\Gyr$), and therefore, mathematically, $r_{\rm COM} \rightarrow 0$. For conveying the information about the on-going disruption of the N-body GC, the curve is colored according to the parameter $f_{\rm GC}$ that roughly defines the bound mass fraction of the GC at a given time instant. Some of the fluctuations in the case of the cuspy subhalo at later times ($T>2\Gyr$) are due to the uncertainty in determining the center of the N-body GC as it has completely disrupted, and under such a scenario the algorithm regards any localised overdensity in stars as the location of the GC. On the other hand, in the cored subhalo case, the GC only slightly disrupts.

Right panels of Figure~\ref{fig:Fig_NumericalConvergence} provide a visual comparison between the evolution of the GC in the cored and cuspy subhalo models. In case of cuspy subhalo, where the GC completely disrupts within finite time, the stellar debris get distributed in the form of a ring. But in the cored case, the tidal disruption of the GC occurs at slower rate, and therefore, the GC survives.

Further, notice that in Section~\ref{subsec:Nbody_simulation_setup} we ensure that the GC survives in the cuspy subhalos for atleast $3-4\Gyr$ (so that the GC can further disrupt in the main halo and form the narrow stream component). This we do by adopting suitable values of $v_t$. Here, for the same initialising values of the GC, we observe that the GC disrupts in merely $T\approx 2\Gyr$. This occurs because the subhalo's potential is fixed in this isolated case. However, in the cases we present in the main text of the paper, there the subhalos disrupt under the Galactic potential of the host. This leads to the scouring of the subhalo, which results in the decrease in mass (or gravitational potential) of the subhalo. Consequentially, this leads to diminution of the tidal forces of the subhalo onto the GC, and the GC survives for a relatively longer period of time. 

\section{Case of GC with small initial orbital radius in the Subhalo}
\label{appendix:small_rp}

Our studies in the main body of the text did not include the case of $r_p << r_0$, where  $r_p$ is the initial orbital radius of the GC in the subhalo and $r_0$ is the core radius of the subhalo.   One might worry that our results might be falsified by this case.  However, we here show that our work applies quite generally.

\begin{figure}
\begin{center}
\includegraphics[width=\hsize]{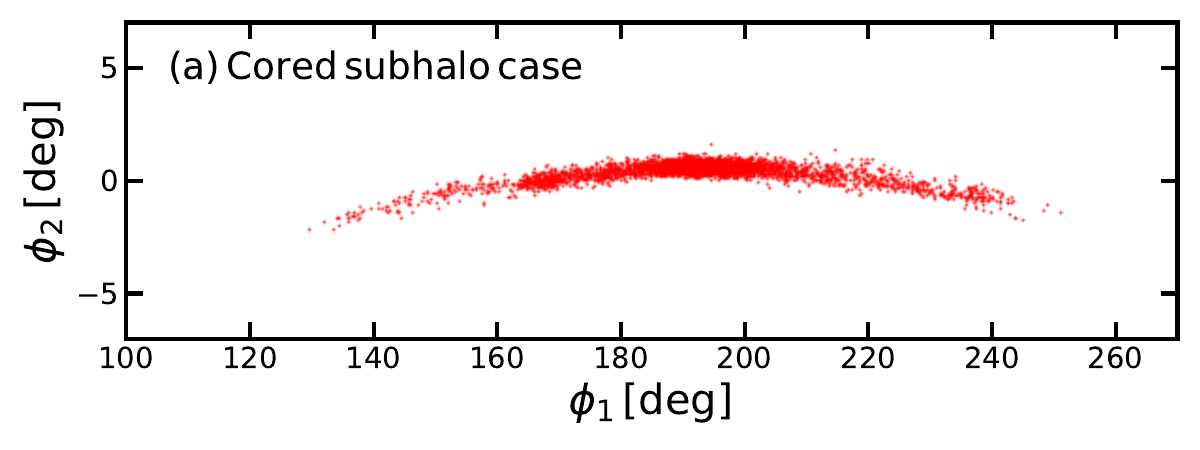}
\includegraphics[width=\hsize]{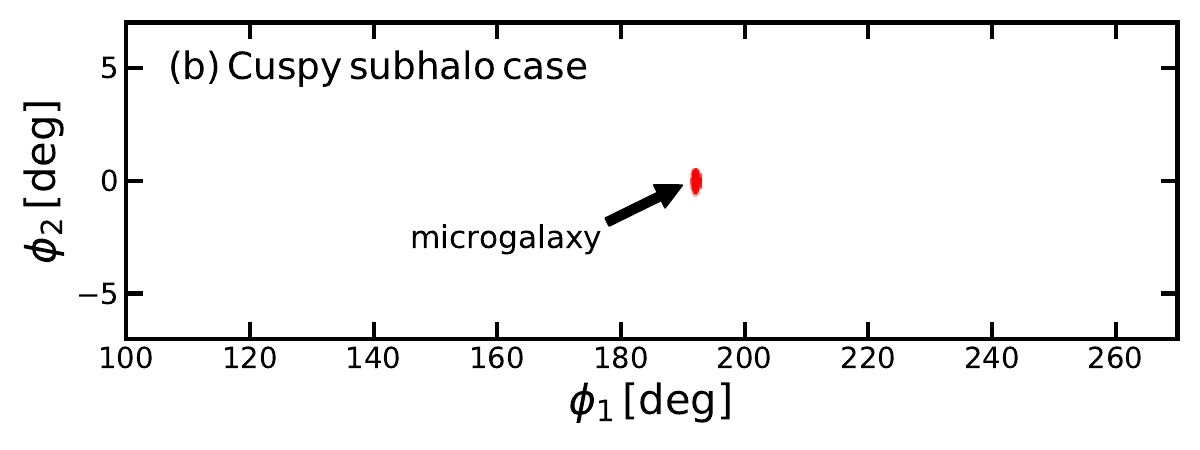}
\end{center}
\vspace{-0.5cm}
\caption{Cases of GCs with small initial orbital radius inside subhalos (i.e., $r_p<<r_0$, see Sec.~\ref{appendix:small_rp}). a) Cored case: The resulting GC stream is only slightly broad and its properties were noted to agree with streams shown in Figures~\ref{fig:Fig_core_streams} and \ref{fig:Fig_comparison}. b) Cuspy case: No stream is formed as GC disrupts entirely inside the subhalo and stars do not escape the potential well of the subhalo. This instead leads to the formation of only a microgalaxy structure.}
\label{fig:Fig_rpssr0}
\end{figure}

We present results here of simulations for the case $r_p << r_0$.
Figure~\ref{fig:Fig_rpssr0} shows the results for the the two cases of 
cored (upper panel) and cuspy (lower panel) subhalos.  In both cases, we took
$M_0=10^8\msun, r_0=0.75\kpc$, $ r_p=0.1\kpc$ and $v_t=10\kms$.  In the cored case of the top panel of Figure~\ref{fig:Fig_rpssr0}, one can see that the resulting stream is only slightly broad in agreement with the results of Figures~\ref{fig:Fig_core_streams} and \ref{fig:Fig_comparison} for GC in cored subhalos.
 Thus cored subhalos produce streams of similar widths and velocity dispersions regardless of the value of the ratio $r_p/r_0$.

In the cuspy case of the lower panel of Figure~\ref{fig:Fig_rpssr0}, there is no stream formation.  Instead, although the GC has disrupted, the stars have been unable to escape the potential of the subhalo.  Instead they remain inside the residual of the cusp of the subhalo.  Thus the case of $r_p << r_0$ for cuspy subhalos produces no streams at all.

In conclusion, although the studies in the main body of the paper did not include the case of $r_p << r_0$, , our results apply more generally to all ratios of $r_p/r_0$ for the analysis of observed streams.

\section{Mass loss of subhalos on orbits of various eccentricities inside the host galaxy}
\label{appendix:mass_loss}

Here, we briefly study the dynamical evolution of subhalo models launched on eccentric orbits in Section~\ref{sec:discussion}.  

\begin{figure}
\begin{center}
\includegraphics[width=\hsize]{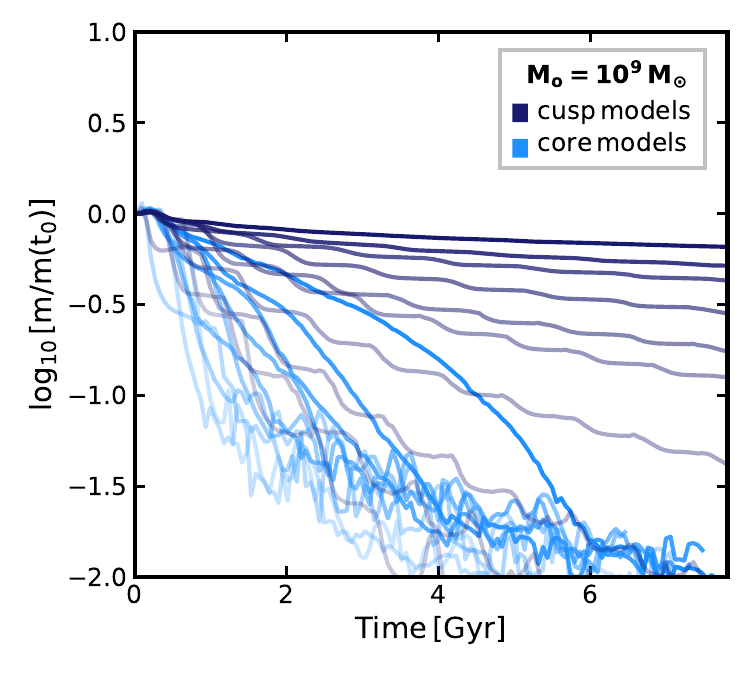}
\end{center}
\vspace{-0.6cm}
\caption{Bound mass fraction as a function of time for the subhalos launched on varied eccentric orbits inside the host galaxy. Different colours are used to differentiate between cuspy and cored subhalo models. Darker (lighter) contrasts represent low (high) orbital eccentricities. Fluctuations can be seen in the cored case which occur due to the dramatic scouring of the subhalos that makes it almost impossible to clearly define its boundary at each time instant. }
\label{fig:Fig_sub-halo_mass_evolution_eccentricites}
\end{figure}

Figure~\ref{fig:Fig_sub-halo_mass_evolution_eccentricites} shows the logarithm of the fraction of mass that remains bound to the subhalos as they orbit inside the host galaxy. This bound mass fraction refers to the ratio of the initialised mass (i.e., $M_0$) to the dynamic mass (calculated at a given time within $r=3\kpc$ of the subhalo). This gives us an approximate means to understand the mass loss in the subhalos launched on different orbital configurations inside the host. The cuspy (cored) subhalo models are shown with dark (light) blue colours. Further, the lower (higher) eccentricity orbits are shown with darker (lighter) shades. 

For nearly circular orbits (low eccentricities), we observed that the cuspy subhalos do not completely disrupt and almost always retain a bound remnant even after losing a substantial fraction of mass. This occurs due to their steep inner density profiles that makes them much resilient to the tidal field of the host (c.f. \citealt{Kazantzidis2004, Penarrubia2010} where this phenomenon has been studied in detail). However, as the orbits become more radial (high eccentricities), the cummulative effects of gravitational shocking and tidal stripping by the host becomes significant, and the subhalo experiences severe disruption. This explains the increase in the slope of the curves with increase in the values of eccentricities in Figure~\ref{fig:Fig_sub-halo_mass_evolution_eccentricites}. For eccentricity=0.9, we found that the cuspy subhalo of $M_0=10^9\msun$ completely disrupted in $\sim 4-5\Gyr$. 

In contrast, cored subhalos launched on nearly circular orbits were observed to undergo complete disruption at finite timescales ($T\sim5\Gyr$). In this case, very steeply decaying curves can be seen in Figure~\ref{fig:Fig_sub-halo_mass_evolution_eccentricites}. The fast decay of these curves is most evident in the cases where the cored subhalos were launched on orbits with higher eccentricities. This happens because radial orbits bring subhalos in the vicinity of the disk where they experience huge amount of tidal mass loss. Soon after their first pericentric passage, the binding energy of the subhalos significantly decreases, causing them to disrupt at lower timescales. We observed that the cored subhalos launched on highly radial orbits disrupted within $T\sim2\Gyr$.

\section{Time evolution of accreted GC streams}
\label{appendix:time_evolution}

\begin{figure}
\begin{center}
\includegraphics[width=1.03\hsize]{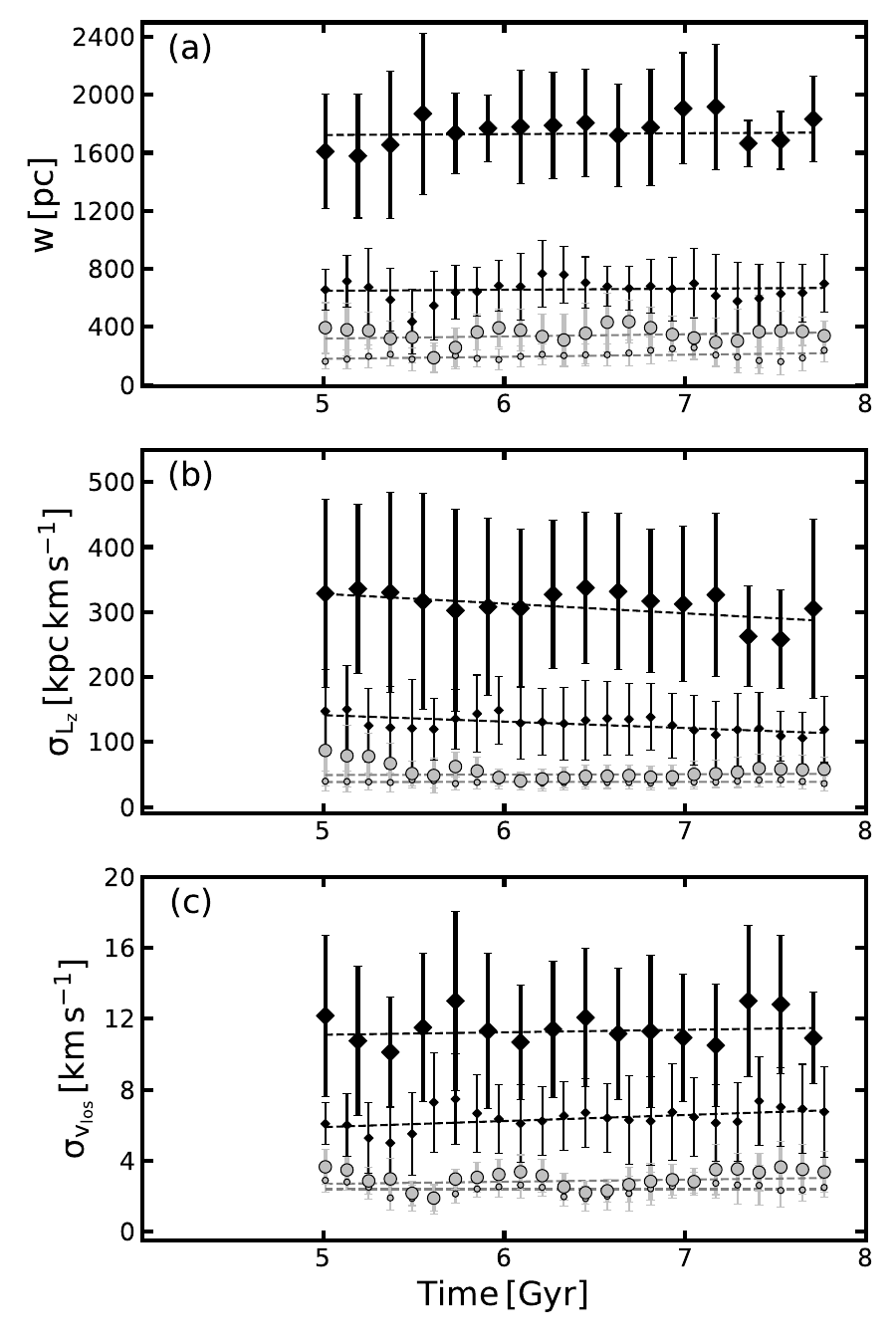}
\vspace{-0.9cm}
\end{center}
\caption{Evolution of the physical parameters of representative streams for each subhalo model. The markers have the same meaning as in Figure~\ref{fig:Fig_comparison}. X axis denotes time, and we start plotting with an instance that is greater than the accretion time of the GCs. The straight lines are linear fits made to the points corresponding to different stream case.}
\label{fig:Fig_properties_time_evol}
\end{figure}

In Figure~\ref{fig:Fig_properties_time_evol}, we pick a random simulation from each of SCo, LCo, SCu and LCu cases, and study the evolution of the structural and dynamical properties of streams. To this end, we wish to understand whether the physical properties of accreted GC streams change significantly with time, and whether they can be used to probe the density profiles inside their parent subhalos at every phase of their evolution. 

For a given subhalo model (shown by specific marker style), $w,\sLz, \sv$ (along with their uncertainties) are calculated at different time steps of the simulation. The straight line fits with nearly zero slopes in Figure~\ref{fig:Fig_properties_time_evol} suggest that  these physical parameters of streams change only little with time. For cored subhalos, we found change in $w,\sLz, \sv$ by $\sim 8\% (5\%),0\% (1\%),1\% (4\%)$ per Gyr for SCo (LCo) model. Similarly, for cuspy cases, we observed the change in the same quantities by $\sim 1\% (1\%),6\% (4\%),5\% (1\%)$ for SCu (LCu) model. 

\section{Computing dynamical width estimators for the GD-1 stream}
\label{appendix:orbit_GD1_cocoon}

Computing the dynamical width estimator $\sLz$  is straightforward for simulated streams since all phase space coordinates for all the star particles are precisely known. However, making this measurement for observed streams is challenging because, despite the availability of parallaxes from Gaia, the estimated distances of individual stars are not accurate (especially if a star is not very bright). In effect, the resulting distance estimates have large uncertainties and this translates into large uncertainties for the velocities in the plane of the sky (derived from proper-motions and distances). Furthermore, due to the unknown distance gradient along a stream, as is measured for GD-1, photometric distances to individual stars may be biased. Consequently, a straightforward computation of $L_z$ for each star in the observed stream, and a computation of $\sLz$, is currently not possible.

An alternative, physically motivated, method is to fit an orbit to a given stream (for e.g., as done in  \citealt{Ibata_Phlegethon_2018} to compute $\sLz$ for the Phlegethon stream). Essentially, an orbit fitting procedure samples orbits in an assumed gravitational potential model of the host galaxy and finds a suitable representative orbital model that fits the data for the observed stream in all the 6 phase space coordinates. The procedure also takes into account the observational uncertainties. Since an orbit is defined by energy $E$ and angular momentum  $L_z$  (for an axisymmetric potential), the dispersion in the sampled orbits can be translated to dispersions in dynamical quantities such as $L_z$. Here, we compute $\sLz$ for the GD-1 structure in a similar manner.

\citet{Malhan2018PotentialGD1} (\citetalias{Malhan2018PotentialGD1} hereafter) implemented an orbit-fitting routine to a sample of GD-1 stars (belonging only to the narrow component) in order to constrain the gravitational potential of the MW. In order to calculate $\sLz$ for the overall GD-1 structure (comprising of the narrow and the \cocoon component, discovered in \citetalias{MalhanCocoonDetection2019}), we use the same Galactic potential that was derived in the \citetalias{Malhan2018PotentialGD1} study. The \citetalias{Malhan2018PotentialGD1} potential is slightly different from the one employed in the present study, and has a circular velocity at the Solar radius of $V_{\rm circ}(R_{\odot})  =  244\kms$, and a density flattening of the dark  halo  as $q_{\rho}=  0.82$. However, we deem that the differences in potential should not significantly affect the comparison with simulated streams in Figure~\ref{fig:Fig_comparison}. We followed the same procedure (with identical likelihood function) as presented in \citetalias{Malhan2018PotentialGD1} to fit an orbit to a sample of GD-1 stars from \citetalias{MalhanCocoonDetection2019} for which complete 6D phase-space information and photometric information is available.

As expected, the best-fit orbit for the GD-1+cocoon structure looked similar to the one shown in Figure~7 of \citetalias{Malhan2018PotentialGD1}. The best-fit orbit has $L_z\approx2930\kpc\kms$, which is consistent with that found for the narrow component of GD-1 ($L_z\sim2950\kpc\kms$, \citetalias{Malhan2018PotentialGD1}). This procedure allowed us to compute the dynamical width estimator for GD-1+cocoon structure as $\sLz \approx 40\kpc\kms$. %

\bsp	
\label{lastpage}
\end{document}